\documentclass{aa}  

\usepackage{graphicx}
\usepackage{txfonts}
\usepackage{amsmath}
\usepackage{bm}

\newcommand{\dif}{\mathrm{d}}

\begin{document}

   \title{Resonant chains in triple-planetary systems}

   \author{Xuefeng Wang\inst{1,2}
          \and
          Li-Yong Zhou\inst{1,2}
        \and
        Cristian Beaug\'e\inst{3} 
         }
        \authorrunning{Wang, Zhou \& Beaug\'e}
        \offprints{L.-Y. Zhou, \email zhouly@nju.edu.cn}

   \institute{School of Astronomy and Space Science, Nanjing University, 163 Xianlin Avenue, Nanjing 210046, PR China
       %\email{zhouly@nju.edu.cn}
         \and
        Key Laboratory of Modern Astronomy and Astrophysics in Ministry of Education, Nanjing University, Nanjing 210046, PR China
        \and
        Instituto de Astronom\'ia Te\'orica y Experimental (IATE), Observatorio Astron\'omico (OAC), Universidad Nacional de C\'ordoba, Laprida 854, C\'ordoba, Argentina
             }

   \date{Received  ; accepted }

% \abstract{}{}{}{}{}
% 5 {} token are mandatory
 
  \abstract
{The mean motion resonance is the most important mechanism that may dominate the dynamics of a planetary system. In a multi-planetary system consisting of $N\ge3$ planets, the planets may form a resonant chain when the ratios of orbital periods of planets can be expressed as the ratios of small integers $T_1: T_2: \cdots :T_N=k_1: k_2: \cdots: k_N$. Due to the high degree of freedom, the motion in such systems could be complex and difficult to depict. }
%aims
{We investigate in this paper the dynamics and possible formation of the resonant chain in a triple-planetary system.}
%methdo
{We define the appropriate Hamiltonian for a three-planet resonant chain and numerically average it over the synodic period. The stable stationary solutions (apsidal corotational resonance, ACR) of this averaged system, corresponding to the local extrema of the Hamiltonian function, can be searched out numerically. The topology of the Hamiltonian around these ACRs reveals their stabilities. We further construct the dynamical maps on different representative planes to study the dynamics around the stable ACRs, and we calculate the deviation ($\chi^2$) of the resonant angle in the evolution from the uniformly distributed values, by which we distinguish the behaviour of critical angles. Finally, the formation of the resonant chain via convergent planetary migration is simulated and the stable configurations associated with ACRs are verified. }
%results
{We find that the stable ACR families arising from circular orbits always exist for any resonant chain, and they may extend to high eccentricity region. Around these ACR solutions, regular motion can be found, typically in two types of resonant configurations. One is characterised by libration of both the two-body resonant angles and the three-body Laplace resonant angle, and the other by libration of only two-body resonant angles. The three-body Laplace resonance seems not to contribute much to the stability of the resonant chain. The resonant chain can be formed via convergent migration, and the resonant configuration evolves along the ACR families to eccentric orbits once the planets are captured into the chain. Ideally, our methods introduced in this paper can be applied to any resonant chain of any number of planets at any eccentricity. }
{}


   \maketitle
%
%________________________________________________________________

\section{Introduction}
Thousands of exo-planetary systems, over $1/4$ of which are multi-planetary systems, have been detected. According to their orbital architectures, these multi-planetary systems are categorized into separable, compact and resonant planetary systems. The separable (secular) systems are protected from mutual collision and close encounter by the low angular momentum deficit (AMD), i.e. the AMD-stability \citep{laskar2017}. The compact multi-planetary systems however, even for those with circular orbits, are generally unstable. \citet{petit2020} show how the slow chaotic diffusion due to the overlap of three-body resonances dominates the timescale leading to the instability for initially coplanar and circular orbits, and such mechanism reproduces very well the qualitative behaviour found in numerical simulations \citep{hussain2020fundamental}.  The resonant systems, in which planets are found in mean motion resonance (MMR) with each other, are likely to be stable and of particular interest because they may bear important clues to the formation and evolution of planetary systems. For the case of two planets in an MMR, the dynamics are well understood through the resonant periodic orbits \citep[e.g.][]{voyatzis2008, voyatzis2009} and the apsidal corotational resonance (ACR) \citep[e.g.][]{michtchenko2008}. But the resonant structure could be much more rich and complex in multi-planetary systems. 

Except for the two-body MMR involving two planets and the hosting star, the three-body resonance (three planets are involved) also plays its roles \citep[e.g.][]{petit2021, cerioni2022}. Particularly, a multi-planetary system with orbital periods of neighbouring planets librating at a sequence of near-integer ratio (2/1, 3/2, 4/3, etc.), often referred as in a resonant chain, is of great interest. The well-known examples include three planets of Kepler-51 in a 1:2:3 resonant chain \citep{masuda2014}, four planets in a 1:2:4:8 resonant chain in HR8799 \citep{gozdziewski2020exact} and a 1:2:5:7 resonant chain in K2-32 \citep{heller2019transit}, five planets of TOI-178 in a 2:4:6:9:12  \citep{leleu2021six} and a 4:6:9:12:18 resonant chain in Kepler-80 \citep{macdonald2021five}, and another five planets composing a resonant chain of 3:2 MMRs in K2-138 \citep{christiansen2018k2, lopez2019exoplanet}. In these systems, in addition to the transit light-curves, transit-timing variation (TTV) and radial velocity data, the dynamics of the resonant chain have also been taken into account to improve the planetary parameters (mass, orbital period, etc.)
%Moreover, the planet-disk interactions along with the tidal effects are included to reproduce the observed configurations (especially the period ratios).

Plenty of numerical studies have been devoted to the formation history and dynamical stability of multi-planetary systems with planets locked in or close to a resonant chain.  At the early stage of planetary system's formation, planets can be captured into a resonant chain through the convergent orbital migration. The tidal effect and the accretion of matter (increasing planetary mass) may influence the further evolution, driving the planetary system away from the nominal positions of resonances, and destabilizing the resonant chain via the secondary resonance. The final surviving systems exhibit complex resonant structures involving in two-body MMR, three-body MMR, resonant chain, secular resonance, etc \cite[e.g.][]{macdonald2018three, pichierri2019role, morrison2020chains, pichierri2020, macdonald2022confirming, Wong2024}. 

Due to the high degree of freedom, the analytical method is difficult to be applied to such systems. \citet{delisle2017} proposes an analytical model for any resonant chain and successfully predicts the resonant configuration of Kepler-223 \citep{mills2016}. And recently, \citet{antoniadou2022} extend the methodology for a pair of resonant exoplanets in coplanar orbits \citep{antoniadou2016orbital, antoniadou2016regular} to the case of triple massive coplanar planets in a resonant chain (general 4 body problem) and particularly, compute the resonant periodic orbits for Kepler-51 system. Based on the stable periodic orbits of the 1:2:3 resonant chain, they demonstrate three possible scenarios that can safeguard the Kepler-51 planetary system. The comprehensive study on the stable periodic orbits can provide an optimum deduction of orbital elements, in addition to the observational data fitting. 

In this paper, we generalize the semi-analytical model for analysing the periodic orbits in the coplanar two-body MMR first developed by \citet{michtchenko2006} to the case of resonant chain. For the sake of explaining our methods clearly, we mainly take the 1:2:3 resonance chain in the Kepler-51 system as an example in our analyses. But we emphasize here that the methods and some conclusions presented in this paper are general and can be applied to any other resonant chains of three or even more planets. In fact, we also show briefly our analyses on some other resonant chains in the end of this paper.  

The rest part of this paper is organized as follows. In Section 2, we revisit the definition of resonant periodic orbits and generalize the definition to arbitrary resonant chains. We introduce in Section 3 our semi-analytical Hamiltonian method and the determination of apsidal corotational resonance (ACR) in the resonant chain. Then this methodology is applied to the case of 1:2:3 resonant chain, and the periodic orbits and ACRs are obtained. In Section 4, the orbital dynamics around the stable ACRs are investigated. We study how the orbital configuration of planets in a resonant chain can be achieved via convergent planetary migration in Section 5. Finally, we draw our conclusion in Section 6.
%__________________________________________________________________

\section{The model} \label{sec:model}

\subsection{Rotational frame and resonant periodic orbits}
The motion of planets in a resonant chain is often described in a rotational frame. Consider three massive planets $m_1, m_2, m_3$ from the inside out orbiting around the central star $M$ on the same plane, with their orbital periods $T_1, T_2, T_3$ satisfying the commensurable condition
\begin{equation}
T_1:T_2:T_3=k_1:k_2:k_3,~~\text{with } k_1,k_2,k_3\in\mathcal{Z}\,.
\end{equation}
Following \citet{antoniadou2022}, a non-inertial frame ($GXY$) is introduced as follows. Let $G$ coincide with the mass centre of $M$ and $m_1$, the axis $GX$ always points to $m_1$ from $M$, and $GY$ be perpendicular to $GX$. In this rotating frame, $m_1$ and $M$ are always on the $GX$ axis, while $m_2$ and $m_3$ revolve about $m_1$ with the relative frequencies of $n_2-n_1$ and $n_3-n_1$, where $n_1, n_2, n_3$ are the mean motions of three planets. The relative frequencies satisfy the near-integer commensurability:
\begin{equation}
\frac{n_2-n_1}{n_3-n_1}=\frac{p}{q}, ~~\text{with }  p,q\in\mathcal{Z}\,.
\end{equation}
The period $T$ of the system, during which $m_2$ revolves $p$ times while $m_3$ revolves $q$ times both around $m_1$, can be easily derived as
\begin{equation}
T=\frac{T_1}{1-T_1/T_2}p=\frac{T_1}{1-T_1/T_3}q\,.
\end{equation}
In practice, such periodicity requires that the planets $m_1, m_2, m_3$ repeat their configuration after time $T$. Therefore, the period $T$ should be related to the least common multiple (LCM) of $T_1, T_2, T_3$, and such periodic condition can be rewritten in a more general form
\begin{equation}
\label{eq:T}
T/T_i=\text{LCM}(k_1,k_2,k_3)/k_i, ~~\text{for } i=1, 2, 3\,.
\end{equation}
For the 1:2:3 resonant chain in the planetary system Kepler-51, Eq.\eqref{eq:T} gives $\text{LCM}(1,2,3)=6$, and $T= 6T_1= 3T_2= 2T_3$, which is equivalent to $p=3, q=4$ and $T=6T_1$ \citep{antoniadou2022}. The periodicity defined in Eq.\eqref{eq:T} can be generalized to resonant chain of $N$ $(N>3)$ planets (e.g. a four-planet resonant system Kepler-233 can be found in Appendix~\ref{appendix2}) as
\begin{equation}
T/T_i=\text{LCM}(k_1, k_2, \cdots, k_N)/k_i, ~~\text{for } i=1,2,\cdots, N\,.
\end{equation}

These periodic orbits can be found in the exact (nominal) location of the MMRs. In addition, according to different orbital configurations, these periodic orbits can be divided into two groups, namely the symmetric periodic orbits and the asymmetric ones. A periodic orbit is symmetric with respect to the $GX$ axis of the rotating frame if it remains invariant under the fundamental symmetry \citep{henon2003generating} $\Sigma:(t,X,Y)\rightarrow (-t,X,-Y)$ 
\begin{equation} \label{eq:symm}
\begin{aligned}
& X_i(-t)=X_i(t), &  & Y_i(-t)=-Y_i(t),   &\\
&\dot{X}_i(-t)=-\dot{X}_i(t), &  & \dot{Y}_i(-t)=\dot{Y}_i(t), &~~i=1,2,3.
\end{aligned}
\end{equation}
This indicates that all the two-body MMRs should take the symmetric configuration. Otherwise, if one of the symmetry condition breaks, i.e. Eq.\eqref{eq:symm} is not satisfied for any $i$, then the configuration is referred as asymmetric. The relationship between symmetric condition and the choice of the resonant angles can also be found in e.g. \citet{voyatzis2008}.

\subsection{Hamiltonian description}
The Hamiltonian of the planar three-planet (from the inside out $m_1, m_2, m_3$) system takes the form \citep[see e.g.][]{laskar1991}
\begin{equation}
\label{eq:hamil}
\begin{aligned}
&\mathcal{H}=\mathcal{H}_0+\mathcal{H}_1\\
&~~~~\mathcal{H}_0=\sum_{i=1}^3\left(\frac{\bm{p}_i^2}{2M}-\mathcal{G}\frac{Mm_i}{|\bm{r}_i|}\right) , \\
&~~~~\mathcal{H}_1=\sum_{1\le i< j\le 3}\frac{\bm{p}_i\cdot\bm{p}_j}{M}-\frac{\mathcal{G}m_im_j}{|\bm{r}_i-\bm{r}_j|}, 
\end{aligned}
\end{equation}
where $\mathcal{G}$ is the gravitational constant, $\bm{r}_i$ is the position vector of the $i$-th planet, and $\bm{p}_i$ is the canonically conjugated momentum. The $\mathcal{H}_0$ is the sum of Keplerian part of each planet, while $\mathcal{H}_1$ describes the interaction between planets.
The Poincar\'{e} canoncial variables in astrocentric coordinate are
\begin{equation} \label{eq:poinvar}
\left\{ \begin{aligned}
&\Lambda_i=\beta_i\sqrt{\mu_ia_i}, & &\lambda_i, \\
&\Gamma_i=\Lambda_i(1-\sqrt{1-e_i^2}),& &\gamma_i=-\varpi_i,
\end{aligned}\right. 
\end{equation}
where $\mu_i=\mathcal{G}(M+m_i)$, $\beta_i=m_iM/(m_i+M)$,  and $a_i, e_i, \varpi_i, \lambda_i$ are the semi-major axis, eccentricity, longitude of  periastron and the mean longitude of  the $i$-th planet $m_i$, respectively. The action variable $\Lambda_i$ is the circular angular momentum and  $\Gamma_i$ is often referred as the AMD \citep[see e.g.][]{petit2017}. 

For two planets $m_i, m_j$ with mean motions $n_i, n_j$, we denote by $k_{ji}/k_{ij}=n_i/n_j=T_j/T_i$ the relative period ratio, and $q_{ij}=q_{ji}=|k_{ji}-k_{ij}|$ the order of the resonance, where $k_{ij}$ and $k_{ji}$ are integers. The two-body resonant angles $\theta_{ij},\Delta\varpi_{ij}$ are defined as 
\begin{equation} \label{eq:critang}
\theta_{ij}=k_{ji}\lambda_j-k_{ij}\lambda_i-(k_{ji}-k_{ij})\varpi_i, ~~~~\Delta\varpi_{ij}=\varpi_i-\varpi_j.
\end{equation}

For planetary systems near MMRs, \citet{delisle2017} introduces the canonical resonant variables for arbitrary resonant chain (with any number of planets, in resonances of any degree). Particularly, the resonant variables of the 1:2:3 resonant chain are
\begin{equation}\label{eq:res_var}
\left\{\begin{aligned}
&\sigma_1=3\lambda_3-2\lambda_2-\varpi_1, & & \Gamma_1,\\
&\sigma_2=3\lambda_3-2\lambda_2-\varpi_2, & & \Gamma_2,\\
&\sigma_3=3\lambda_3-2\lambda_2-\varpi_3, & & \Gamma_3,\\
&\varphi_0=3\lambda_3-4\lambda_2+\lambda_1, & & L_0=\Lambda_1,\\
&\varphi_G=3\lambda_3-2\lambda_2, & & G=\sum_{i=1}^3\Lambda_i-\Gamma_i, \\
&Q=Q_{23}=\lambda_2-\lambda_3, & & J=6\Lambda_1+3\Lambda_2+2\Lambda_3.
\end{aligned}\right.
\end{equation}
The first three angles in Eq.\eqref{eq:res_var} are resonant angles conjugated to each planet's  AMD, and $\varphi_0$ is the angle of Laplace resonance (0-th) between three planets. Since $\varphi_G$ does not appear explicitly in the Hamiltonian \citep[see details e.g. in][]{delisle2017}, and
 the corresponding conjugated momentum $G$ is constant, implying the preservation of angular momentum. 
 Thus, the problem is in fact of five degree-of-freedom (DoF).
The last angle $Q$ is the fast angle with respect to the first four resonant angles $(\sigma_1, \sigma_2, \sigma_3, \varphi_0)$, thus it can be removed by numerical averaging \citep[see e.g. Gauss-Legendre formula, ][]{press1996}. The averaging process guarantees that $J$ is another constant, which is similar to the ``spacing parameter'' in the planar two-body MMR. Moreover, the integral constant $J$ in triple-planet resonant chain $T_1:T_2:T_3=k_1:k_2:k_3$ can be rewritten in a more general form $L$ (hereafter we refer $L$ as the ``spacing parameter'' in a resonant chain)
\begin{equation}
L=\sum_{i=1}^3\frac{\Lambda_i}{k_i}.
\end{equation}

For the $k_1:k_2:k_3=1:2:3$ resonant chain, during the period $T$ (see the definition in Eq.~\eqref{eq:T}), the angle $Q$ evolves  
\begin{equation}
Q^T=2\pi \times \text{LCM}(k_1, k_2, k_3) \left(\frac{1}{k_2}-\frac{1}{k_3}\right)=2\pi\,.
\end{equation}
Therefore, the averaged Hamiltonian $\bar{\mathcal{H}}$ can be defined as
\begin{equation}
\label{eq:aver_hamil}
\bar{\mathcal{H}}=\frac{1}{Q^T}\int_0^{Q^T}\mathcal{H}\dif Q\,.
\end{equation}
The resonant periodic orbits will appear as stationary solution of the averaged Hamiltonian Eq.~\eqref{eq:aver_hamil}.
After the numerical averaging process, the final Hamiltonion $\bar{H}$ is of 4 DoF in canonical variables $(\sigma_1,\sigma_2,\sigma_3,\varphi_0,\Gamma_1,\Gamma_2,\Gamma_3,L_0)$ parametrized by the two constants $(L,G)$. 

Seemingly, the resonance between $m_2$ and $m_3$ plays an essential role in Eq.\eqref{eq:res_var}, but it is
only due to the arbitrary choice of the canonical variables. For any different choice of resonant angles, the expression of the synodic 
angle $Q$ would be different. For example, the choices $Q=Q_{12}=\lambda_1-\lambda_2$, or $Q=Q_{13}=\lambda_1-\lambda_3$, the averaged Hamiltonian should be respectively calculated by:
\begin{equation}
\begin{aligned}
\bar{\mathcal{H}}=\frac{1}{Q_{12}^T}\int_0^{Q_{12}^T}\mathcal{H}\dif Q_{12}=
\frac{1}{Q_{13}^T}\int_0^{Q_{13}^T}\mathcal{H}\dif Q_{13}, \\
		Q_{12}^T=2\pi\times\text{LCM}(k_1,k_2,k_3)\left(\frac{1}{k_1}-\frac{1}{k_2}\right)=6\pi, \\
		Q_{13}^T=2\pi\times\text{LCM}(k_1,k_2,k_3)\left(\frac{1}{k_1}-\frac{1}{k_3}\right)=8\pi.
\end{aligned}
\end{equation}

For a better characterization of the resonant configuration, we perform the linear transformation to angular variables defined in Eq.\eqref{eq:res_var}, and the resonant angles $(\theta_{12}, \Delta\varpi_{12}, \theta_{23}, \Delta\varpi_{23})$ between consecutive planets can be obtained as:
\begin{equation}
\label{eq:angle123}
\left\{
\begin{aligned}
& \theta_{12}=2\lambda_2-\lambda_1-\varpi_1, & 
\Delta\varpi_{12}=\varpi_1-\varpi_2, \\
& \theta_{23}=3\lambda_3-2\lambda_2-\varpi_2, & 
\Delta\varpi_{23}=\varpi_2-\varpi_3.
\end{aligned} \right.
\end{equation}
The first two angles $\theta_{12}, \Delta\varpi_{12}$ are for the MMR between the inner two planets $m_1, m_2$ while $\theta_{23}, \Delta\varpi_{23}$ are for the outer planets $m_2, m_3$ respectively, just as defined in Eq.\eqref{eq:critang}. 

For isolated two-body MMRs, the resonant angles $(\theta_{ij}, \Delta\varpi_{ij})$ are all $2\pi$-periodic, i.e., the geometric configuration of MMR characterized by $(\theta_{ij}, \Delta\varpi_{ij})$ is identical to the one by $(\theta_{ij}+2k'\pi, \Delta\varpi_{ij}+2k''\pi)$, with $k', k'' \in\mathcal{Z}$.
For the 1:2:3 resonant chain, the averaged Hamiltonian Eq.\eqref{eq:aver_hamil} is $2\pi$-periodically dependent on all the angular variables defined in Eq.\eqref{eq:angle123}. According to the symmetry in the co-rotational frame $GXY$ with respect to $m_1$, the configuration, in which the resonant angles $(\theta_{12}, \Delta\varpi_{12}, \theta_{23}, \Delta\varpi_{23})$ take either $0$ or $\pi$, is symmetric, while other configurations are asymmetric. It should be noted that the above conditions may not always guarantee the proper periodicity and symmetric configuration for a general three-planet resonant chain if we choose the resonant angles between adjacent planets (rather $\theta_{i,i+1}$, $\Delta\varpi_{i,i+1}$ than  $\theta_{i,i+2}$, $\Delta\varpi_{i,i+2}$) to characterize the system. We leave a detailed discussion for this problem to Appendix \ref{appendix1}.

We note that the aforementioned procedure can be simply extended to systems containing $N$ $(N>3)$ planets, in which the two constants take the form:
\begin{equation}
\begin{aligned}
&G=\sum_{i=1}^N\beta_i\sqrt{\mu_ia_i(1-e_i^2)},\\
&L=\sum_{i=1}^N\frac{\beta_i\sqrt{\mu_ia_i}}{k_i}.
\end{aligned}
\end{equation}
And the resonant configuration for an $N$-planet system can be represented by $(\theta_{12}, \Delta\varpi_{12}, \cdots, \theta_{n-1,n}, \Delta\varpi_{n-1,n})$.
 
Now, searching for the resonant periodic orbits is equivalent to find the stationary solution of the averaged Hamiltonian with two free constants $L$ and $G$.

\section{Topology of Hamiltonian and stationary solutions}

The averaged Hamiltonian Eq.\eqref{eq:aver_hamil} has been reduced to 4 DoF. Due to the complex structure in high dimensional phase space (8 dimensional phase space for a planar three-planet system), searching for stationary solutions in the entire phase space is very difficult. Fortunately, the topology of the Hamiltonian $\bar{\mathcal{H}}$ can be understood from several representative planes (section of phase space), namely, the period ratio plane $(n_1/n_2, n_2/n_3)$, the eccentricity plane $(e_1, e_2)$ or $(e_2, e_3)$, and the angular planes $(\theta_{12}, \Delta\varpi_{12})$ and $(\theta_{23},\Delta\varpi_{23})$. 

Hereafter, as an example, we focus on a 1:2:3 resonant chain with the mass parameters of $M=1.0, m_1=2.1\times 10^{-6}, m_2=4.0\times 10^{-6}, m_3=7.6\times 10^{-6}$. The mass ratios between the planet pairs are $\rho_\text{i}=m_1/m_2=0.525$ and $\rho_\text{o}=m_2/m_3=0.526$ (subscripts ``i'' for inner and ``o'' for outer), which in fact is similar to the ones in the Kepler-51 system \citep{masuda2014}. We set the gravitational constant $\mathcal{G}=1$ and fix the semi-major axes of $m_2$ as $a_2=1$\,AU. As a result, the orbital period of $m_2$ is 1 year.

\subsection{Topology on period ratio plane $(n_1/n_2,n_2/n_3)$}

In a three-planet system, the planetary semi-major axes $(a_1, a_2, a_3)$ are constrained by the spacing parameter $L$, so that they can be uniquely determined by the period ratio pair $(n_1/n_2, n_2/n_3)$ when $L$ is given. Therefore, we can study the topology of the Hamiltonian on the period ratio plane $(n_1/n_2, n_2/n_3)$ with $e_1, e_2, e_3, \theta_{12}, \Delta\varpi_{12}, \theta_{23}, \Delta\varpi_{23}$ being appropriately fixed. 

\begin{figure}
   \centering
   \includegraphics[width=\hsize]{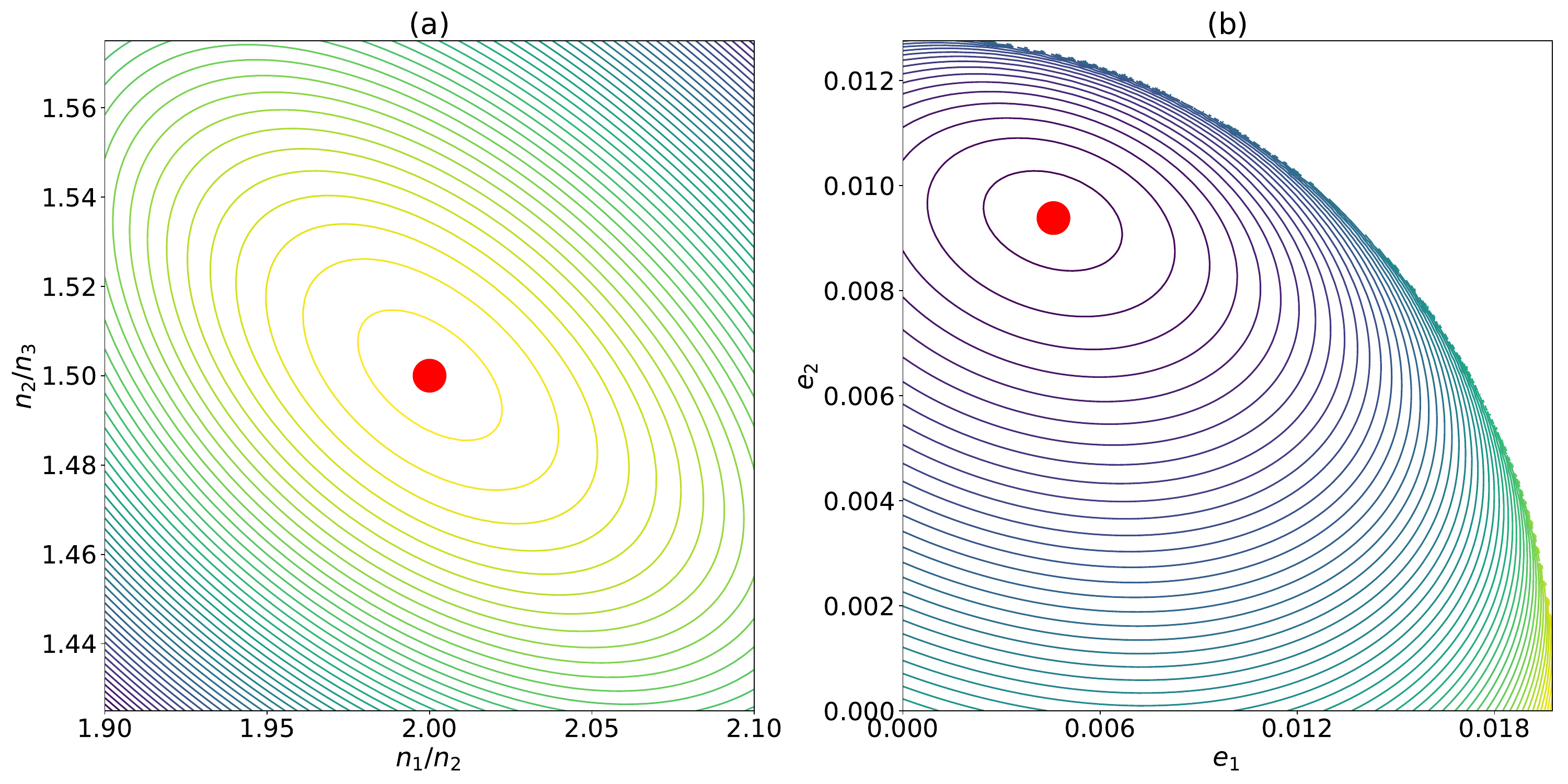}
      \caption{Topology of Hamiltonian on different representative planes. 
		(a) The topology on period ratio plane $(n_1/n_2, n_2/n_3)$ with fixed eccentricities. (b) The topology on eccentricity plane $(e_1, e_2)$. The stationary solution located at $(n_1/n_2, n_2/n_3)=(2,1.5)$ and $(e_1, e_2, e_3)=(0.00458, 0.00938, 0.00564)$ is marked by a red dot on both panels. The corresponding critical angles are $(\theta_{12}, \Delta\varpi_{12}, \theta_{23}, \Delta\varpi_{23}) = (\pi, 0, \pi, \pi)$. The colour indicates the value of Hamiltonian, light yellow for high and dark blue for low. The same colour code is adopted in all figures in this paper, unless otherwise stated. 
              }
         \label{fig:topo_ele}
   \end{figure}

An example of the topology on $(n_1/n_2, n_2/n_3)$ plane is shown in Fig.~\ref{fig:topo_ele}(a), for which the resonant angles $(\theta_{12}, \Delta\varpi_{12}, \theta_{23}, \Delta\varpi_{23}) = (\pi, 0, \pi, \pi)$ and eccentricities $(e_1, e_2, e_3)$ are fixed to $(0.00458,0.00938,0.00564)$.
The local extremum (maximum) is indicated by a red dot, and it is very close to the nominal value $(n_{1c}/n_{2c}, n_{2c}/n_{3c})=(2,1.5)$. We note that such structure always holds for other parameters and even in other resonance chains, thus we can fix the period ratios at the nominal value $(2, 1.5)$ as a first approximation and further discuss the topology on the eccentricity plane.

\subsection{Topology on eccentricity plane $(e_1,e_2)$}
Once the semi-major axes $a_1, a_2, a_3$ have been determined, the conservation of total angular momentum $G$ restricts the eccentricities $e_1, e_2, e_3$ to a two-dimensional plane, therefore we can consider the topology on a representative eccentricity plane, e.g. $(e_1, e_2)$ plane. An example is illustrated in Fig.~\ref{fig:topo_ele}(b). From this plot, a node-like stationary solution located at $(e_1, e_2) = (0.00458, 0.00938)$ can be found easily. We note that such stationary solutions are just ACRs.

\begin{figure}
	\centering
	\includegraphics[width=\hsize]{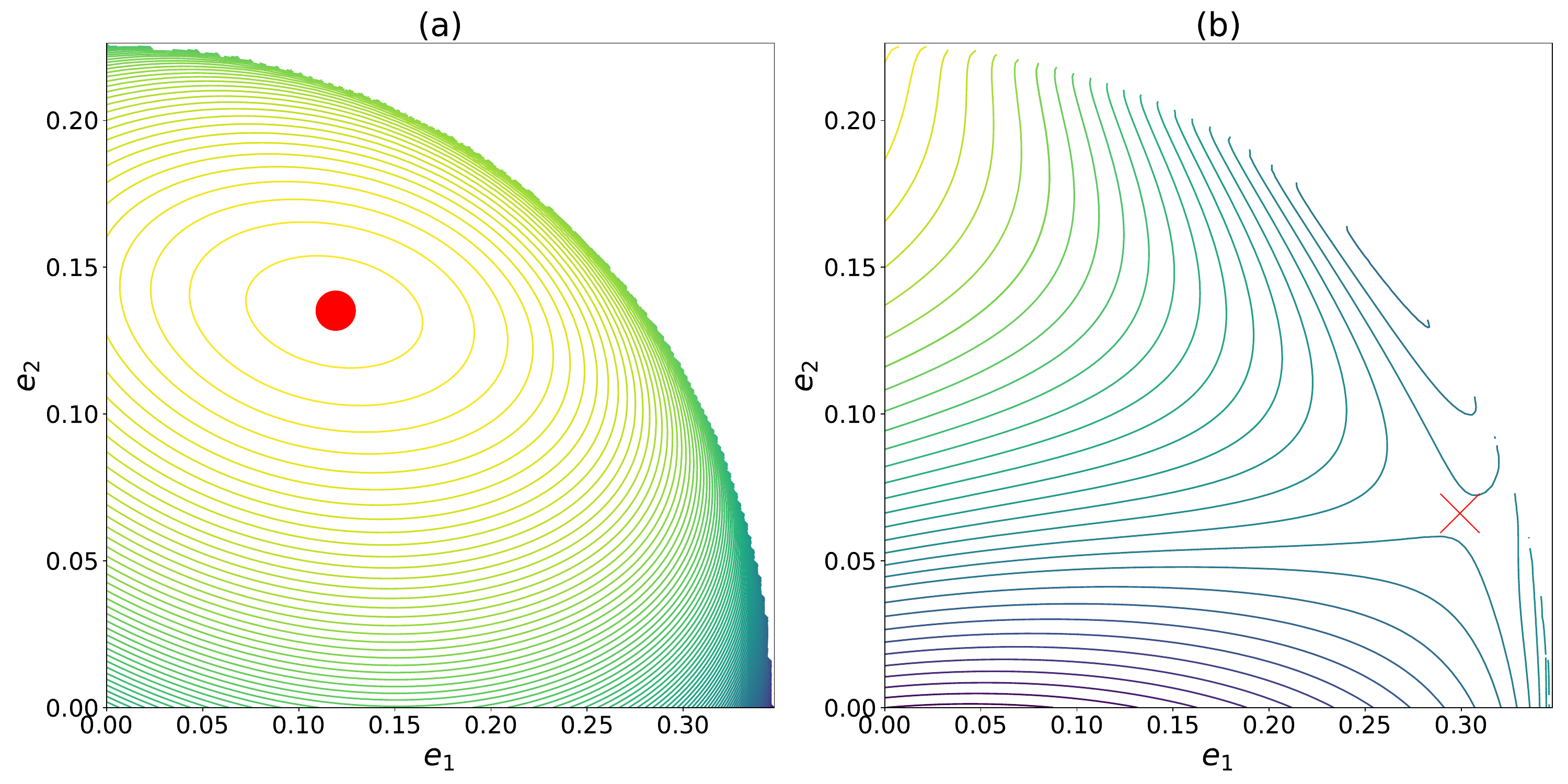}
	\caption{Different topologies on the $(e_1, e_2)$ plane for different symmetric configurations (a) $(\theta_{12}, \Delta\varpi_{12}, \theta_{23}, \Delta\varpi_{23}) = (0, \pi, 0, \pi)$ and (b) $(\pi, \pi, 0, 0)$. The local extrema are $(e_1, e_2, e_3)=(0.119, 0.1352, 0.119)$ (red solid circle) and $(0.300, 0.0662, 0.0671)$ (red cross), respectively. 
	}
	\label{fig:topo_ecc}
\end{figure}

In fact, for different resonant configurations, i.e. different values of $(\theta_{12}, \Delta\varpi_{12}, \theta_{23}, \Delta\varpi_{23})$, as well as the total angular momentum $G$, the Hamiltonian has different structures on the $(e_1, e_2)$ plane. 
\begin{enumerate}
\item
For resonant configuration $(\pi, 0, \pi, \pi)$ as shown in Fig.~\ref{fig:topo_ele}, the stationary solution is the local maximum in the $(n_1/n_2, n_2/n_3)$ plane, while is the minimum in the $(e_1, e_2)$ plane, as indicated by colour of the contours. Therefore, the solution is a saddle-like point in the 4-dimensional phase space $(n_1/n_2, n_2/n_3, e_1, e_2)$ and should be considered as unstable.
\item For resonant configuration $(0, \pi, 0, \pi)$,  with appropriate eccentricities, the topology on $(n_1/n_2, n_2/n_3)$ plane is similar to Fig.~\ref{fig:topo_ele}(a), but the topology on $(e_1, e_2)$ plane, as shown by an example in  Fig.~\ref{fig:topo_ecc}(a), has different features (from Fig.~\ref{fig:topo_ele}(b)).  The stationary solution located at $(e_1, e_2, e_3)=(0.119, 0.135, 0.119)$ now appears as the local maximum and is surrounded by robust energy level curves. The stability of this type ACR will be determined by the topological behaviour on the representative angular plane(s).
\item Except for the node-like structure on eccentricity plane, we also observe the saddle-like ACR for resonant configuration $(\pi, \pi, 0, 0)$ as shown in Fig.~\ref{fig:topo_ecc}(b). It's worth noting that such ACR appears only for high eccentricities $e_1\sim0.3$. Since the saddle-like structure always reveals the orbital instability, these ACRs will not be taken into our consideration.
\end{enumerate}

\subsection{Symmetric ACRs}
In a symmetric ACR, the critical angle could take either 0 or $\pi$. For the 1:2:3 resonant chain, the four critical angles make totally $2^4=16$ symmetric configurations. With fixed symmetric resonant angles, period ratios $(n_1/n_2, n_2/n_3)$ and the total angular momentum $G$, we find and locate all the ``node-like'' (local minimum as in Fig.~\ref{fig:topo_ele}(b) and local maximum as in Fig.~\ref{fig:topo_ecc}(a)) stationary solution (i.e. ACR) on the eccentricity plane. By varying $G$, we obtain the symmetric ACR families. The stability of an ACR solution can be determined by analysing the Hamiltonian structure in the two angular planes $(\theta_{12}, \Delta\varpi_{12})$ and $(\theta_{23}, \Delta\varpi_{23})$.

\begin{figure}
	\centering
	\includegraphics[width=\hsize]{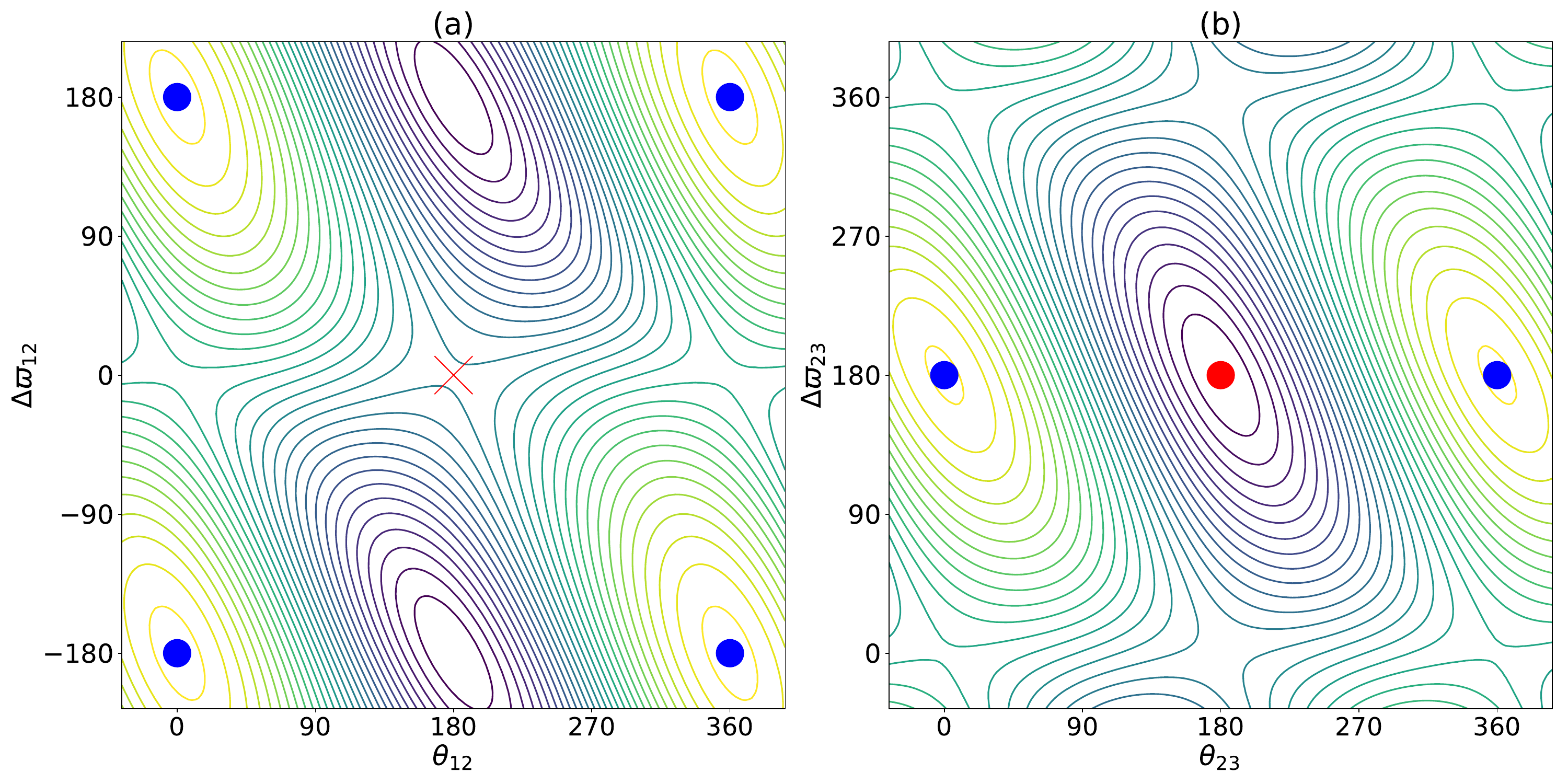}
	\caption{Topology on representative angular planes. The eccentricities of the symmetric ACR are $(e_1, e_2, e_3)=(0.00458, 0.00938, 0.00564)$. (a) Hamiltonian level curves on $(\theta_{12}, \Delta\varpi_{12})$ plane with $(\theta_{23}, \Delta\varpi_{23})$ fixed to $(\pi, \pi)$, the saddle-like point at $(\theta_{12}, \Delta\varpi_{12}=(\pi,0)$ is marked by a red cross.  (b) Topology on $(\theta_{23}, \Delta\varpi_{23})$ plane with $(\theta_{12}, \Delta\varpi_{12})$ fixed to $(\pi,0)$. The node-like point at $(\theta_{23},\Delta\varpi_{23})=(\pi,\pi)$  is marked by a red dot. In addition, the stable symmetric configuration  $(\theta_{12}, \Delta\varpi_{12}, \theta_{23}, \Delta\varpi_{23}) = (0,\pi,0,\pi)$ is shown as blue dots.   }
	\label{fig:topo_angles_symm}
\end{figure}

In Fig.~\ref{fig:topo_angles_symm}, on the representative angular planes, we show examples of the Hamiltonian structure for small eccentricity region, where only the symmetric ACRs exist. The saddle-like resonant configuration $(\theta_{12}, \Delta\varpi_{12}, \theta_{23}, \Delta\varpi_{23})=(\pi,0,\pi,\pi)$ marked by the red cross in Fig~.\ref{fig:topo_angles_symm}(a) is unstable, while the node-like configuration $(0,\pi,0,\pi)$ marked as blue dots is stable.

\begin{figure}
	\centering
	\includegraphics[width=\hsize]{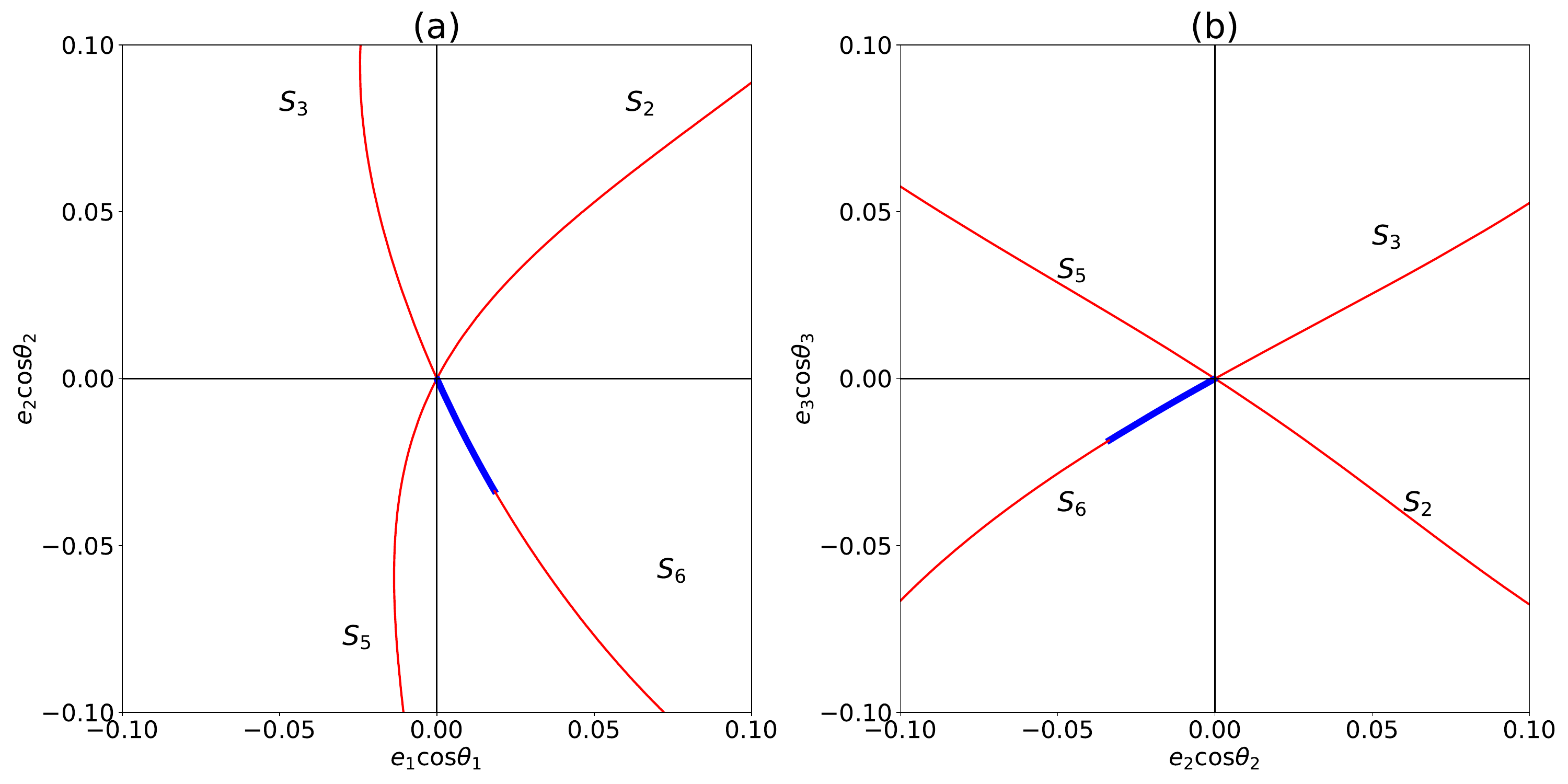}
	\caption{Symmetric ACR families $S_2,S_3,S_5,S_6$ in the 1:2:3 resonant chain for small-eccentricity configurations on (a) $(e_1\cos\theta_1, e_2\cos\theta_2)$ plane, and (b) ($e_2\cos\theta_2, e_3\cos\theta_3$) plane. The stable symmetric ACRs are in blue while the unstable ones in red. The notations of the families are the same as in \citet{antoniadou2022}, and the resonant angles of different configurations are presented in Table~\ref{table:1}. 
	}
	\label{fig:symm123}
\end{figure}

We calculate the symmetric ACRs for small eccentricity and summarize them in Fig.\ref{fig:symm123}. We note that similar results have been obtained by \citet{antoniadou2022} using different methods. For the convenience of comparison, we adopt the same notations of the families and use the following resonant angles:
\begin{equation}
\left\{ \begin{aligned}
&\theta_1=2\lambda_2-\lambda_1-\varpi_1, \\
&\theta_2=2\lambda_2-\lambda_1-\varpi_2, \\
&\theta_3=3\lambda_3-2\lambda_2-\varpi_2, \\
&\theta_4=3\lambda_3-2\lambda_2-\varpi_3.
\end{aligned}\right.
\label{eq:theta1234}
\end{equation}
To distinguish the ACR families belonging to different configurations, we present these symmetric families on the $(e_1\cos\theta_1, e_2\cos\theta_2)$ and $(e_2\cos\theta_2, e_3\cos\theta_3)$ plane. The positive or negative value on the plane indicates that the corresponding resonant angle takes the value of $0$ or $\pi$. 

The ACRs we obtained in Fig.\ref{fig:symm123} are just the periodic orbits in the rotating frame. Starting from the circular orbit configuration, \cite{antoniadou2022} find six symmetric families of periodic orbits, and they denote them by $S_i$, $i=1,\cdots,6$ (``S'' for symmetric) . We find four of them, namely $S_2, S_3, S_5$ and $S_6$, of which the critical angles are detailed in Table~\ref{table:1}.

\begin{table*}   
	\caption{The critical angles (resonant angles, longitudes of pericenter $\varpi_i$, and mean anomalies $M_i$) of different symmetric resonant configurations.}     
	\label{table:1}      
	\centering          
	\begin{tabular}{c| c c c c c c c c c c}     % 7 columns
		\hline\hline       
		% To combine 4 columns into a single one
		Family & $\theta_{12}$ & $\Delta\varpi_{12}$ & $\theta_{23}$ & $\Delta\varpi_{23}$
		& $\varpi_1$  & $\varpi_2$  & $\varpi_3$ & $M_1$ & $M_2$ & $M_3$\\
		\hline
		$S_2$ & 0  & 0 & 0 &$\pi$ 
		& 0& 0 &$\pi$ & 0 & 0  &$\pi$  
		\\
		$S_3$ & $\pi$  & $\pi$ & $\pi$ & $\pi$
		& $\pi$& 0 &$\pi$ & $\pi$& 0  &$0$ 
		\\
		$S_5$  & $\pi$  &0 & $\pi$ & $\pi$
		& $\pi$ &$\pi$ &$0$ & $\pi$ & $\pi$  &$0$ 
		\\
		$S_6$ & 0 & $\pi$ & 0 & $\pi$ 
		& 0& $\pi$ &$0$ & 0 & 0  &$\pi$ \\
		\hline
	\end{tabular}
\end{table*}

As for the two missing families, $S_1$ and $S_4$ in \citet{antoniadou2022}, their stable segments appear only very shortly when the eccentricities are very small (typically $e_i\sim 10^{-6}$), which makes them less significant since a real planetary system is unlikely to be trapped in such a small-eccentricity configuration and survive for long time span. We failed to detect these two families probably because our semi-analytical Hamiltonian only contains the perturbation up to the first order in the planet-star mass ratio, and the contribution from high-order terms are neglected.

In addition, the symmetric family $S_5$ of configuration ($\theta_{12}, \Delta\varpi_{12}, \theta_{23}, \Delta\varpi_{23})= (\pi, 0, \pi, \pi)$  labelled as stable in \citet{antoniadou2022} is topologically unstable in our calculation. In fact, the $S_5$ ACR is the local maximum on $(n_1/n_2, n_2/n_3)$ plane but local minimum on $(e_1, e_2)$ plane (e.g. Fig.~\ref{fig:topo_ele}), and such structure holds for the entire family. And we did not find any example that reaches such orbital configuration in our numerical simulations of the convergent migration (see Section~\ref{sec:migrate}).

\subsection{Bifurcation and asymmetric ACRs}

\begin{figure}
	\centering
	\includegraphics[width=\hsize]{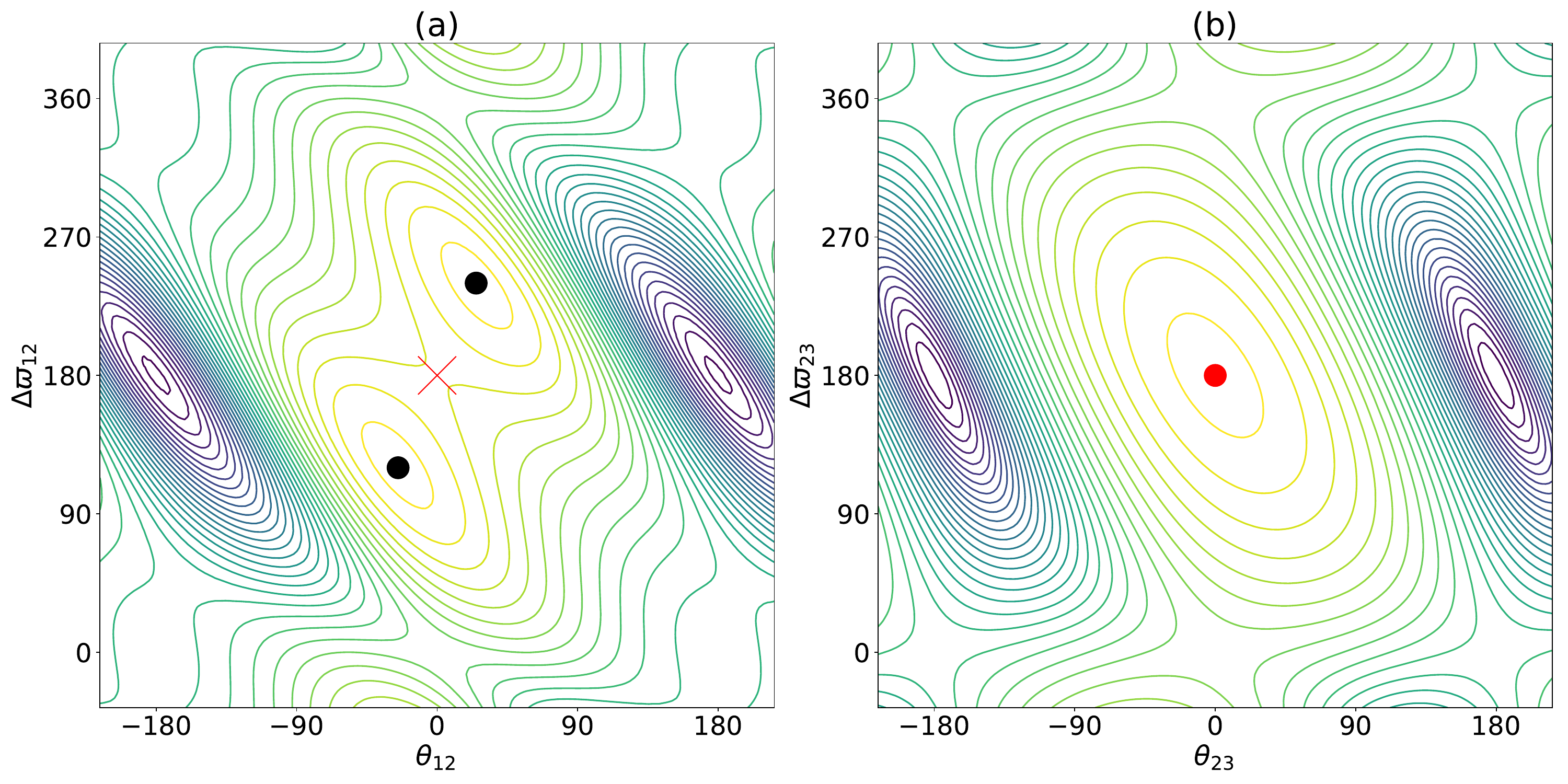}
	\caption{Topology on representative angular planes for symmetric ACR $S_6$ located at $(e_1, e_2, e_3)=(0.0569, 0.0846, 0.0531)$. (a) Topology on $(\theta_{12}, \Delta\varpi_{12})$ plane with fixed	$(\theta_{23}, \Delta\varpi_{23}) = (0, \pi)$. The saddle-like point at $(\theta_{12}, \Delta\varpi_{12} = (0, \pi))$ marked by a cross indicates the instability of this solution, while two asymmetric ACRs bifurcated from this unstable symmetric one are indicated by black dots. (b) The topology on $(\theta_{23}, \Delta\varpi_{23})$ plane with $(\theta_{12}, \Delta\varpi_{12})$ fixed to $(0, \pi)$. The node-like point at $(\theta_{23}, \Delta\varpi_{23})=(0,\pi)$ is marked by a red dot. }
	\label{fig:topo_angle_asym}
\end{figure}

As shown in Fig.~\ref{fig:symm123}, only the $S_6$ family is stable, and it loses the stability as the eccentricity increases. Along with the increasing eccentricity,  the topology around the $S_6$ family varies in a complicated way. An example is illustrated in Fig.~\ref{fig:topo_angle_asym}. 

On the $(\theta_{12}, \Delta\varpi_{12})$ plane with fixed $(\theta_{23}, \Delta\varpi_{23})=(0,\pi)$, the node-like extreme point at $(0,\pi)$ changes to a saddle-like point while two additional asymmetric ACRs,  namely $A_6$ (``A'' for asymmetric), arise as the eccentricity increases (Fig.~\ref{fig:topo_angle_asym}(a)). Meantime, we note that the feature of the node-like structure around $(0,\pi)$ on the  $(\theta_{23}, \Delta\varpi_{23})$ plane with fixed $(\theta_{12},\Delta\varpi_{12})=(0,\pi)$ remains the same  (Fig.~\ref{fig:topo_angle_asym}(b)).

The stable asymmetric ACRs are the local extrema of the full phase space ($n_1/n_2, n_2/n_3, e_1, e_2, \theta_{12}, \Delta\varpi_{12}, \theta_{23}, \Delta\varpi_{23}$). We adopt the geometrical approach \citep{michtchenko2006} to search these asymmetric ACRs in the 8-dimensional space.

\begin{figure}
	\centering
	\includegraphics[width=\hsize]{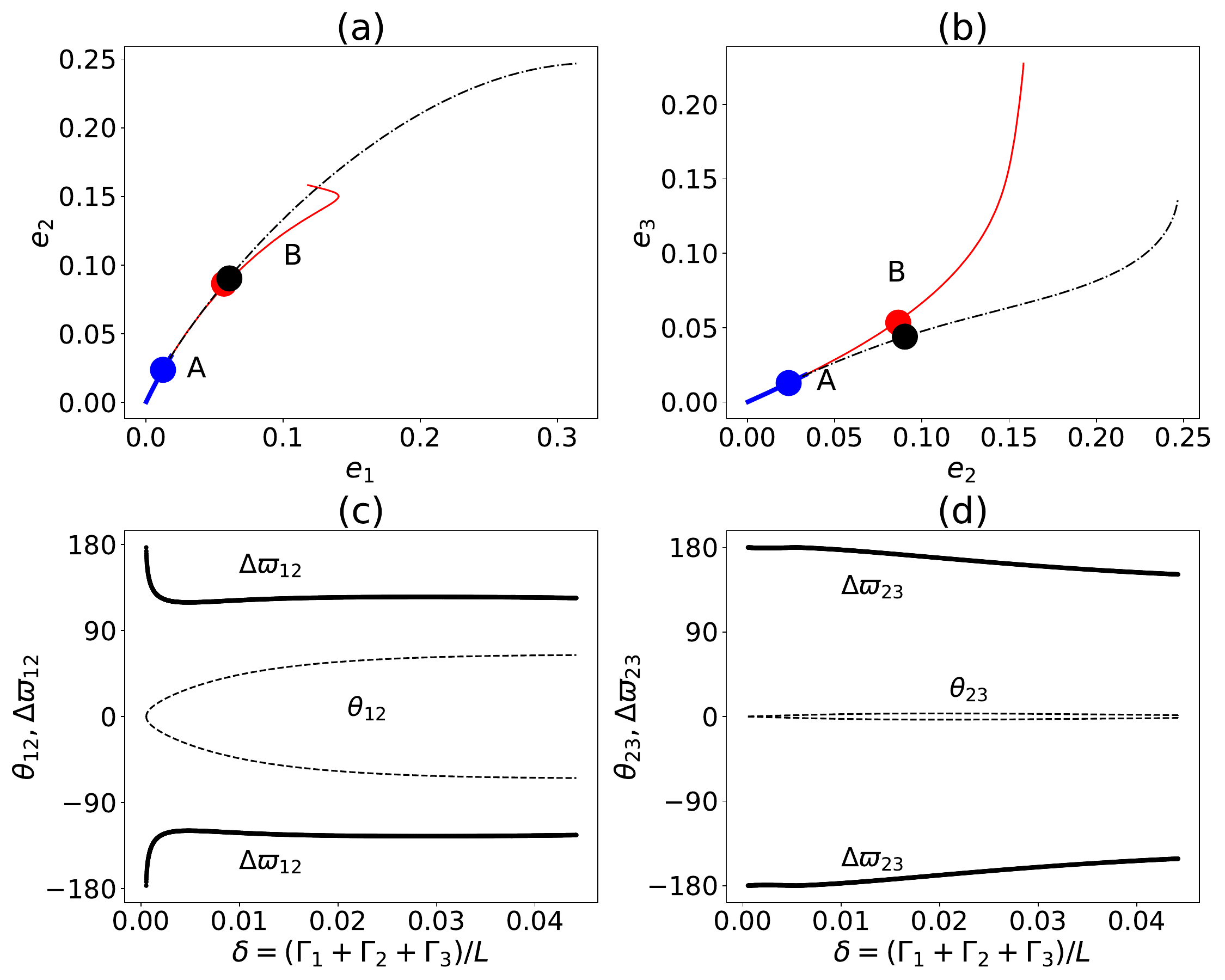}
	\caption{Locations and resonant configurations of the symmetric ACR family ($S_6$) and the asymmetric ACR family ($A_6$). In the upper two panels (a) and (b), the stable symmetric,  unstable symmetric, and stable asymmetric ACRs are indicated by blue solid, red solid, and black dashed lines, respectively. Three solid circles on these lines indicate the initial conditions that will be analysed in detail later (see text). The lower two panels (c) and (d) show the evolutions of resonant angles along the asymmetric ACR as the function of $\delta$ (scaled AMD, see text). The resonant angles $\theta_{12}, \theta_{23}$ and the secular angles $\Delta\varpi_{12}, \Delta\varpi_{23}$ are labelled by the corresponding lines.}
	\label{fig:s6a6}
\end{figure}

All the ACRs are computed under two fixed constants $L$ and $G$. Since the spacing parameter $L$ only sets the global scale of the system and does not influence the resonant dynamics (except for changing the scales of distance, time, and energy), the evolution of ACRs is in fact parametrized by the constant $G$, or equivalently, the scaled AMD $\delta$ defined as
\begin{equation} \label{eq:delta}
\delta=\frac{\Gamma_1+\Gamma_2+\Gamma_3}{L}.
\end{equation}
We summarize in Fig.~\ref{fig:s6a6} both the symmetric ACR $S_6$ and the asymmetric ACR $A_6$, in which the evolutions of eccentricities $(e_1, e_2, e_3)$ and resonant angles along these ACR solutions are illustrated.

Along the $A_6$ family, the resonant angles $\theta_{12}, \Delta\varpi_{12}$ between the inner adjacent planets $m_1, m_2$ shift significantly from the (unstable) symmetric ACR, while the configuration of the outer planet pair $(\theta_{23}, \Delta\varpi_{23})$ experiences only small variations. Particularly,  the angle $\Delta\varpi_{23}$ remains tightly close to $\pi$ as $\delta<0.01$. 

Starting from near circular orbits ($e_i\sim 0$), the periodic solution extends to high eccentricity region along the stable branch, blue solid lines first and black dashed lines later in Fig.~\ref{fig:s6a6}(a)(b). In a real system, planets in the resonant chain may not revolve exactly along these periodic solutions to high eccentricity, instead, they are expected to be around (but not exactly in) these solutions.

In practice, the numerically averaged Hamiltonian $\bar{\mathcal{H}}$ works for arbitrary planetary eccentricities because it does not depend on the literal expansion. Consequently, all the stable ACRs should appear as local extrema in the entire 8-dimensional phase space, and they can be numerically determined by the downhill simplex method under the constraints of fixed $L$ and $G$. For the 1:2:3 resonant chain, we find more than ten families of stable ACRs at relatively high eccentricity region and they form very complicated structures on the $(e_1,e_2)$ and $(e_2,e_3)$ planes. However, further investigations reveal that the topology and the dynamical portraits around these ACR families are similar to each other. Therefore, hereafter we simply focus on the ACR families starting from near circular orbits, namely the $S_6, A_6$ family. 

\subsection{Law of structure}
\begin{figure}
	\centering
	\includegraphics[width=\hsize]{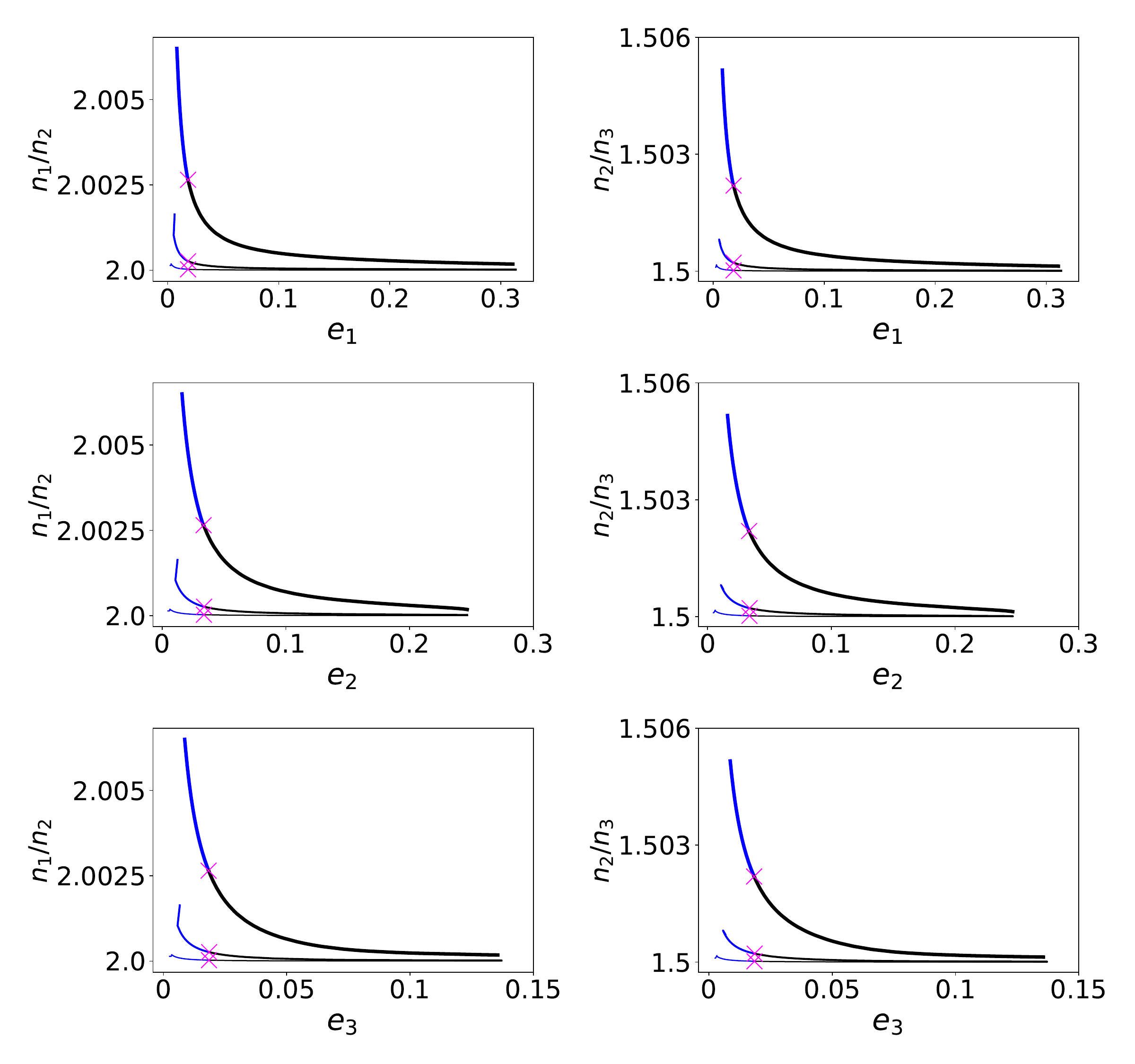}
	\caption{Law of structure for the 1:2:3 resonant chain. The left (right) column of panels show the relation between the mean motion ratio $n_1/n_2$ ($n_2/n_3$) and the eccentricities ($e_1, e_2, e_3$ from top to bottom). Blue lines are for the stable symmetric ACR families ($S_6$) while the black lines for the stable asymmetric family ($A_6$). On each panel, the lowest line is for the planets in the Kepler-51 system, while the middle and the top lines are for the cases of 10 and 100 times more massive planets, respectively (see text). The critical points, at which the $S_6$ family becomes unstable and the asymmetric solution appears, are denoted by red crosses.
	}
	\label{fig:los}
\end{figure}

As we have shown, for the 1:2:3 resonant chain, the $S_6$-$A_6$ family is the only family that starts from circular orbits and then evolves to highly eccentric configuration. During the evolution from low to high eccentricity along the ACR solution, the mean motion ratios ($n_1/n_2, n_2/n_3$) of the adjacent planet pairs may change. Since these ACR solutions are found by searching for the extrema in the 8-dimensional full space, $n_1/n_2$ and $n_2/n_3$ are obtained automatically in our calculations. The evolution of these ratios as function of the planets' eccentricities are summarised in Fig.~\ref{fig:los}. 

As shown in Fig.~\ref{fig:los}, for the ACR solution, the ratios $n_{i}/n_{i+1}$ $(i=1,2)$ are always above the nominal values (2 and 1.5 in the 1:2:3 resonant chain). Therefore, the planet pairs near the stationary resonant configuration, i.e. in resonance with small oscillations, always have orbital period ratio at one side of the nominal value. The deviation from the nominal value is large for near circular orbits and it decreases for larger  eccentricities (thus larger AMD).

The mean motion ratio of the ACR solution depends on planetary masses. Except for the planetary masses as in the Kepler-51 system, we also test two additional systems, in which the planets are 10 and 100 times heavier than the ones in Kepler-51, i.e., keep the $m_i/m_j$ ($i,j=1,2,3$) values but set $m_i/M$ 10 and 100 times larger than in the Kepler-51 system. Assume they are in the same 1:2:3 resonance chain, we calculate the mean motion ratios, and plot the results in Fig.~\ref{fig:los}, too. For near circular orbits, the stable ACR solution is symmetric (blue segments in Fig.~\ref{fig:los}), and from a critical $e_i$ it is continued by asymmetric ACR (black segments). Our calculations demonstrate that the critical points for three cases of different total planet masses are the same $(e_1, e_2, e_3)= (0.0185, 0.0336, 0.0186)$, implying the independence of the resonant configuration on the total  planetary mass. Indeed, only the mass ratio between adjacent planets influences the resonant configuration \citep{antoniadou2022}.

Apparently, for a resonant chain with given mass ratios between planets, the larger the total mass of planets the stronger the short-term perturbations the system would experience. Thus, the resonant chain with higher total planetary mass is more likely to be unstable. The main dynamical mechanism inducing such orbital instability is the 1:1 secondary resonance, which occurs when one of the libration period is almost equal to the period of synodic angle \citep{pichierri2020,goldberg2022}. This secondary resonance excites the orbital eccentricity and finally destabilizes the orbits. In other words, the short term perturbation is too large to be averaged, if the total mass of planets is large enough. 
%\textcolor{red}{between the fastest libration frequency and a difference in synodic frequencies.}
%The libration frequency of the triple planetary system is characterized as $f_\text{lib} = \max \{f_{\theta_{12}}, f_{\theta_{23}}\}$ ($f_\theta$ corresponds to the quasi-libration frequency of $\theta_{12}, \theta_{23}$ around the ACR), the synodic frequency in the Hamiltonian system reads $f_Q=\frac{k_i}{\text{LCM}(k_1,k_2,k_3)} \times f_i=\frac{k_i}{\text{LCM}(k_1,k_2,k_3) } \times \frac{1}{T_i}$. 
%It's simple to prove that such relationship holds for the 1st order resonant chain.\\
%In \cite{goldberg2022criterion}, they also generalize their criterion to resonant chain of N (N>3) planets. Since they only \textcolor{red}{focus on the resonant chain composed of 1st order MMRs}, based on our averaged model, here we provide a generalized explanation to arbitrary three-planet resonant chain:

\section{Dynamics around stable ACRs}

For the 1:2:3 resonant chain, the $S_6$-$A_6$ is the only stable family that could evolve from circular orbits to regime of high eccentricity. It is possible that a planetary system may evolve along a ``path'' guided by the $S_6$-$A_6$ family. In this part, by constructing dynamical maps on the appropriate representative planes, we explore the motion in the vicinity of this evolution path along the $S_6$-$A_6$ ACRs. 

\subsection{Initial conditions}
All the ACR solutions were obtained in the averaged Hamiltonian system, but to construct the dynamical maps we have to use the osculating orbital elements to numerically integrate the orbits. Therefore, we should choose the initial conditions carefully to build a proper association between the averaged system and the real system. Generally, the osculating elements are related to averaged elements by the canonical transformation \citep[see e.g.][for two-planet and three-planet cases, respectively]{ramos2015,charalambous2018}. But in this paper, we use a simple transformation based on the orbital configuration, particularly the symmetric orbits, as follows.

Typically, the period ratios $(n_1/n_2, n_2/n_3)$ of planets inside a resonant chain have only very small oscillation around their nominal value $(n_{1c}/n_{2c}, n_{2c}/n_{3c})$, otherwise the period commensurability and resonance between planets would break. Therefore, the period ratios, thus the osculating semi-major axes $a_i$, can be fixed at the values of the corresponding stable ACRs $\bar{a}_i$, that is, $a_i=\bar{a}_i$ ($i=1,2,3$). The same treatment can be applied to eccentricities and resonant angles, $e_i=\bar{e}_i$ ($i=1,2,3$), $\theta_{i,i+1} = \bar{\theta}_{i,i+1}, \Delta\varpi_{i,i+1} = \overline{\Delta\varpi}_{i,i+1}$ ($i=1,2$). In fact, we construct dynamical maps on the $(e_1, e_2)$ plane with fixed constants $L$ and $G$ defined in Eq.~\eqref{eq:res_var}. For each point on the $(e_1, e_2)$ plane, the conversation of $L$ is automatically satified when the semi-major axes $a_1, a_2, a_3$ were given (in following maps, they were fixed to the values of the exact ACRs), while the eccentricity of the outer planet $e_3$ can be determined through the total angular momentum $G$. After $a_1, a_2, a_3, e_1, e_2, e_3, \theta_{12}, \Delta\varpi_{12}, \theta_{23}, \Delta\varpi_{23}$ have been given as the initial conditions, only one angular variable is left free. We choose the synodic angle $Q$ as the last variable to be determined.  

Let $(X_i, Y_i, \dot{X}_i, \dot{Y}_i)$ be the position and velocity of the $i$-th body $m_i$ in the rotational frame. An orbit starting from the $GX$ axis at time $t_0=0$ and satisfying $X_i(t_0)\neq 0, \dot{X}_i(t_0)=0, Y_i(t_0)=0, \dot{Y}_i(t_0)\neq 0$ ($i=1,2,3$) is symmetric with respect to the $GX$ axis, because the symmetric condition Eq.\eqref{eq:symm} is apparently fulfilled. The algebraic manipulation reveals that the symmetry in the rotational frame leads to
\begin{equation}
	\begin{aligned}
		& \left|\bm{r}_i(t)-\bm{r}_j(t)\right|=\left|\bm{r}_i(-t)-\bm{r}_j(-t)\right|, \\
		& \bm{p}_i(-t) \cdot \bm{p}_j(-t)=\bm{p}_i(t) \cdot \bm{p}_j(t).
	\end{aligned}
\end{equation}
Thus, the Hamiltonian $\mathcal{H}$ should follow the same symmetry $\mathcal{H}(-t) = \mathcal{H}(t)$. Since $\mathcal{H}$ depends explicitly on rather $Q$ than $t$, while $Q$ is a linear function of $t$, up to the 1st order of the planet-star mass ratio $(m_1+m_2+m_3)/M$, we have $\mathcal{H}(-Q)=\mathcal{H}(Q)$. Therefore, the Hamiltonian must be symmetric with respect to $Q_0=Q(t=0)$ and $\mathcal{H}(Q_0)$ must be an extremum. To determine the proper osculating orbital elements, we need to find the appropriate $Q$, and now it is equivalent to find the $Q$ at which $\mathcal{H}(Q)$ reaches the extrema. 

\begin{figure}
	\centering
	\includegraphics[width=\hsize]{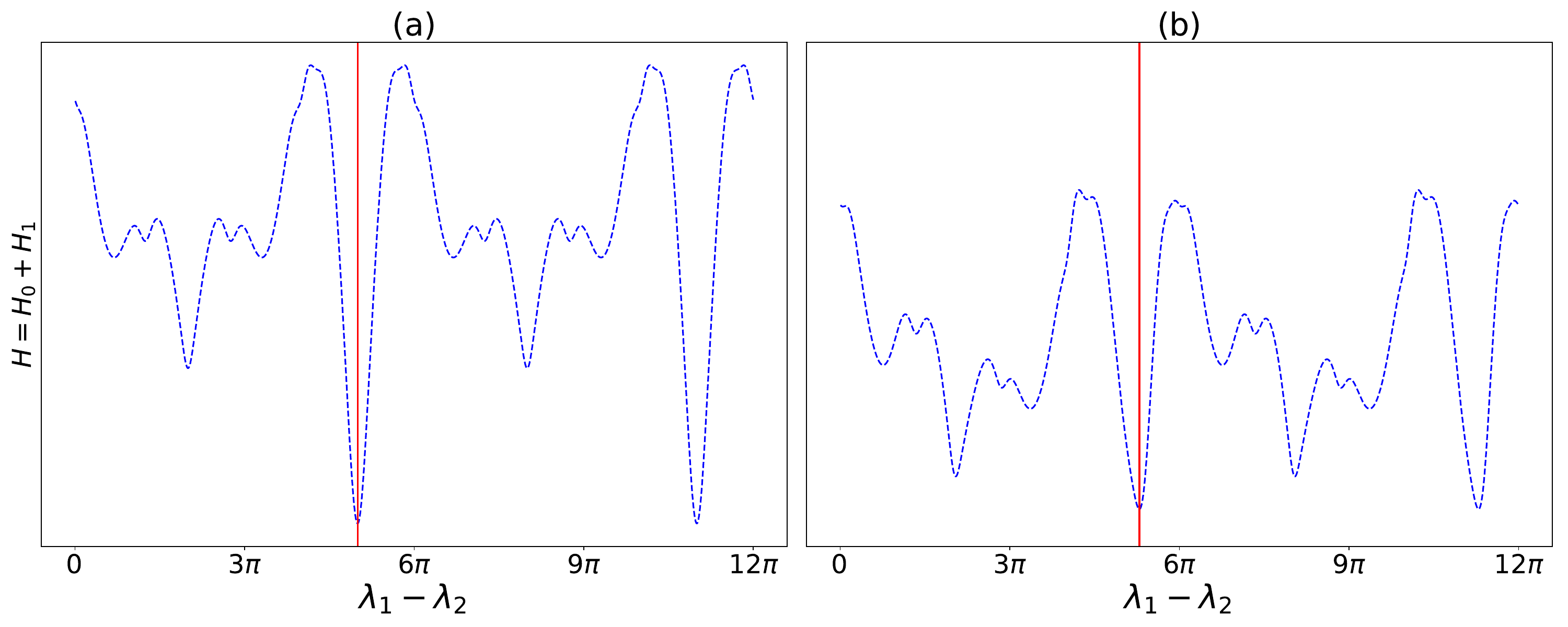}
	\caption{The variation of Hamiltonian in two synodic periods $2Q_{12}^T=12\pi$ for (a) symmetric configuration, and (b) asymmetric configuration. The vertical line denotes the position when the Hamiltonian attains the minimum. 
	}
	\label{fig:a2o}
\end{figure}

We plot in Fig.~\ref{fig:a2o} the variation of the Hamiltonian with respect to the synodic angle $Q$ for both symmetric and asymmetric resonant configurations. It is clear that the Hamiltonian is symmetric with respect to the red vertical line in Fig.~\ref{fig:a2o}(a) which corresponds to the global minimum. The above calculations can be applied to all the symmetric ACRs, and all the angles $(\varpi_i, M_i)$ $(i=1,2,3)$ at the symmetric solutions have been summarized in Table~\ref{table:1}.

For asymmetric configurations, as shown in Fig.~\ref{fig:a2o}(b), the symmetry condition cannot be fulfilled. The $Q$ value, at which the Hamiltonian attains the minimum, is calculated and adopted as the initial configuration.

\subsection{ Max($\Delta e$) indicator and $\chi^2$ criterion}
\label{sec:chi2}
For each point on the $(e_1, e_2)$ plane, we numerically integrate the equations of motion up to $10^5$ years from the corresponding initial conditions (osculating elements). We note that the ``year'' here is defined as the orbital period of the middle planet $m_2$. During the integration, we keep tracking the maximal and minimal values attained by the planetary eccentricities and adopt the variation range $(\Delta e_1, \Delta e_2, \Delta e_3)$ as the indicator to characterise the orbital stability. As mentioned in \citet{ramos2015}, although the $\Delta e$ is not a rigorous measure of chaos of the motion, it's a useful tool for the estimation of both the possible fixed points and the dynamical separatrix of the resonant region. In practise, we may consider the orbits with small $\Delta e$ as ``stable'' and the ones with large $\Delta e$ as ``chaotic'' (close to the dynamical separatrix).

In a two-body MMR, after the numerical averaging over the synodic angle $Q$, the system is reduced to a Hamiltonian of only 2 DoF. The existence of the resonant invariant torus, or dynamical separatrix \citep{alves2016}, forbids the transition between the resonant and the non-resonant region. Inside the resonant zone, the resonant angles librate around the ACR solutions, while they circulate outside the resonant region. Thus, different types of motion can be easily distinguished from each other. Unfortunately, in the three-planet resonant system, the resonant angles behave in a much more complicated way. Due to the perturbation from the third planet, a two-body resonant angle may librate irregularly with a large amplitude and appear like in a complex of both libration and circulation. As a result, it's difficult to determine accurately whether the system is in a resonance (or dominated by the resonance). Following the idea by \citet{gallardo2014}, another indicator, $\chi^2$ of the resonant angles, is used in this paper.

The indicator $\chi^2$ is defined to characterize different effects of the resonance in a three-planet system: deep in resonance (resonant angles librating), at the boundary of resonance (resonant angles circulating or librating with large amplitudes), and outside resonance.

The basic idea is to analyse the resonant angle over a period of time $T^*$. The statistical difference $\chi^2$ between the distributions of a resonant angle and a uniform distribution in $[-\pi,\pi]$ is adopted as an indicator to determine whether the system is in a resonance: the larger the $\chi^2$, the more likely the resonance angle is in libration, and \textit{vice versa}. 

\begin{figure}
   \centering
   \includegraphics[width=\hsize]{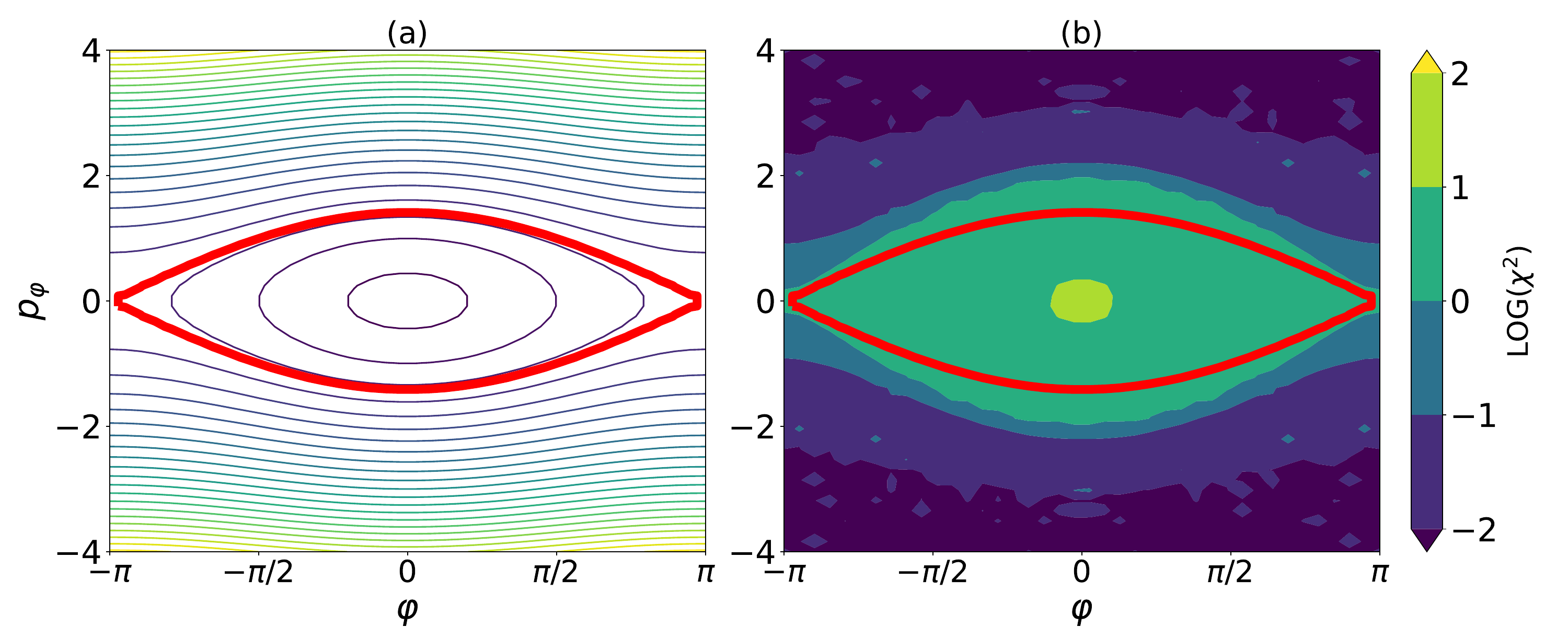}
   \caption{$\chi^2$ map for the standard pendulum model. (a) Phase structure (energy level) on $(\varphi, p_\varphi)$ plane. The broad red curve marks the separatrix between the libration and circulation region. (b) $\chi^2$ value in logarithm indicated by colour.  }
   \label{fig:chi2}
 \end{figure}

The semi-quantitative criterion will be given through a standard pendulum model with Hamiltonian:
\begin{equation}
\mathcal{H}_{\text{pend}}=\frac{p_\varphi^2}{2}-\cos\varphi.
\end{equation}
From each point of a $100\times 100$ grid on the phase space $[p_\varphi,\varphi] \in [-4,4]\times [-\pi,\pi]$, we integrate the corresponding Hamiltonian equations and output  $N=2^{13} = 8192$ points $\{\varphi_i\}_{i=1}^{N=8192}$ evenly in a time step $\Delta t=5$. The interval $[-\pi,\pi]$ is divided into 180 equal bins, and the $\chi^2$ of such an orbit is computed as 
\begin{equation}
\chi^2=\sum_{k=1}^{180}\frac{(O_{k}-E_{k})^2}{E_{k}},
\end{equation}
where $O_{k}$ is the number of $\varphi_i$ observed in the $k$-th bin, and  $E_{k}$ is the expected number in the same bin if the 8192 values are in a uniform distribution. We show in Fig.~\ref{fig:chi2} the phase space of the standard pendulum and the $\chi^2$ map.

Apparently, all the points located inside the libration region in Fig.~\ref{fig:chi2} have $\chi^2>1$ $(\log\chi^2>0)$, while those points with $\chi^2 <0.1$ are surely in the circulation region. For initial conditions in the close vicinity of the separatrix, their orbital periods are so long that the total integration time could not cover several times of their periods and therefore, some orbits in the circulation region might be misunderstood as ``libration''. Empirically, the resonant behaviours can be estimated by the value of $\chi^2$:
\begin{itemize}
\item $\chi^2>1$, the resonant angle is in libration and the system is deeply inside the resonance;
\item $\chi^2<0.1$, the resonant angle is in circulation and the systems is outside the resonance;
\item $0.1<\chi^2<1$, the system is affected by the resonance, but the resonant angles may experience both the libration and circulation in a long time span. We classify this behaviour as the motion near the separatrix.
\end{itemize}

For the MMR between planets $m_i, m_j$, the resonance is characterized by the libration of at least one of the resonant angles
\begin{equation}
\label{eq:res_angles}
\theta^l_{ij}=k_{ji}\lambda_j-k_{ij}\lambda_i-l\varpi_i-(q_{ij}-l)\varpi_j,  ~~l=0,1,\cdots,q_{ij}, 
\end{equation}
where $q_{ij}=|k_{ij}-k_{ji}|$ is the order of the resonance, and the librating angle is referred as the true resonant angle \citep{alves2016}. Therefore, the criterion $\chi^2$ of an MMR is defined as the largest $\chi^2$ value of all resonant angles, and we may adopt the same empirical criterion obtained in the pendulum model.

\subsection{Motion around symmetric ACRs at low eccentricity}

Along the $S_6$-$A_6$ family, the symmetric ACR solution is stable at low eccentricity (Fig.~\ref{fig:s6a6}). Arbitrarily, we select on the ACR one initial condition $(e_1, e_2, e_3)=(0.0124, 0.0237, 0.0129)$, which is labelled by `A' in Fig.~\ref{fig:s6a6}. The resonant angles of this solution are $(\theta_{12}, \Delta\varpi_{12}, \theta_{23}, \Delta\varpi_{23}) = (0,\pi, 0, \pi)$, and the period ratios are set as the nominal values in the 1:2:3 resonant chain. Keeping the  constant $L$ and $G$ calculated from the above $a_i, e_i$ $(i=1,2,3)$, we generate a set of initial conditions both for the averaged Hamiltonian system and for the corresponding osculating system. We denote the case with these initial conditions by ``Case A'', and investigate the motion.  

\begin{figure}
	\centering
	\includegraphics[width=\hsize]{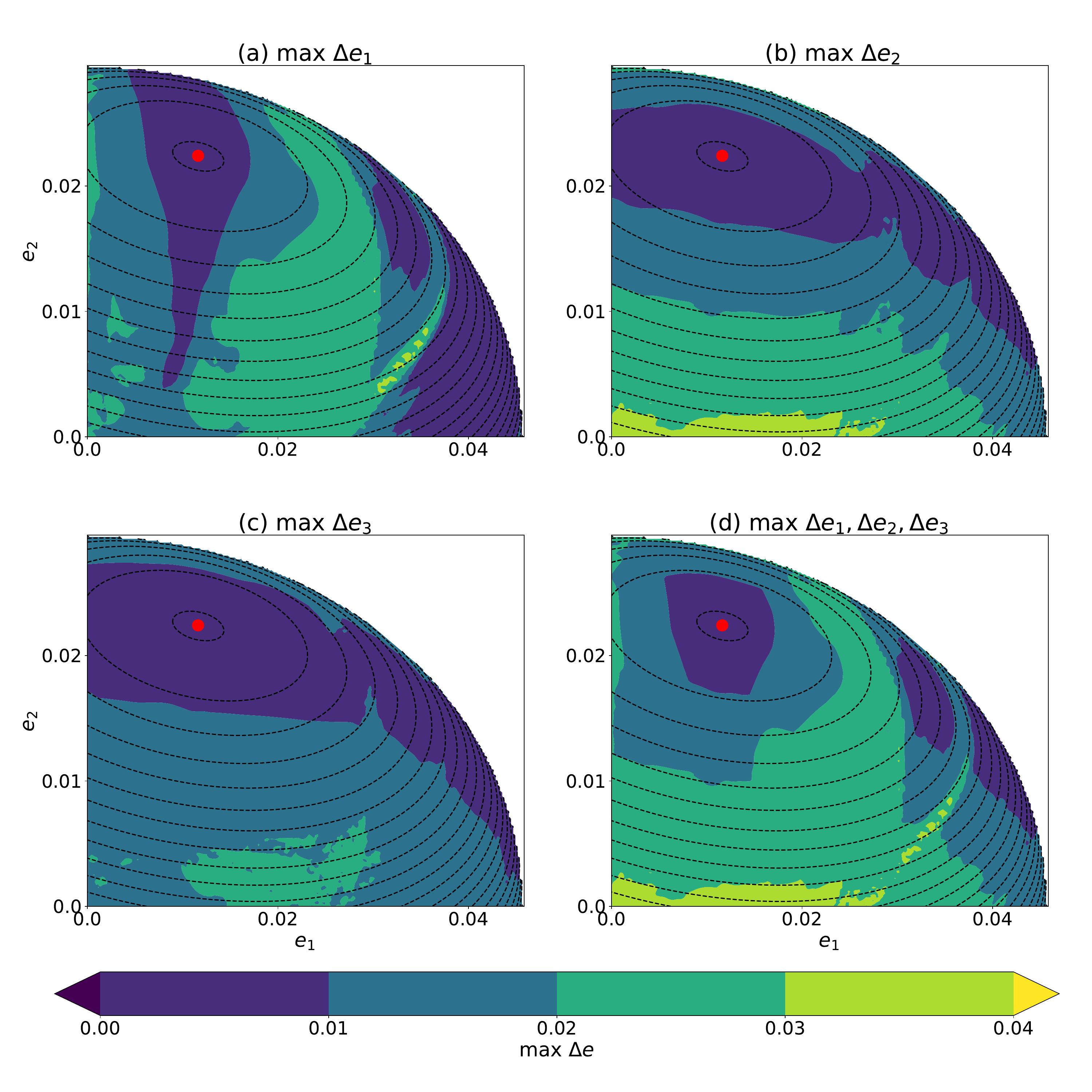}
	\caption{Topology (contour line) and dynamical map (colour map) on the $(e_1, e_2)$ plane for Case A, with constants $L$ and $G$ determined by the initial condition on the stable ACR solution $(e_1, e_2, e_3, \theta_{12}, \Delta\varpi_{12}, \theta_{23}, \Delta\varpi_{23}) = (0.0124, 0.0237, 0.0128, 0, \pi, 0, \pi)$, which is indicated by a red dot on all maps. The eccentricity variations $\Delta e_i$ are used as the stability indicator as labelled in each panel. 
	}
	\label{fig:de_s6a}
\end{figure}

We present in Fig. \ref{fig:de_s6a} on the $(e_1,e_2)$ plane both the topologies (contour curves) of the corresponding averaged Hamiltonian and the dynamical maps (coloured map) obtained from numerical integration of the osculating orbital elements. The ACR solution appears as the local maximum and is surrounded by ellipse-like energy curves, implying its stability. The maximum of the eccentricity variation $\Delta e_i$ is adopted as the indicator of stability as labelled on each panel in Fig.\ref{fig:de_s6a}. And particularly, the maximum of $(\Delta e_1,\Delta e_2,\Delta e_3)$ is used  in the last panel.

As clearly shown in Fig.\ref{fig:de_s6a}, all the four maps reveal that the motion is very stable around the ACR with very small eccentricity variations. Further analyses show that these stable orbits are characterized by the libration around the stable ACR, performing the so called quasi-periodic motion. As a matter of fact, such regular region can occupy a large area on the $(e_1,e_2)$ plane.  

Away from the stable ACR, the dynamical maps of different indicators may show different patterns. The region of small $\Delta e_1$ and $\Delta e_2$ extends roughly along either the $e_1$ (horizontally) or the $e_2$ direction (vertically). At about $(e_1,e_2)=(0.025,0.02)$ in Fig.\ref{fig:de_s6a}(a), the planet $m_1$ would suffer relatively large eccentricity excitation $(\Delta e_1>0.02)$ and the corresponding orbit is much less regular.
In addition, starting from about $(e_1,e_2)=(0.032,0.006)$ we also notice a very narrow strip with relative large value of $\Delta e_1$.  
%the structure of small $\Delta e_1$ shows a strong dependence on $e_1$ while a weak dependence on $e_2$, appearing as a nearly vertical stripe on $(e_1,e_2)$ plane.

 The dynamical features in panels (b) and (c) of Fig.\ref{fig:de_s6a} for $\Delta e_2$ and $\Delta e_3$ are similar. At the bottom of the $(e_1,e_2)$ plane, $\Delta e_2$ attains its maximum value ($\sim$0.04), implying (relatively) the least stability of motion there. And in all three panels (a), (b) and (c), deviated from the ACR, regular region can also be found at the right corner of $(e_1,e_2)$ plane, where motions are distinct from the quasi-libration around the ACR.
%Since Fig.\ref{fig:de_s6a}(b) unveils more details, thus we mainly focus on the dynamical structure revealed by $\Delta e_2$. On the contrary of the behaviours shown in Fig.\ref{fig:de_s6a}(a), the shape of $\Delta e_2$ appears as the horizontal strips on $(e_1,e_2)$, i.e. the $\Delta e_2$ depends strongly on the initial $e_1$ instead of $e_2$.

Finally, we find that all these dynamical features can be revealed by the maximum of the three eccentricity variations, i.e.  $\max(\Delta e)$= $\max (\Delta e_1,\Delta e_2,\Delta e_3)$, as shown in Fig.\ref{fig:de_s6a}(d). Particularly, we see  in Fig.\ref{fig:de_s6a}(d) that the most stable motions occur in three regions, around the ACR, around $(e_1,e_2)=(0.032,0.017)$, and around $(0.04,0.01)$, respectively. Below in this paper, we adopt $\max(\Delta e)$ as the stability indicator and construct the dynamical maps. 

\begin{figure}
	\centering
	\includegraphics[width=\hsize]{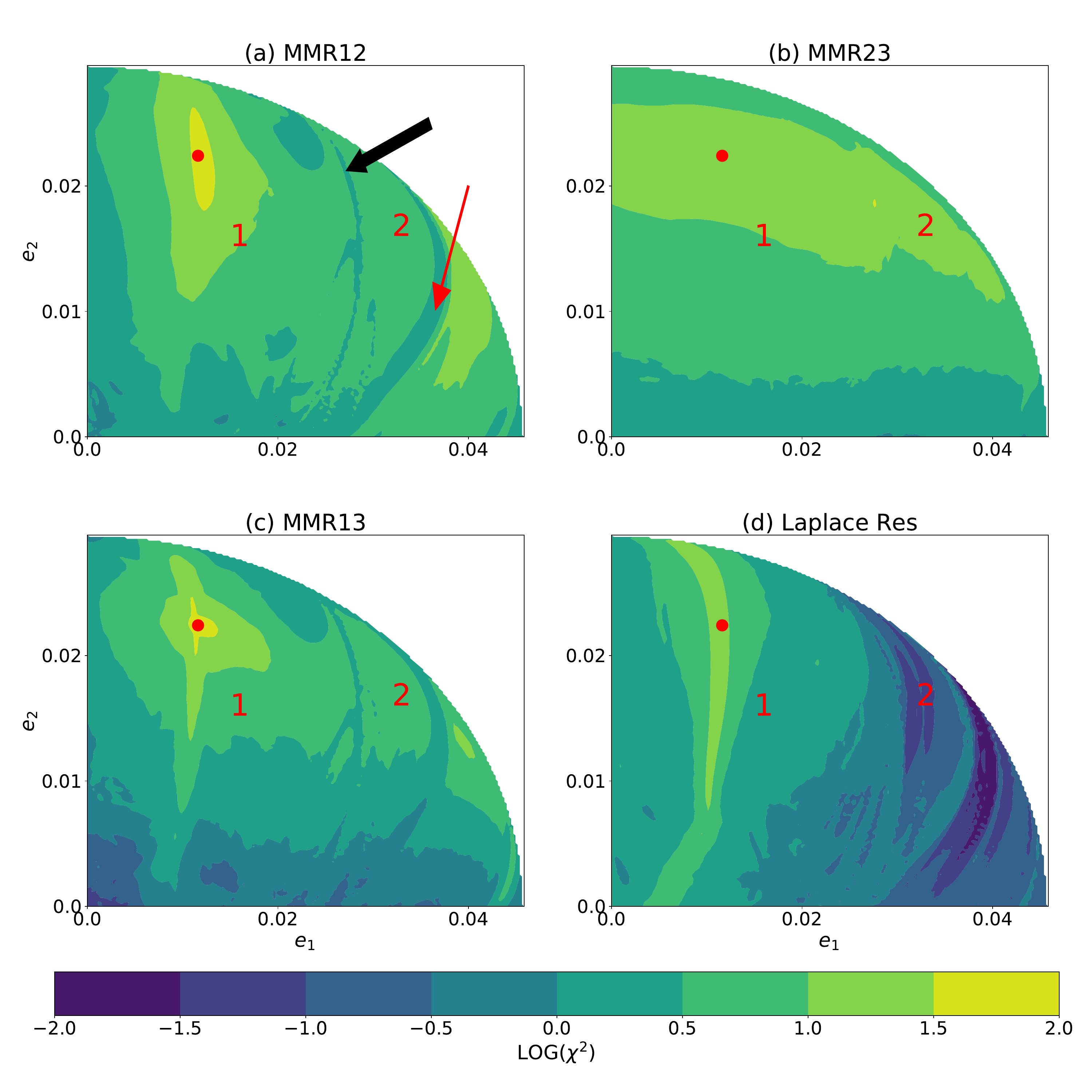}
	\caption{Resonant structure indicated by $\chi^2$ on the $(e_1, e_2)$ plane for Case A. Four panels are for the two-body MMRs between (a) $m_1$ and $m_2$, (b) $m_2$ and $m_3$, (c) $m_1$ and $m_3$, and (d) the three-planet Laplace resonance with resonant angle $\varphi_0$. The large value of $\chi^2$ (light colour) indicates that the resonant angle is in libration, while the small $\chi^2$ (dark tones) indicates circulation. The narrow strip-like structure indicated by an black arrow in (a) divides the map into two parts, namely Region 1 and Region 2. Moreover, another  narrow strip (indicated by the red arrow) can be seen in Region 2. }
	\label{fig:chi_s6a}
\end{figure}

The $\chi^2$ maps might provide more details on the resonant dynamics. In Fig.~\ref{fig:chi_s6a}, we depict the $\chi^2$ maps for all the MMRs inside the 1:2:3 resonant chain. For all MMRs, the closer to the stable ACR, the larger the $\chi^2$, implying that in the close vicinity of the stable ACR solution, all the resonant angles are librating, as well as the three-body Laplace resonant angle. The regular motion, characterized by the quasi-periodic libration around the ACR, is common around the ACR solution. 

The $\chi^2$ maps in Fig.~\ref{fig:chi_s6a} reveal that, as $e_1$ increases (while $e_2$ remains the value of ACR), the 2:3 two-body resonance (MMR23) in the outer planet pair persists and seems to play an essential role to the resonant stability, while other resonances in the resonant chain are weaken or even destroyed. The region with small $\Delta e_2$ in Fig.~\ref{fig:de_s6a}(b) coincides with the region of $\chi^2>1$ in Fig.~\ref{fig:chi_s6a}(b). When $e_2$ deviates from the ACR value (with $e_1$ almost being frozen), the inner planet pair's 1:2 resonance (MMR12), the 1:3 resonance (MMR13) between $m_1$ and $m_3$, and the three-planet Laplace resonance, persist in a range of $e_2$, as shown by the vertical strip-like structures of large $\chi^2$ in Fig.~\ref{fig:chi_s6a}(a)(c)(d). Therefore, away from the ACR solution along the direction of $e_1$, although the MMR23 is stable, the strength of MMR12, MMR13, and Laplace resonance becomes weaker, implying that the corresponding resonant angles may librate with large amplitudes and eventually the Laplace angle $\varphi_0$ will be in circulation. 

%As shown in Fig.~\ref{fig:chi_s6a}(a)(c), two separatrix-like sturcture divide the stable region of MMR23 into three parts: namely Region 1, Region 2a and Region 2b. We check carefully the $\chi^2$ values in these three regions and find that the orbits in Region 1 are in both the two-body MMRs and the Laplace resonance, while in Region 2a and Region 2b, although all the two-body MMRs (i.e. MMR12, MMR23, and MMR13) occur, the Laplace angle $\varphi_0$ circulates, as indicated by the small $\chi^2$ value in the area of Region 2a and 2b in Fig.~\ref{fig:chi_s6a}(d)}

\begin{figure}
	\centering
	\includegraphics[width=\hsize]{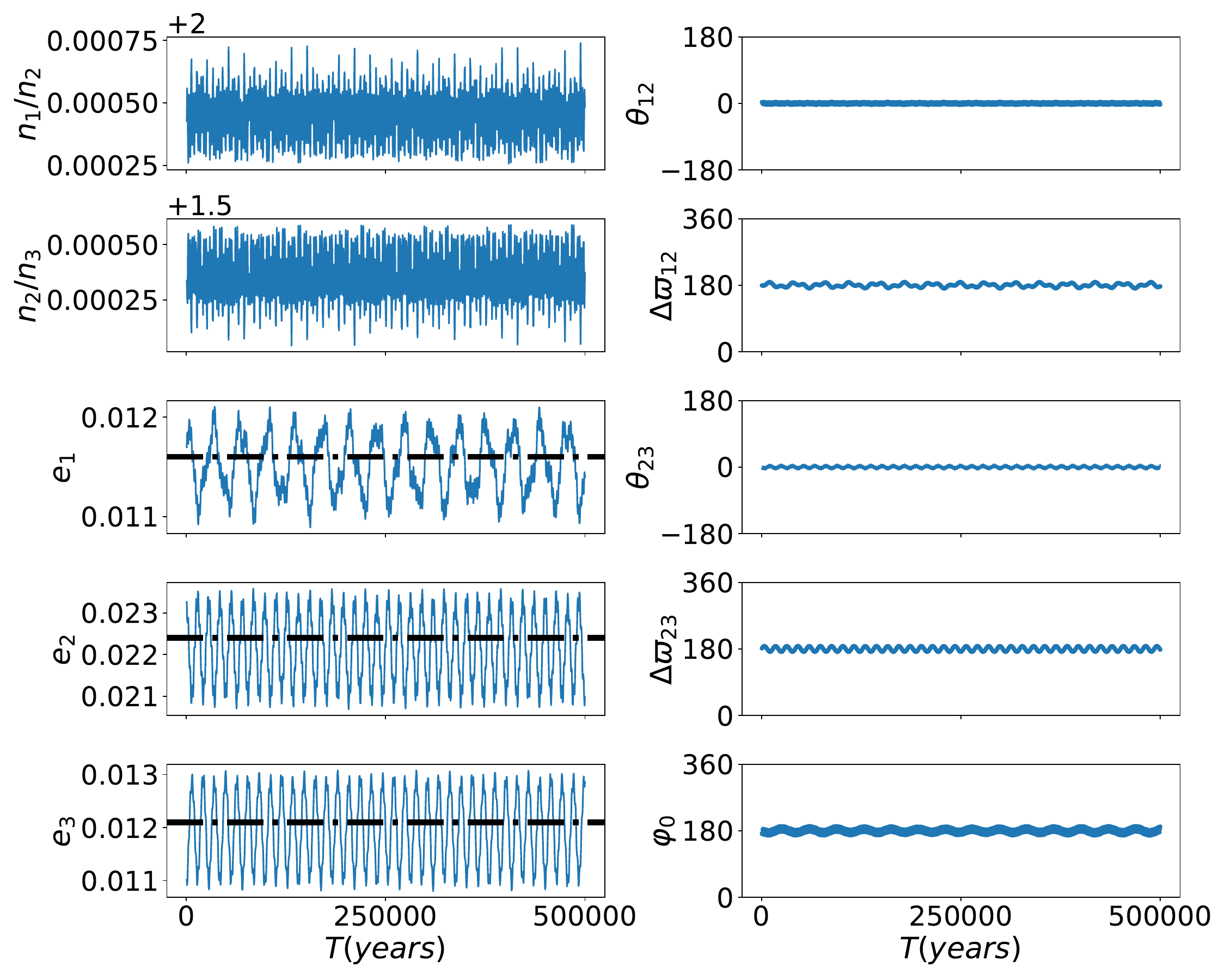}
	\caption{Typical behaviour of orbits in Region 1 (Fig.~\ref{fig:chi_s6a}(a)). The temporal variations of $n_1/n_2, n_2/n_3, e_1, e_2, e_3$ are in left panels, while right panels are for resonant angles $\theta_{12}, \Delta\varpi_{12}, \theta_{23}, \Delta\varpi_{23}, \varphi_0$. The orbits are librating quasi-periodically around the stable symmetric ACR solution $S_6$. The eccentricities of $S_6$ are indicated by horizontal dashed lines. }
	\label{fig:s6a1}
\end{figure}

\begin{figure}
	\centering
	\includegraphics[width=\hsize]{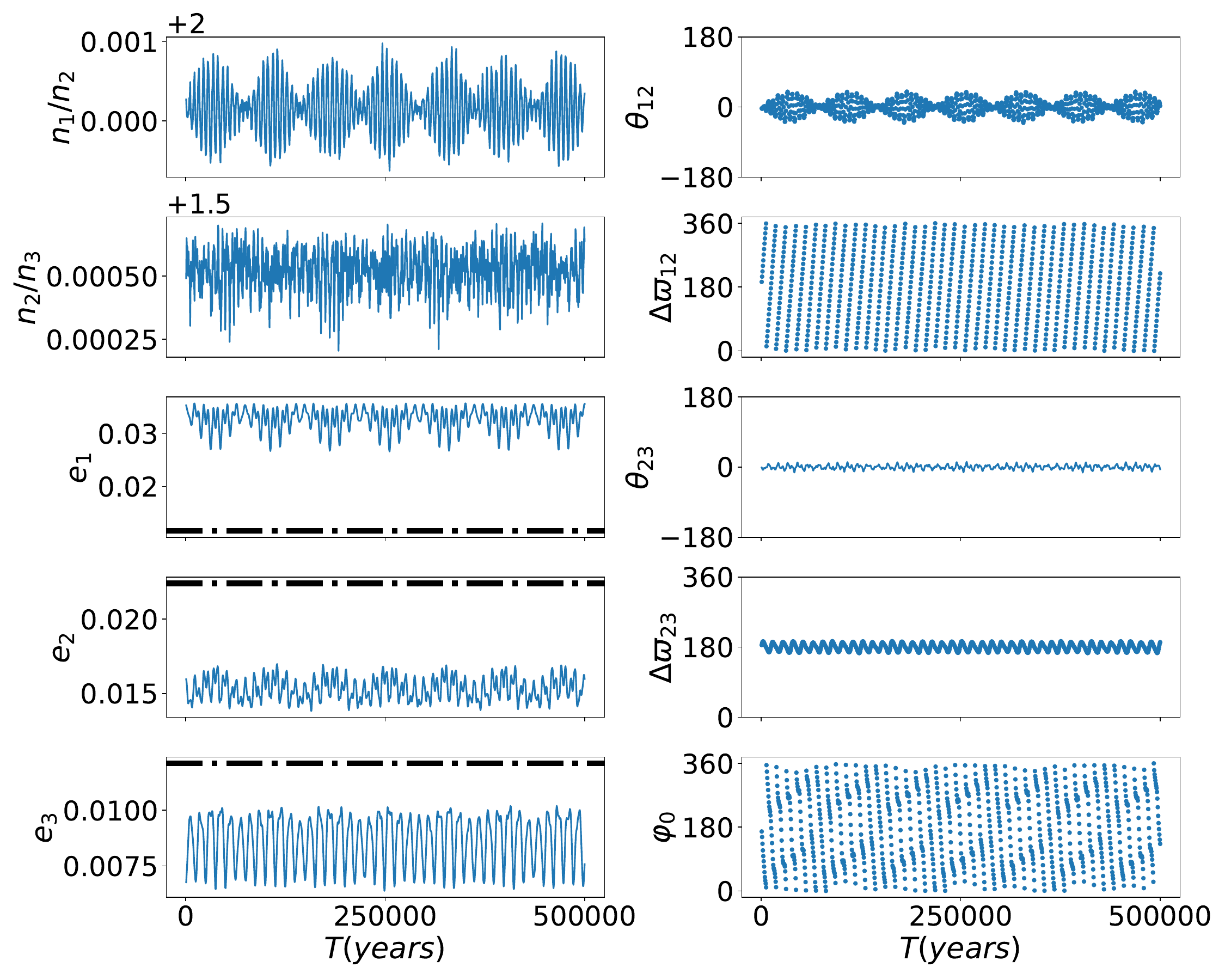}
	\caption{The same as Fig.~\ref{fig:s6a1} but for Region 2 in Fig.~\ref{fig:chi_s6a}(a).  }
	\label{fig:s6a2}
\end{figure}

A nearly vertical narrow strip of small $\chi^2$ that divides the stable region into two parts,  namely Region 1 and Region 2, can be seen in Fig.~\ref{fig:chi_s6a}(a)(c), the separatrix-like structure corresponds to the broad region with large $\Delta e_1$ on the the dynamical map (Fig.~\ref{fig:de_s6a}(a)). We check carefully the motion in these two regions and find that the orbits in Region 1 are librating around the stable ACR (Fig.~\ref{fig:s6a1}), while in Region 2, although all the two-body MMRs (i.e. MMR12, MMR23, and MMR13) occur, the Laplace angle $\varphi_0$ circulates (Fig.~\ref{fig:s6a2}), as indicated by the small $\chi^2$ value in the area of Region 2 in Fig.~\ref{fig:chi_s6a}(d).

A typical example of orbits in Region 1 is displayed in Fig.~\ref{fig:s6a1}. The mean motion ratios $n_1/n_2, n_2/n_3$ librate with very small amplitudes around the nominal values (2 and 1.5), implying that the semi-major axes are almost constants. The eccentricities $e_1, e_2, e_3$ oscillate around the periodic solution (symmetric ACR, indicated by horizontal lines in the corresponding panels). And the tiny libration of critical resonant angles (including the Laplace angle) around $0^\circ$ or $180^\circ$ reveals that the motion is deeply in the resonant chain, even we know that the initial condition is not exactly on the ACR solution. 

But for orbits in Region 2, as shown in Fig.~\ref{fig:s6a2}, the resonant angles of MMR12 and MMR23, i.e. $\theta_{12}$ and $\theta_{23}$, librate around $0^\circ$ with very small amplitude, as well as the apsidal difference  $\Delta\varpi_{23}$ librates around $180^\circ$. As a result, the resonant angle of MMR13, $\theta^1_{13}= \theta_{12}+ \theta_{23}+ \Delta\varpi_{23} = 3\lambda_3- \lambda_1- \varpi_1- \varpi_3$ (see Eq.\eqref{eq:res_angles} for the definition) librates and the system is surely in the MMR between the non-consecutive planets $m_1$ and $m_3$. However, due to the circulation of $\Delta\varpi_{12}$, the Laplace angle $\varphi_0 = \theta_{12} - \theta_{23} - \Delta\varpi_{12}$ eventually circulates, indicating that the system is not in the three-body MMR. The eccentricities of the stable ACR solution are also plotted in Fig.~\ref{fig:s6a2}, in which it can be seen that all $e_1, e_2$ and $e_3$ are far away from these ACR values.
This implies that the typical orbit in Region 2 is no longer librating around the stable ACR. The motion in Region 2 in fact follows a new stable mode in the resonant chain, in which the dynamics is actually dominated by the individual two-body MMRs. Since in such configuration each pair of planets is in the corresponding two-body MMR but the three-planet Laplace resonance does not happen, we refer to such configuration as ``joint two-body resonance''.

It is worth noting that a separatrix-like structure indicated by the red arrow in Fig.~\ref{fig:chi_s6a}(a) further splits Region 2 into two subregions. But the orbits from both sides of this structure are very similar to each other, characterised by libration of  $\theta_{12},\theta_{23},\theta_{13}^1$ and circulation of $\varphi_0$. The only difference between orbits from two sides is the oscillation centres of eccentricities,  $(e_1, e_2, e_3) \approx (0.032, 0.017, 0.009)$ for the left side (as the one shown in Fig.~\ref{fig:s6a2}) and  $(e_1, e_2, e_3) \approx (0.040, 0.012, 0.007)$ for the right side. Most probably, such fine structure in Region 2 attributes to the nonlinear terms in the perturbation \citep{rath2022}. We also note that the corresponding structure can also be found in the map of maximum of eccentricity variation (Fig.~\ref{fig:de_s6a}).

%\textcolor{red}{In Region 2b, we observe the similar patterns as the ones in Region 2a, i.e. the angles $\theta_{12},\theta_{23},\theta_{13}^1$ librate while the Laplace angle $\varphi_0$ circulates. The only difference between the motion of Region 2a and Region 2b is the oscillation centers of eccentricites:  $(e_1,e_2,e_3)=(0.0312,0.0167,0.00883)$ for Region 2a and $(e_1,e_2,e_3)=(0.0395,0.0115,0.00613)$ for Region 2b, respectively. Generally, we refer to Region 2a and Region 2b as Region 2 in the following text.}

%\textcolor{red}{According to the detailel analysis of the motion within three regions, we then obtain that: 
%\begin{itemize}
%\item The three regions in Fig.~\ref{fig:chi_s6a}(a) are corresponding to the most stable ones in Fig.~\ref{fig:de_s6a}(d);
%\item 
%The nearly vertical strip as indicated by the black arrow in Fig.~\ref{fig:chi_s6a}(a) is related to the broad region with large $\Delta e_1$ on the the dynamical map (Fig.~\ref{fig:de_s6a}(a));
%\item The separatrix-like structure as indicated by the red arrow in Fig.~\ref{fig:chi_s6a}(a) is associated to the large value of $\max \Delta e$ at about $(e_1,e_2)=(0.032,0.006)$ in Fig.~\ref{fig:de_s6a}(d).
%\end{itemize}}

Finally, at the bottom of the dynamical map (Fig.~\ref{fig:de_s6a} at $e_2\sim 0$) we observe a thin region with large eccentricity variation, where the eccentricities $(e_2, e_3)$ are excited mainly because the outer two-body MMR (MMR23) is close to its boundary, as indicated by the small $\chi^2$ value in the bottom area in  Fig.~\ref{fig:chi_s6a}(b).

\subsection{Motion around asymmetric ACRs at moderate eccentricity}

The symmetric ACR solution loses its stability at a specific point, and it bifurcates into two asymmetric ACR solutions. We select again arbitrarily an initial condition from the asymmetric ACR branch, and perform the similar calculations as for the symmetric case. The eccentricities at this initial point, indicated by the black solid circle labelled ``B''  in Fig.~\ref{fig:s6a6}, are $(e_1, e_2, e_3)=(0.0602, 0.0903, 0.0440)$. We call this case  ``Case B''. 

\begin{figure}
	\centering
	\includegraphics[width=\hsize]{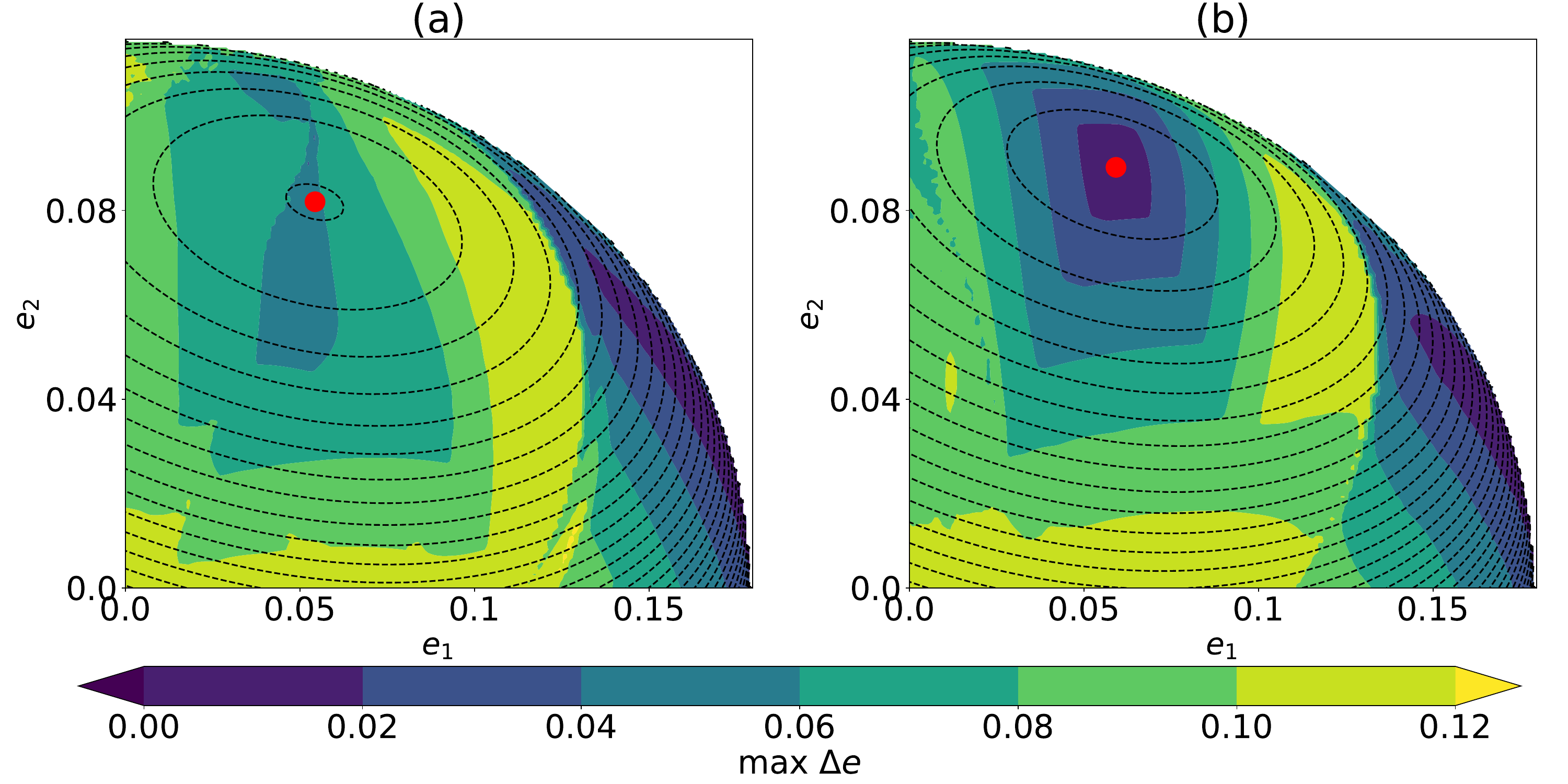}
	\caption{The same as Fig.~\ref{fig:de_s6a} but for Case B. The constants $L$ and $G$ are determined by the initial condition on the stable asymmetric ACR solution  (black solid circle in Fig.~\ref{fig:s6a6}). 
	(a) Around the unstable symmetric ACR solution with the same $L, G$. The corresponding eccentricities $e_1, e_2, e_3$ are $0.0569, 0.0846, 0.0531$ (red solid circle in Fig.~\ref{fig:s6a6}). (b) Around the stable asymmetric ACR solution. The colour in both panels indicates the value of $\max(\Delta e)$.}
	\label{fig:de_s6b}
\end{figure}

The resonant angles at this initial point of asymmetric ACR are $(\theta_{12}, \Delta\varpi_{12}, \theta_{23}, \Delta\varpi_{23}) = (26.7^\circ, -119.5^\circ, 1.16^\circ, 179.5^\circ)$. From these initial values, we obtain the corresponding constants $L$ and $G$. For comparison we also calculate the unstable symmetric ACR solution with the same $L$ and $G$, of which the eccentricities and resonant angles are $(0.0569, 0.0846, 0.0531, 0, 180^\circ, 0, 180^\circ)$, i.e. the red solid circle in Fig.~\ref{fig:s6a6}. Then, for both of these two initial conditions, we calculate the topology and dynamical maps, and present them in Fig.~\ref{fig:de_s6b}.
In this figure, the $\max(\Delta e)$ is chosen as the stability indicator.

Comparing the two panels in Fig.~\ref{fig:de_s6b}, we find that the topology of the unstable symmetric ACR is similar to the one of the stable asymmetric ACR, but their dynamical maps show different patterns. 
The non-zero values of $\max(\Delta e)\sim 0.04$ at the close proximity of the unstable symmetric ACR in Fig.~\ref{fig:de_s6b}(a) implies the orbits there deviate from the unstable stationary solution in the evolution.
%\textcolor{red}{\footnote{If the stationary solution $\bm{x}_0$ is Lyapunov stable, for all the initial condition $\bm{x}$ start from the vicinity of $\bm{x}_0$, the solution $\bm{x}(t)$ should also stay in its vicinity.}}. 
On the contrary, the stable asymmetric ACR in Fig.~\ref{fig:de_s6b}(b) is surrounded by regular orbits as indicated by the $\max(\Delta e) \sim 0$. 

\begin{figure}
	\centering
	\includegraphics[width=\hsize]{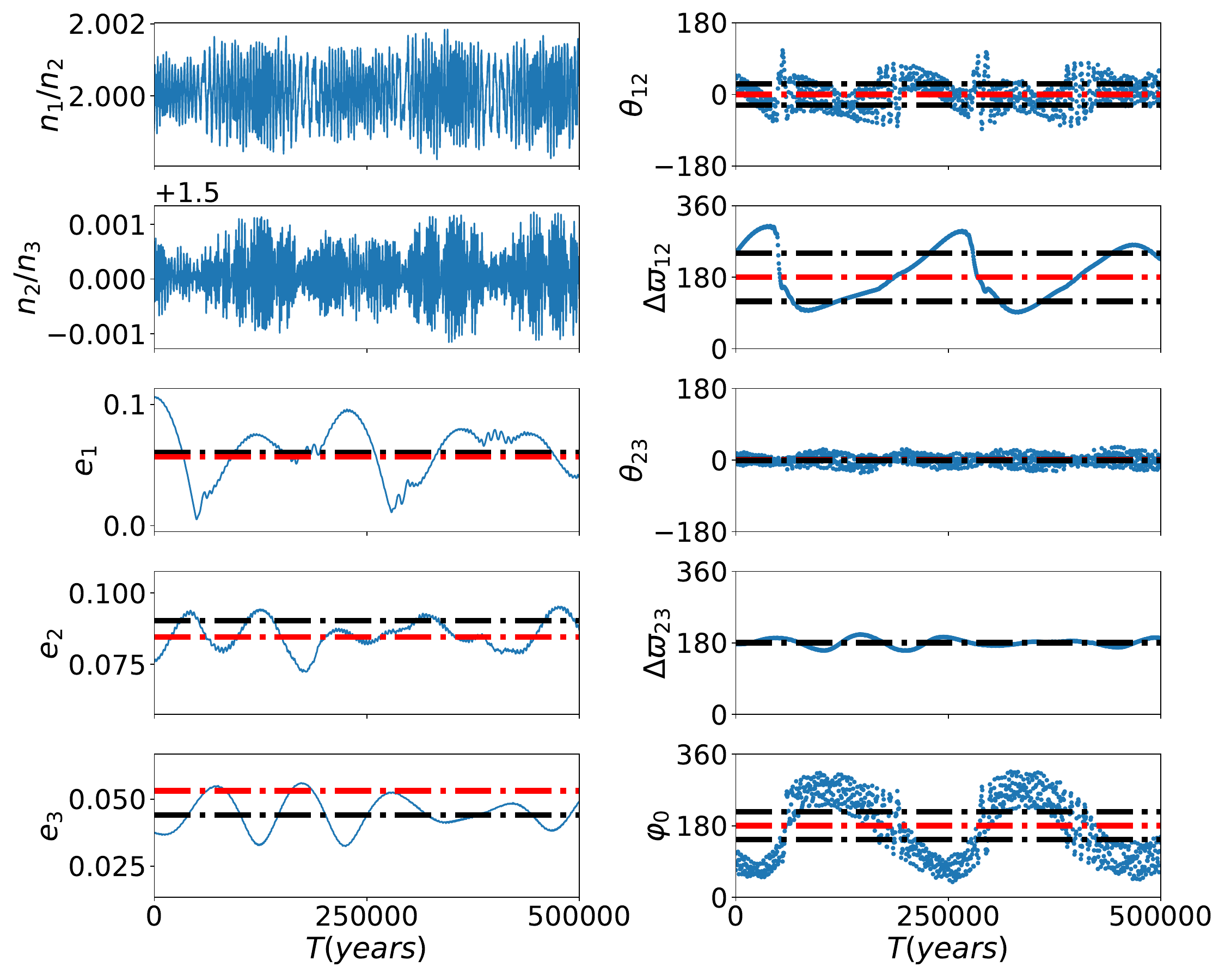}
	\caption{Horseshoe-like orbit near the unstable symmetric ACR. The unstable symmetric ACR and the stable asymmetric ACRs are indicated by the red and black horizontal lines, respectively.}
	\label{fig:s6b1}
\end{figure}

A closer look at the orbits around the unstable symmetric ACR reveals that they are still in  resonance, librating around both the asymmetric ACRs and the symmetric ACR with large amplitude, in a way similar to the horseshoe orbit around the Lagrange points $L_4-L_3-L_5$ in the circular restricted three-body problem \citep[see e.g.][]{murray2000}. One typical orbit in that region is shown in Fig.\ref{fig:s6b1}. Although it behaves chaotically due to the instability of symmetric ACR, the libration of the resonant angles indicate that the motion is still inside the resonant chain after the end of the numerical integration. Such horseshoe-like orbits can surely survive in the resonant chain in long term.

\begin{figure}
	\centering
	\includegraphics[width=\hsize]{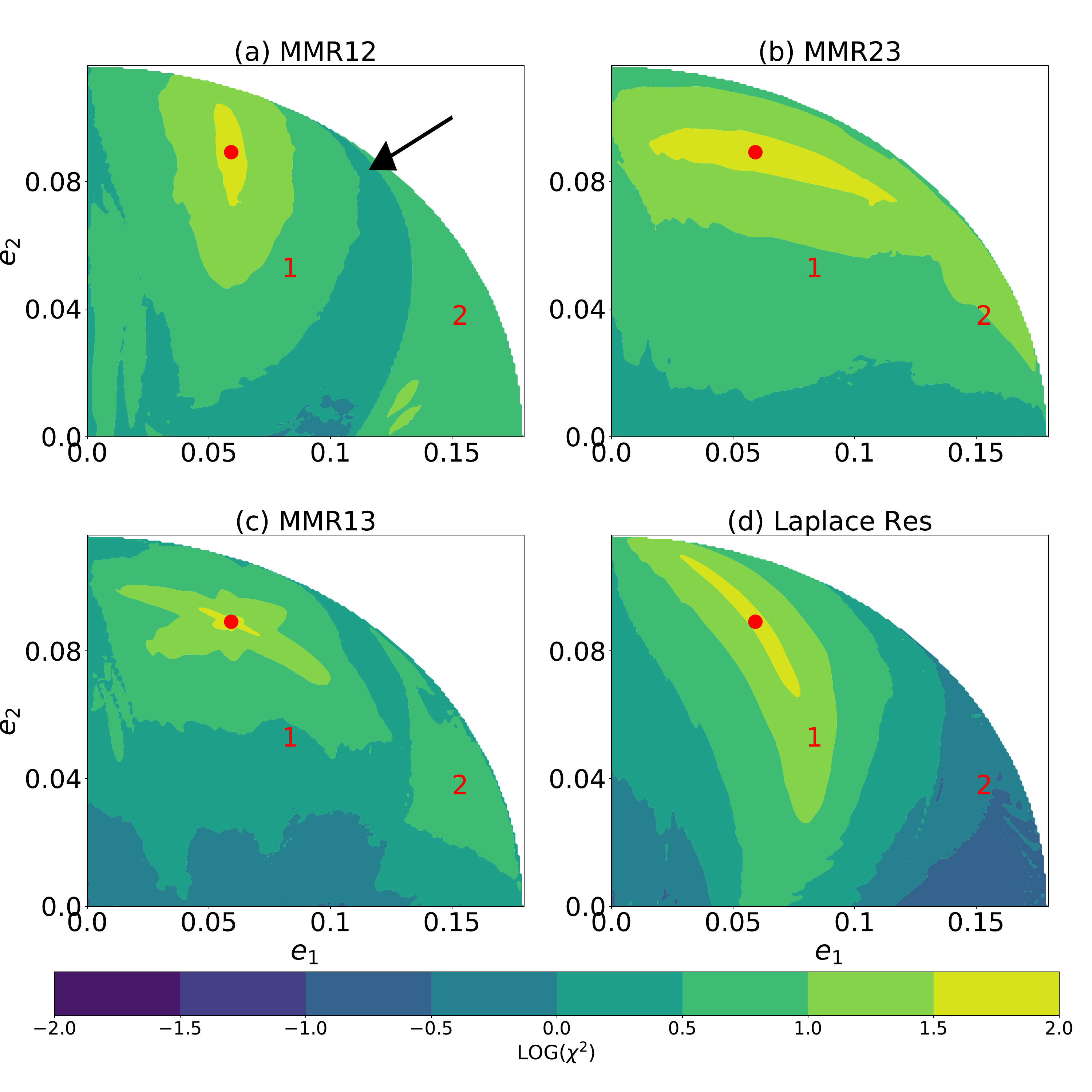}
	\caption{The same as Fig.~\ref{fig:chi_s6a}, but for Case B. The map is divided into two parts (Region 1 and Region 2) by a stripe-like structure of relatively smaller $\chi^2$ value, as indicated by the arrow in panel (a).  } 
	\label{fig:chi_s6b}
\end{figure}

On the other hand, the orbits near the stable asymmetric ACR solution are in a regular area (Fig.~\ref{fig:de_s6b}(b)), and they are protected by the resonant chain. We present in Fig.~\ref{fig:chi_s6b} the $\chi^2$ map on the representative plane where all the initial resonant angles are fixed as the ones of the asymmetric ACR solution. As can be seen in Fig.~\ref{fig:chi_s6b}, the close proximity around the stable ACR is deeply involved in the MMRs, as indicated by the large $\chi^2$ value (yellow colour). Consistent with the dynamical map in Fig.~\ref{fig:de_s6b}(b), such regular region occupies large phase space, with the eccentricities $e_1, e_2$ extending to $e_1\rightarrow 0.1, e_2\rightarrow 0.1$. For orbits inside the yellow region $(\chi^2>10)$ in Fig.~\ref{fig:chi_s6b}, the libration amplitudes of resonant angles $\theta_{12}, \theta_{23}$ are smaller than $10^\circ$. As a consequence, some three-body resonant angles $\Psi_{[M,N]}=M\theta_{12}+N\theta_{23}$ librate. We note that the libration of $\Psi$ in this case does not indicate a ``real'' three-planet resonance but is the necessary consequence of the libration of two-body resonance angles. 

Similar to the resonant structure near the stable symmetric ACR (Fig.~\ref{fig:chi_s6a}), the librating region of MMR23 extends in the $e_1$ (but not $e_2$) direction while the one of MMR12 extends along the $e_2$ direction. The MMR between planets $m_1, m_3$ (MMR13) and the Laplace resonance appear as the intersection of MMR12 and MMR23. With increasing displacement from the ACR, we notice a broad stripe with small $\chi^2$ in Fig.~\ref{fig:chi_s6b}(a) that divides the libration zone into two parts, namely Region 1 and Region 2 again. Such ``chaotic border''  is similar to the one in Fig.~\ref{fig:chi_s6a}(a) for Case A. And it has apparent correspondence in the dynamical map in Fig.~\ref{fig:de_s6b}(b) for Case B. 

\begin{figure}
	\centering
	\includegraphics[width=\hsize]{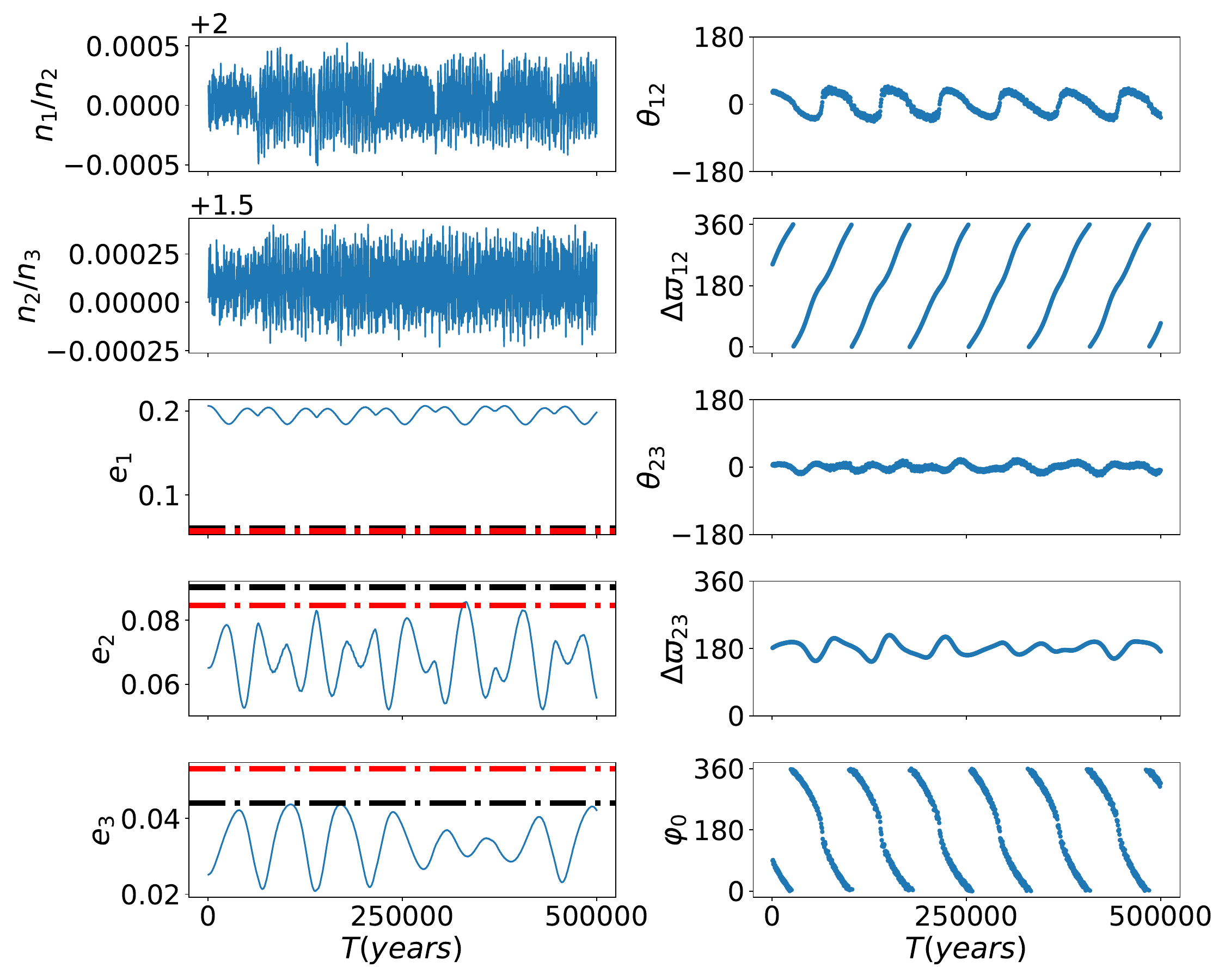}
	\caption{The same as Fig.\ref{fig:s6a2}, but the initial conditions are from the high eccentricity region of Fig.~\ref{fig:de_s6b} (see text).  }
	\label{fig:s6b2}
\end{figure}

Judging from the $\chi^2$ value, the orbits in Region 1 are involved in all three two-body MMRs and the three-body Laplace resonance. The motion is nearly the same as the one in Case A shown in Fig.~\ref{fig:s6a1}, only except that the resonance centre is at the asymmetric values. Close to the ACR solution, the motion is regular, and departing away from the ACR the irregularity increases, as shown by the $\max(\Delta e)$ value in Fig.~\ref{fig:de_s6b}(b). However, in Region 2, which is far away from the ACR solution, the small $\max(\Delta e)$ value indicates that the motion is regular, although the $\chi^2$ value implies that some resonances have broken. To show the orbital behaviour in Region 2, we illustrate a typical example in Fig.~\ref{fig:s6b2}. 

As shown in Fig.~\ref{fig:s6b2}, the period ratios $n_1/n_2, n_2/n_3$ still stay close to the nominal values (2,1.5). The eccentricities oscillate regularly around the centre $(e_1^*, e_2^*, e_3^*) = (0.19, 0.072, 0.035)$ that is quite far away from the ACR solution. Instead of librating around the ACRs (the unstable symmetric one and the two stable asymmetric ones), the resonant angles behave in a similar way as the ones in Case A illustrated in Fig.~\ref{fig:s6a2}, that is, all the two-body MMRs occur (at least one of the corresponding resonant angles librates) while the three-body Laplace resonance does not ($\varphi_0$ circulates). Therefore, the ``joint two-body resonance'' happens in this region of (relatively) high eccentricity.  

The orbits of ``joint two-body resonance'' found in both Cases A and B (Figs.~\ref{fig:s6a2}\&\ref{fig:s6b2}) reveal that this configuration can occur in a wide range of initial conditions. 
%The characteristics of this configuration may be parameterised by integral constants $L$ and $G$, or equivalently, the scaled AMD, $\delta$ in Eq.~\eqref{eq:delta}.
We note that such mode of stable ``joint two-body resonance'' is missed in our averaged model, i.e. the motion is not the resonant periodic orbits, but it is a  common outcome from planetary migration and tidal dissipation, as shown in \citet{morrison2020chains}.

\subsection{Stable motion inside the 1:2:3 resonant chain}

Basically, in the close neighbourhood of the ACR solutions as we have investigated in this paper, two types of stable motion of the 1:2:3 resonant chain can be found.  One is the stable ACR (resonant periodic orbits) and the associated quasi-periodic motion, where all the resonant angles, including the two-body MMRs and the Laplace resonance angle, are librating ($\chi^2>1$). The other one is the ``joint two-body MMR'', where $\theta_{12}, \theta_{23}$ and $\theta_{13}^1$ are librating while the Laplace angle $\varphi_0$ circulates. Judging from the dynamical maps and our long-term numerical simulations, the difference in stability between these two types of motion is negligible. Therefore, in the close vicinity of the nominal resonance, we might conclude that the individual two-body MMRs dominate the stability of the 1:2:3 resonant chain, while the libration of the three-body Laplace angle is not a necessary condition of the resonant chain but a necessary outcome of the two-body resonant angles' librating. 

%\textcolor{red}{We also note that both ACRs and ``joint two-body MMR'' exist for other mass parameters (i.e. different $m_1/m_2,m_2/m_3$) of 1:2:3 resonant chain, but the position as well as the strength (stability region on $(e_1,e_2)$ plane) depend strongly on the mass parameters}.

This feature can be explained through the Hamiltonian model. The perturbed part $\mathcal{H}_1$ of Hamiltonian in Eq.~\eqref{eq:hamil} only contains the interactions between each two planets. After the numerical averaging process (eliminating all the non-resonant term), up to the 1st order in planet-star mass ratio, the final averaged Hamiltonian $\bar{\mathcal{H}}$ in Eq.~\eqref{eq:aver_hamil} only contains the terms induced by the two-body MMRs but not the three-body resonance. Consequently, for a planetary system close to a resonant chain, the dynamics is mainly dominated by the two-body MMRs.

Therefore, if we can verify that a planetary system locates inside all the two-body MMRs (at least one critical resonant angle librates for each MMR), no matter how the Laplace angle $\varphi_0$ behaves, we will know for sure that the resonant chain is formed. In the practice of dynamical studies, the $\chi^2$ indicator can distinguish libration from circulation and thus provide the priori estimation of the stability in short-time numerical integration (only up to tens of resonant periods, empirically $10^4$-$10^5$ synodic periods), which is in fact easy to perform.
%\textcolor{red}{Just as we have shown in section 4.3 and 4.4, the $\chi^2$ indicator can distinguish libration from circulation. In the practice of dynamical studies, combined with the orbital stability indicated by $\max\Delta e$, we can identify how the resonant chain (or individual two body MMRs) affect the long-term stability.}

In addition, the three-body Laplace resonant angle $\varphi_0$ can be easily detected in transit data. Once the orbital periods and semi-major axes of the planets are determined, the behaviour of $\varphi_0$ is often adopted as an indicator to probe whether the planetary system is involved in a resonant chain \citep[e.g. ][]{jontof2016secure}. According to our analyses of the 1:2:3 resonant chain, the libration of the three-body resonant angle can almost ensure that the system is librating around a certain ACR. In this case, the dynamics and the possible behaviours of the orbital elements can be explained and predicted by studying the Hamiltonian topology and the dynamical maps on appropriate representative planes.

If a system is not in the close vicinity of the nominal resonances, i.e. if the offset from the exact period commensurability is considerable, the pure three-body MMRs (with absence of any two-body MMR) may play its role. Recently, \citet{rath2022} propose an analytical model (perturbed pendulum) of the chaotic dynamics near the resonant chain, and they demonstrate that the pure three-body resonances should appear as the secondary resonance with the critical angle defined as $\Psi_{[M,N]}=M\phi+N\psi$, where $\phi$ and $\psi$ are the critical angles of individual two-body MMRs: 
\begin{eqnarray}
\phi=k_{21}\lambda_2-k_{12}\lambda_1,~~~~\psi=k_{23}\lambda_2-k_{32}\lambda_3. 
\end{eqnarray}
Note the $\varpi$ terms have been neglected. Obviously, the [M,N]=[1,1] type is the largest secondary resonance \citep[for the example of the 1st order three-body resonance, see][]{2023ApJ...954...57C}. For the 1:2:3 resonant chain, it's just the Laplace resonance $\varphi_0=\lambda_1-4\lambda_2+3\lambda_3$. Specifically for the 1:2:3 resonant chain, \citet{antoniadou2022} find the stable region associated with this 0th-order three-body resonance. However, we did not detect these configuration, perhaps because our initial condition are always chosen at exact nominal resonance. More efforts are needed to investigate how the pure three-body resonance affects the resonant configuration and the long term stability.

\section{Resonant configuration via convergent migration} 
\label{sec:migrate}
So far, we have shown both by the Hamiltonian method and by numerical simulations what the orbital configurations might be in a planetary resonant chain. Generally, an MMR between two planets can be easily attained via planetary convergent migration, which can also lead multiple planets to form a resonant chain and then to achieve orbital configuration of high-eccentricity \citep[see e.g.][]{voyatzis2016}. In this section, we study whether the orbital configurations we have analysed above can be attained through convergent migration of planets. 

Due to the interaction between the planet and the circumstellar disk (gaseous disk or planetesimal disk), the migration of planet is quite common. For simplicity, in this paper the migration is simulated by adding an artificial dissipative force on the planet. Specifically, the torque acting on the $i$-th planet is assumed as \citep[see e.g.][]{delisle2015}:
\begin{equation}
\ddot{\bm{r}}_i=-\left[\frac{\bm{v}_i}{2\tau_{a,i}}+\frac{2(\bm{v}_i\cdot\bm{r}_i)\bm{r}_i}{r_i^2\tau_{e,i}}\right],
\end{equation}
where $\bm{r}_i,\bm{v}_i$ are the position and velocity vector,  and $\tau_{a,i}, \tau_{e,i}$ are timescales of orbital migration and eccentricity damping, for the $i$-th planet. Up to the first order in eccentricity, the effects of the dissipative force on the planetary semi-major axis and eccentricity read:
\begin{equation}
a_i(t)=a_i(t=0)\exp\left(-\frac{t}{\tau_{a,i}}\right),~~ e_i(t)=e_i(t=0)\exp\left(-\frac{t}{\tau_{e,i}}\right).
\end{equation}
Constant timescales were adopted in our simulations, and in order to fulfil the global convergent condition \citep[e.g.][]{beauge2022conditions,Wong2024}, different migration parameters will be chosen for different resonant chains.

\subsection{1:2:3 resonant chain}
To reach the 1:2:3 resonant chain, following \citet{beauge2022conditions}, we assumed that all three planets migrated inward and the migration timescales $\tau_{a,i}$ were set to be $0.5, 0.2, 0.1$\,Gyr for planets $m_1, m_2$ and $m_3$. The eccentricity damping timescales $\tau_{e,i}$ were set in such a way that for each planet $\kappa=\tau_{a,i}/\tau_{e,i}=50$. As we did for the dynamical maps, the initial semi-major axis of $m_2$ is set to be $a_2=1$\,AU. All planets start from coplanar and circular orbits and the initial period ratios $T_2/T_1$ and $T_3/T_2$ are set in a $15\times 15$ grid in the ranges of $[2.01,2.02]$ and $[1.51,1.53]$. The initial angles $\varpi_i, \lambda_i$ were all set to be $0^\circ$. This choice corresponds to a collinear configuration, in which the mutual distances between planets are minimum. All the orbits starting from the 225 initial conditions were then integrated for a total time span of $T=10^7$\,yr.

During the numerical integrations, we keep tracking the behaviour of the resonant offset $\Delta_{ij}=n_i/n_j-k_j/k_i$ between the adjacent planets. Once all the resonant offsets are smaller than a given level $\Delta^c_{ij}$, we then monitor the behaviour of resonant angles by checking their $\chi^2$ and classify the motion. It should be noted that the critical $\Delta^c_{ij}$ can be chosen empirically, and after some tests, we set $\Delta^c_{ij} = 20 \times (m_1+m_2+m_3)/M\times k_j/k_i$. 

\begin{figure}
	\centering
	\includegraphics[width=\hsize]{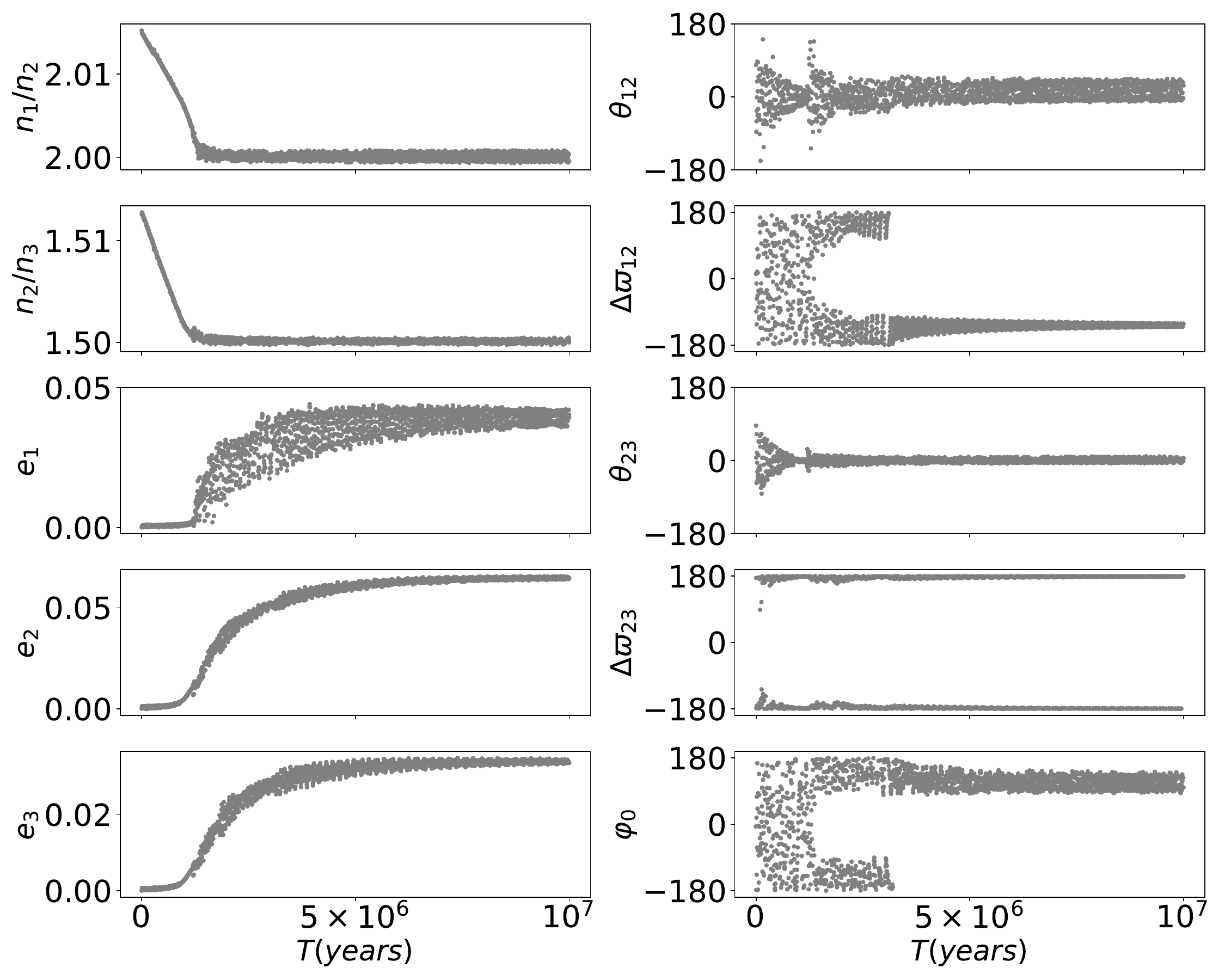}
	\caption{Formation and evolution of the 1:2:3 resonant chain via convergent planetary migration. The left panels show the temporal evolutions of period ratios and eccentricities, and the critical angles are shown in the right panels. The planetary system is captured in symmetric configuration $S_6$ firstly, subsequently evolves along the $A_6$ family and finally librates around the asymmetric configuration.}
	\label{fig:123a}
\end{figure}

For the 1:2:3 resonant chain, we found that only two types of resonant configuration are achievable in our numerical simulations. The first one is that the planetary system attains the resonant chain and eventually librates around the stable ACR solutions ($S_6, A_6$). One such example is shown in Fig.~\ref{fig:123a}. As the planets migrate, both period ratios $n_1/n_2$ and $n_2/n_3$ decrease, while the eccentricities remain close to zero. After the system reaches the 1:2:3 resonant chain at $t\approx 10^{6}$\,yr, the period ratios are subsequently locked into the resonance and the eccentricities are excited from then on. Meanwhile, all the angles begin to librate around the values of the symmetric ACR,  $(\theta_{12}, \Delta\varpi_{12}, \theta_{23}, \Delta\varpi_{23}) = (0,\pi,0,\pi)$, i.e. the $S_6$ family. Once the bifurcation point is reached at $t\approx 2.4\times10^6$\,yr, the system follows the asymmetric family $A_6$ and finally librates around the asymmetric ACR at about $(e_1^*, e_2^*, e_3^*, \theta_{12}^*, \Delta\varpi_{12}^*, \theta_{23}^*, \Delta\varpi_{23}^*)= (0.0371, 0.0623, 0.0345, 24.6^\circ, -133.5^\circ, 0.5^\circ, 178.9^\circ)$.

\begin{figure}
	\centering
	\includegraphics[width=\hsize]{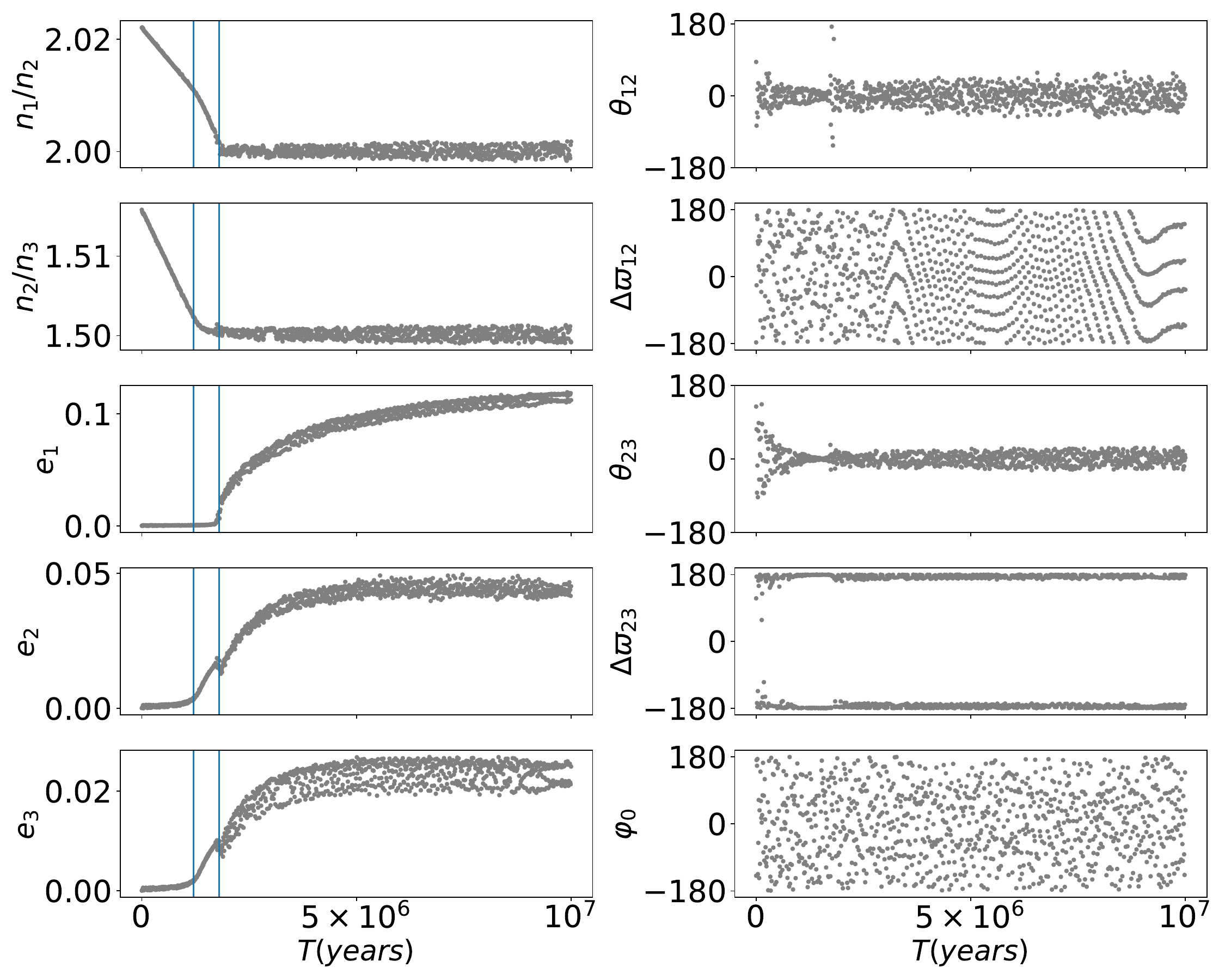}
	\caption{The same as Fig.~\ref{fig:123a} but for a ``joint two-body MMR'' case (see text). In the final resonance configuration, $\Delta\varpi_{12}$ and $\varphi_0$ circulate while  $\theta_{12}, \theta_{23}$, and $\Delta\varpi_{23}$ librate. The vertical lines in the left panels indicate two moments $t_1=1.2$\,Myr and $t_2=1.8$\,Myr. }
	\label{fig:123b}
\end{figure}

The second one is the ``joint two-body MMR'', of which an example is shown in Fig.~\ref{fig:123b}. Similar patterns as in Fig.~\ref{fig:123a} can be observed for $(n_1/n_2, n_2/n_3, e_1, e_2, e_3)$: once the system encounters the 1:2:3 resonant chain, the eccentricities monotonously increase while the period ratios remain at the nominal values. However, the resonant angles seem not be affected by the ACRs of the resonant chain, and the three-body Laplace resonant angle ($\varphi_0$) always circulates but not librates. All these regular dynamics inside the 1:2:3 resonant chain are just as predicted by our previous analyses.

Comparing the orbital evolution in Fig.~\ref{fig:123a} and in Fig.~\ref{fig:123b}, we find that in the latter case the period ratio $n_2/n_3$ has been trapped to 1.5 (3:2) at $t_1\approx 1.2\times 10^6$\,yr. At this moment, $n_1/n_2$ has not reached the nominal value of 2.0, but a change of the relative migration rate can be seen clearly in the top left panel of Fig.~\ref{fig:123b}. The 1:2 commensurability between the inner two planets however seems to occur later at $t_2\approx 1.8\times 10^6$\,yr. Its influence on the resonant chain is reflected by the small ``spikes'' in the $e_2$ and $e_3$ variation in two bottom left panels.

 The evolution in Fig.~\ref{fig:123b} seemingly implies that the dynamical configuration a resonant chain finally takes might be influenced by the forming sequence of individual two-body MMRs. But in fact, which configuration the resonant chain will occupy depends on the masses of planets, the initial conditions and the migration rates in a very complicated way.  We check carefully all the 225 planetary systems in our simulations, and find that totally 180 out of 225 systems are trapped in the motion around the ACR as in Fig.~\ref{fig:123a} and the rest 45 systems are finally trapped in the joint two-body MMR. Unfortunately, we were unable to find out a rule regarding which type of motion a system will finally take in the resonant chain, libration around ACR or joint two-body MMR.

% about which type of motion a system finally will take in the resonant chain, libration around ACR or joint two-body MMR, was found to be random, i.e. it does not depend on the initial period ratios. 
%of motion type of ``joint two-body MMR'' is more likely to be formed if the 2:3 MMR between the outer two planets occurs first and the 1:2 MMR between the inner two planets join in later.
%\textcolor{blue}{Unfortunately, for other planetary masses than the ones as in the Kepler-51 system, our calculations show that the configuration of joint two-body MMR occupies only a small fraction of area in the representative $(e_1, e_2)$ plane, and it can hardly be attained via convergent migration. We would like to leave the probability and formation of such resonant configuration for a separate paper in future. }

%\begin{figure}
%	\centering
%	\includegraphics[width=\hsize]{caprslt.jpg}
%	\caption{The dependence of final configuration of the resonant chain on the initial period ratios. The $15\times 15$ initial grid of $n_1/n_2$ and $n_2/n_3$ is coloured according to the final configuration of the resonant chain: big blue solid circles indicate the motion around ACR and small red dots for the joint two-body MMR.  }
%	\label{fig:caprslt}
%\end{figure}

Aiming to check whether the 1:2:3 resonant chain can be formed through planets' migration and whether the periodic solutions can also be obtained in this way, we have set the initial period ratios very close to the target resonant chain in our simulations. We performed some extra simulations from initial period ratios far from the resonant chain ($n_1/n_2>2.1$, $n_2/n_3>1.6$), and found that the capture into the resonance chain can also succeed, and all the captured systems are in either stable ACR or the joint two-body MMRs. The eccentricities may also affect the capture and the subsequent evolution. We note that the eccentricities will be inhibited by the eccentricity damping effect in the numerical model. However, for cases where the system encounters the 1:2:3 resonant chain in eccentric orbits (typically $e\in [0.05, 0.1]$), the resonant chain can also be achieved but the system may temporarily stay around the resonant chain with relatively large libration amplitudes. Such a large eccentricity variation may drive the system out of the resonant chain and eventually leaves only individual two-body MMRs.
Therefore, in the following investigations, we only focus on the capture starting near the resonant chain and from near circular orbits.

\subsection{Other resonant chains}
So far, our investigations are on the 1:2:3 resonant chain. The consistence between the averaged Hamiltonian and the numerical simulations motivates us to extend our semi-analytical model to other resonant chains. As far as we know, previous works, e.g. \citet{morrison2020chains, kajtazi2023mean}, focus mainly on the 1st order resonant chains with $(k_{21}-k_{12} = k_{32}-k_{23}=1)$. To show how our methods may work for general resonant chains, below we present our analyses on some other resonant chains. We also note that starting from circular orbits, the ACR solutions of a resonant chain may emerge as symmetric configurations with critical resonant angles being $0, \pi$, or as asymmetric ones.

\subsubsection{Cases starting from symmetric configuration}

We will show two cases, a 2:3:5 resonant chain and a 3:5:7 resonant chain.
 The former is a resonant chain of mixed orders consisting of the 1st and 2nd order MMRs between adjacent planets, while the latter is a resonant chain with two second order MMRs. Although no exoplanetary system has been confirmed in these resonant configuration, we present these two examples to demonstrate the availability of methods introduced in this paper for higher-order resonances. Meanwhile the existence of such configuration but absence of such real systems also implies that the stability of such configuration is weak. 

For the 2:3:5 resonant chain, we consider the fictitious planetary system with central star of mass $M=1$ and three planets $m_1= m_2= m_3= m= 10^{-6}$, and as before, the semi-major axis of $m_2$ is set as $a_2=1$\,AU. Using the same technique we have applied to the 1:2:3 resonance chain, we obtain the symmetric and asymmetric ACR solutions and their stabilities for the 2:3:5 resonant chain. Following the definition of critical angles in Eq.\eqref{eq:critang}, we define
\begin{equation}
\left\{
\begin{aligned}
&\theta_{12}=3\lambda_2-2\lambda_1-\varpi_1, & \Delta\varpi_{12}=\varpi_1-\varpi_2, \\
&\theta_{23}=5\lambda_3-3\lambda_2-2\varpi_2, & \Delta\varpi_{23}=\varpi_2-\varpi_3,
\end{aligned}
\right.
\end{equation}
and we find that the symmetric family with $(\theta_{12}, \Delta\varpi_{12}, \theta_{23}, \Delta\varpi_{23}) = (0,\pi,\pi,\pi)$ is the unique stable ACR arising from the circular configuration. Along this family, the planets may evolve to the configuration of high eccentricities when the angular momentum $G$ decreases. This stable symmetric family finally halts at $(e_1, e_2, e_3)=(0.311,0.275,0.294)$ where the planetary system may experience close encounters. Our detailed analyses on the dynamical maps reveal that all the regular motions we found in this 2:3:5 resonant chain are characterised by quasi-periodic libration around the ACR.
 %Since the resonant dynamics near close encounters is beyond the scope of this paper, we mainly focus on the stable resonant motion at small eccentricities. 
 
To investigate the formation and evolution of this resonant chain, we perform numerical simulations of the convergent migration and resonance capture again. The initial period ratio $T_2/T_1$ is uniformly distributed in the range $[1.51,1.53]$, and $T_3/T_2$ in $[1.68,1.71]$. All planets' initial orbits are assumed to be coplanar and circular, and the angles $\varpi_i=0, \lambda_i=0$. The migration timescales for three planets $\tau_{a,1}, \tau_{a,2}, \tau_{a,3}$ are set as 0.3, 0.2, 0.1\,Gyr, respectively, and the eccentricity damping time scale $\tau_{e,i}$ is set to satisfy $\kappa= \tau_{a,i}/\tau_{e,i}=100$.

\begin{figure}
	\centering
	\includegraphics[width=\hsize]{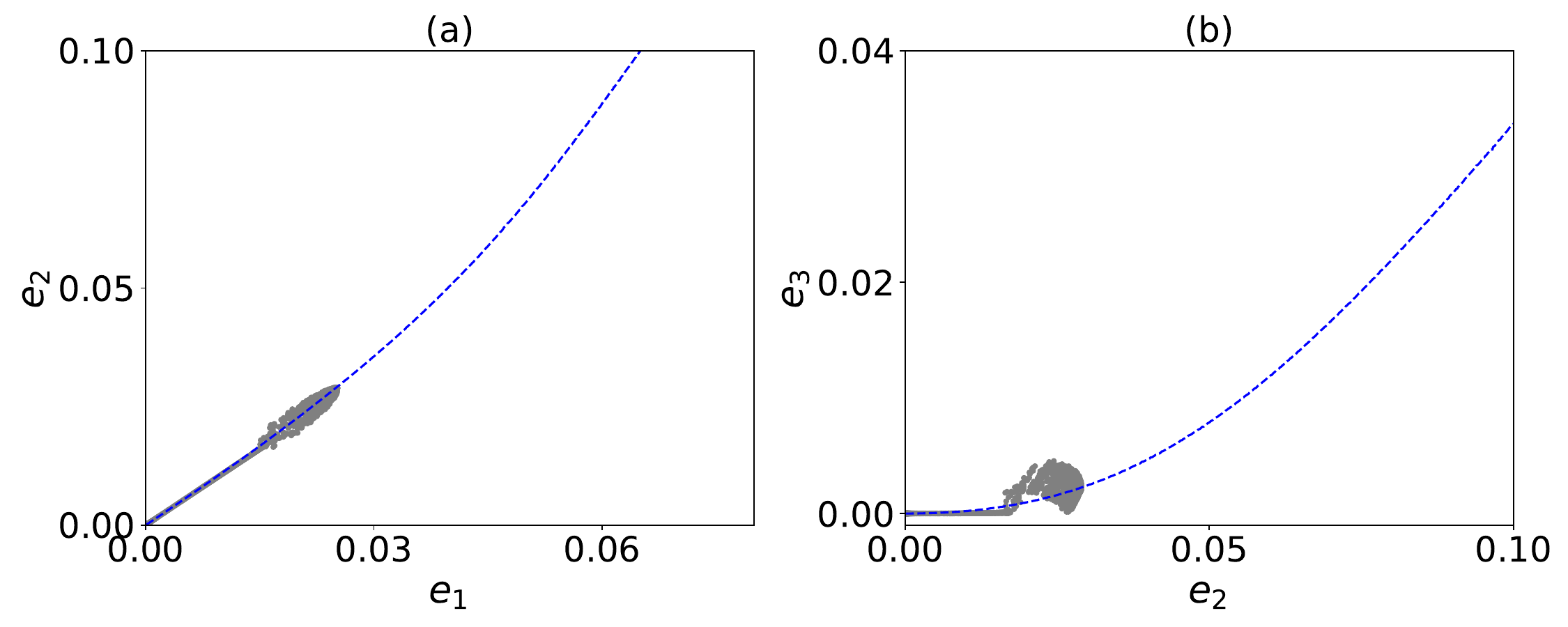}
	\caption{Resonant capture and evolution of the 2:3:5 resonant chain on the eccentricities plane. The stable $(\theta_{12}, \Delta\varpi_{12}, \theta_{23}, \Delta\varpi_{23}) = (0,\pi,\pi,\pi)$ ACR is indicated by blue dashed lines while the grey dots show the migrating path of the system evolving to higher eccentricities. }
	\label{fig:235}
\end{figure}

We find that all of 225 fictitious systems are captured into the 2:3:5 resonant chain and evolve along the stable symmetric family $(\theta_{12}, \Delta\varpi_{12}, \theta_{23}, \Delta\varpi_{23}) = (0,\pi,\pi,\pi)$. Instead of plotting the evolutions of orbital elements and the resonant angles, we depict on eccentricities planes the migrating path along this ACR family in Fig.~\ref{fig:235}. It's clear that this stable ACR family guides the migration of the planetary system. The evolution eventually stops at $(e_1^*, e_2^*, e_3^*)=(0.0460, 0.0612, 0.0122)$ in this example, but surely the final status depends on the migrating parameters. If all the torques on planets disappear at this moment, the system will oscillate around the stable ACR at $(e_1^*,e_2^*,e_3^*)$ in long term. 

\begin{figure}
	\centering
	\includegraphics[width=\hsize]{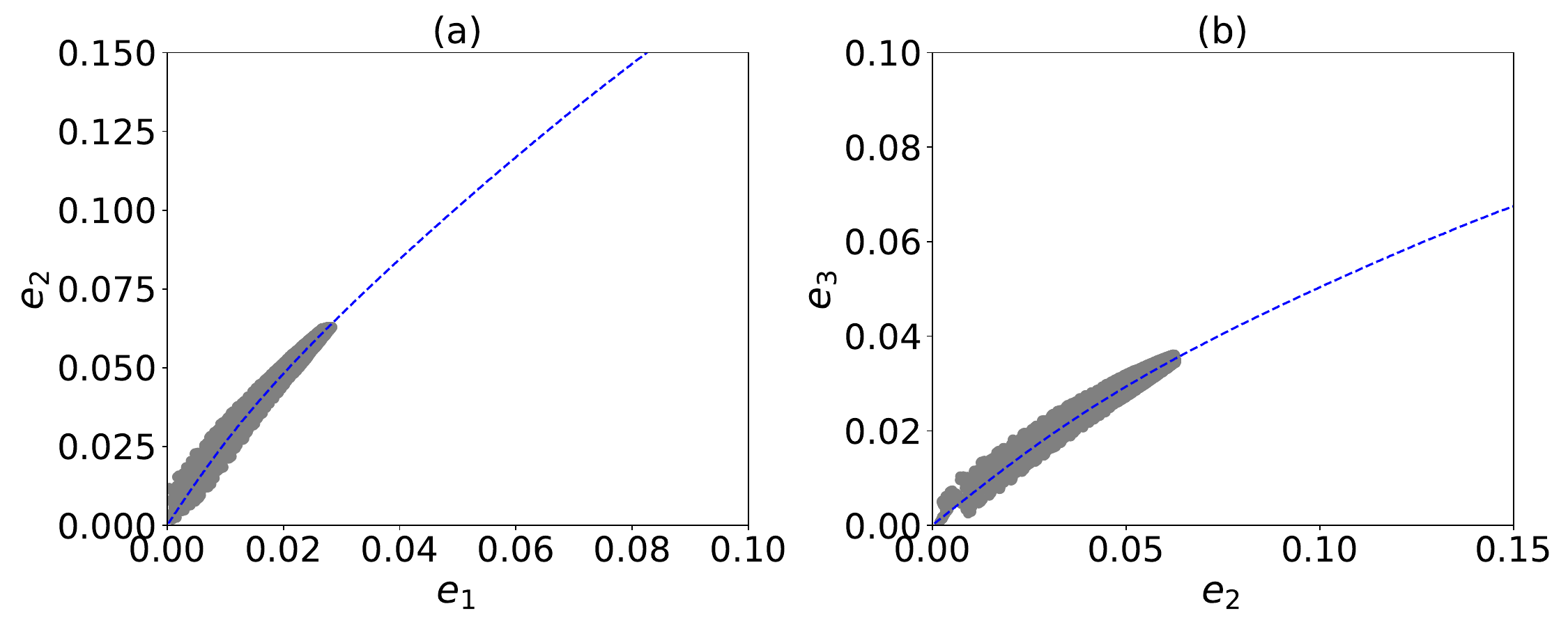}
	\caption{The same as Fig.~\ref{fig:235} but for the 3:5:7 resonant chain. The blue dashed lines indicate the location of the stable ACR of $(\theta_{12}, \Delta\varpi_{12}, \theta_{23}, \Delta\varpi_{23}) = (\pi,\pi,\pi,\pi)$.}
	\label{fig:357}
\end{figure}

For the case of the 3:5:7 resonant chain, the critical angles are defined as in Eq.\eqref{eq:critang}
\begin{equation}
\left\{
\begin{aligned}
&\theta_{12}=5\lambda_2-3\lambda_1-2\varpi_1, & \Delta\varpi_{12}=\varpi_1-\varpi_2, \\
&\theta_{23}=7\lambda_3-5\lambda_2-2\varpi_2, & \Delta\varpi_{23}=\varpi_2-\varpi_3,
\end{aligned}
\right.
\end{equation}
and the stable symmetric ACR of $(\theta_{12}, \Delta\varpi_{12}, \theta_{23}, \Delta\varpi_{23})= (\pi,\pi,\pi,\pi)$ was found to dominate the regular motion when the eccentricities are small. This symmetric ACR solution loses its stability at $(e_1, e_2, e_3)= (0.196, 0.238, 0.0845)$ and it bifurcates into two branches of asymmetric families in a similar way as in the 1:2:3 resonant chain. Along the asymmetric family, the eccentricities might finally attain $(e_1, e_2, e_3)= (0.267,0.568,0.617)$. Since both the 3:5 and 5:7 MMRs are relatively high order resonances, the resonant capture can be observed only if the migration is very slow. We show an example of the formation of 3:5:7 resonant chain in Fig.~\ref{fig:357}, where the migration timescales $\tau_{a,1}, \tau_{a,2}, \tau_{a,3}$ for three planets are 5, 2, and 1\,Gyr, respectively. In fact, if a faster migration rate is applied, most of the systems will skip the 3:5:7 resonance chain and fall into a more compact configuration. We note that such a phenomenon has been reported in \cite{kajtazi2023mean}. 
%Just like the 2:3:5 resonant chain, we are only interested in the resonant capture and evolution in low-eccentricity region.}
%only the libration around the stable ACR can be safeguarded by the resonant chain.

To produce the 3:5:7 resonant chain as in Fig.~\ref{fig:357}, we started our simulations from initial $(n_1/n_2, n_2/n_3) = (1.668, 1.402)$,  which are very close to the nominal values in the resonant chain. And an eccentricity damping effect with $\kappa=100$ was imposed in our simulations. The system evolves along the stable symmetric ACR $(\theta_{12}, \Delta\varpi_{12}, \theta_{23}, \Delta\varpi_{23})=(\pi,\pi,\pi,\pi)$ and the eccentricity damping effect finally halts the system around the configuration $(e_1^*, e_2^*, e_3^*) = (0.0272, 0.0627, 0.0356)$.

\subsubsection{Cases starting from asymmetric configuration}

For all the aforementioned cases (1:2:3, 2:3:5, and 3:5:7 resonant chains), the stable ACR solutions arising from the circular configuration are always symmetric ones. As a result, the three-body Laplace angle $\varphi_0$ is always librating around the symmetric value, 0 or $\pi$. In this sense, if the $\varphi_0$ is found to librate around 0 or $\pi$, the system must be in such a resonant chain. However, this rule breaks for the 2:3:4 resonant chain. \citet{siegel2021resonant} find that the unreduced three-body critical angle $2\varphi_0$ (they adopt $2\varphi_0=2\lambda_1+4\lambda_3-6\lambda_2$ instead of $\varphi_0$) for the 2:3:4 resonant chain shift to asymmetric values $\sim$162$^\circ$ and $\sim$198$^\circ$, instead of the symmetric value $2\pi$. 

\begin{figure}
	\centering
	\includegraphics[width=\hsize]{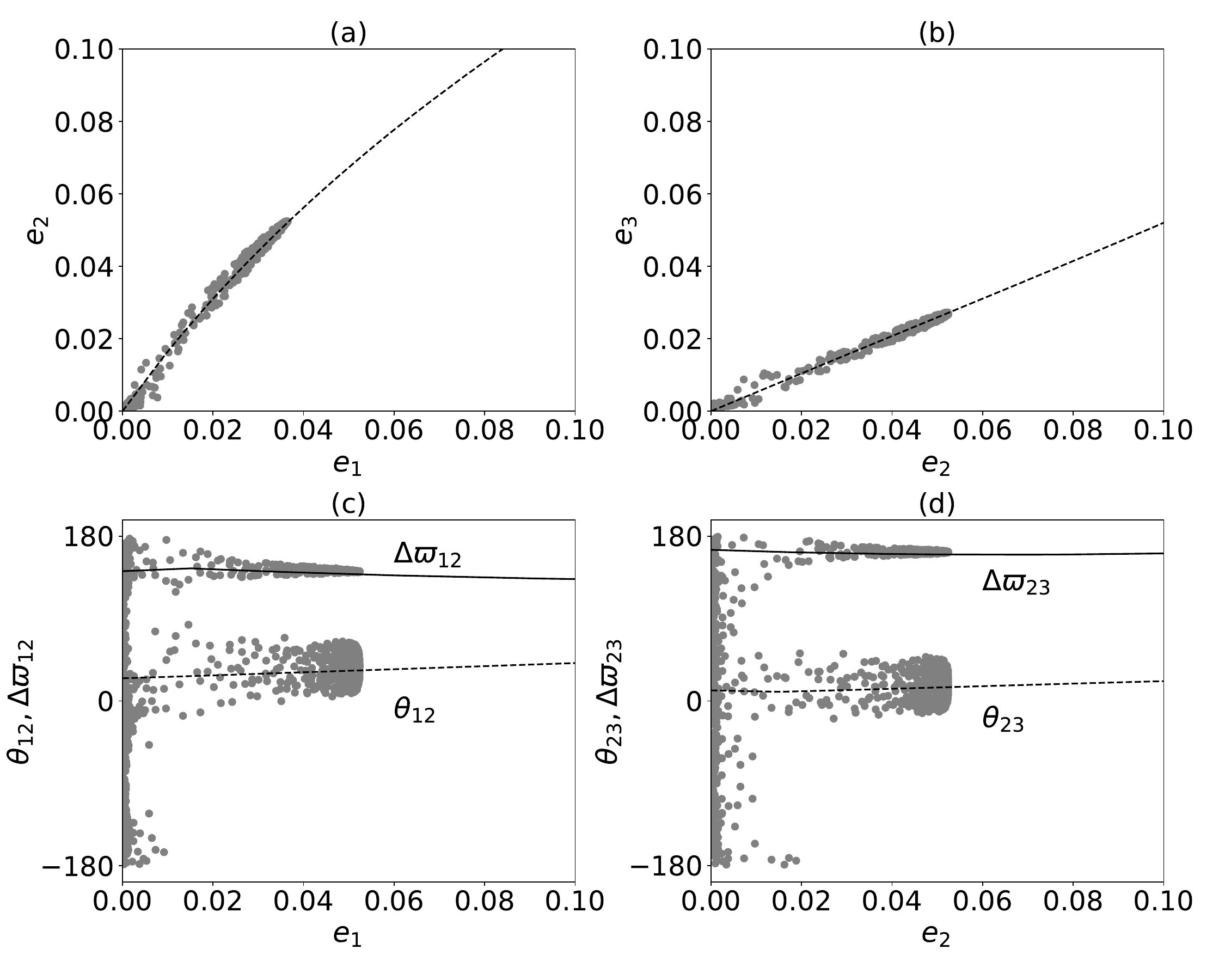}
	\caption{Resonant capture and evolution of the 2:3:4 resonant chain. The black lines indicate the location of the asymmetric ACR solution arising from the circular configuration. The grey dots stand for the capture and evolution of planets by the resonant chain in the numerically simulated convergent migration. }
	\label{fig:234}
\end{figure}

Motivated by this interesting result, we apply our model to the 2:3:4 resonant chain with mass parameters $M=1, m_1=m_2=m_3=10^{-5}$, similar to the ones in \citet[][]{siegel2021resonant}. The the resonant angles are
\begin{equation}
\left\{
\begin{aligned}
&\theta_{12}=3\lambda_2-2\lambda_1-\varpi_1, & \Delta\varpi_{12}=\varpi_1-\varpi_2, \\
&\theta_{23}=4\lambda_3-3\lambda_2-\varpi_2, & \Delta\varpi_{23}=\varpi_2-\varpi_3.
\end{aligned}
\right.
\end{equation}
Different from the 1:2:3 case, the final averaged Hamiltonian $\bar{\mathcal{H}}$ for the 2:3:4 resonant chain is not $2\pi$-periodically dependent on the angles $(\theta_{12}, \Delta\varpi_{12}, \theta_{23}, \Delta\varpi_{23})$. Following the procedure in Appendix \ref{appendix1}, we find that the degeneracy of these four resonant angles are $(2, 2, 2,1)$, respectively. Instead of the symmetric ACRs, we found two stable asymmetric ACRs in the near-circular configuration. We present these ACRs by black lines in Fig.~\ref{fig:234}. For better view, only one of the two asymmetric families is plotted in Fig.~\ref{fig:234}, the other family simply takes the same values in terms of $(e_1, e_2, e_3)$ and the symmetrically opposite values of the resonant angles.

Starting from circular orbits and initial period ratios $T_2/T_1$ uniformly in [1.51,1.53] and $T_3/T_2$ in [1.34,1.36], three planets $m_1, m_2, m_3$ in 225 fictitious systems with different initial period ratios are forced to migrate with migration timescales $\tau_{a,i} =1, 0.5, 0.2$\,Gyr, while the eccentricity damping effect satisfies $\kappa=\tau_{a,i}/\tau_{e,i}=50$. 
All these systems are eventually captured into either one of the two asymmetric ACR families. And a typical example of resonant capture and evolution is overlaid on the stable ACRs in Fig.~\ref{fig:234}.

As shown in Fig.~\ref{fig:234}(c)(d), once the system reaches the 2:3:4 resonant chain just after the migration begins when the orbits are still near circular, the resonant angles are immediately locked into the asymmetric value  $(\theta_{12}, \Delta\varpi_{12}, \theta_{23}, \Delta\varpi_{23}) = (27.0^\circ, 144.8^\circ, 9.9^\circ, 162.7^\circ)$ and the Laplace angle librates around $2\varphi_0=162.7^\circ$ (Mod[$360^\circ$]). The system evolves subsequently along the asymmetric ACR family and eventually librates around the ACR located at $(e_1^*, e_2^*, e_3^*) = (0.0357, 0.0512, 0.0265)$ and $(\theta_{12}^*, \Delta\varpi_{12}^*, \theta_{23}^*, \Delta\varpi_{23}^*) = (33.4^\circ, 138.4^\circ, 15.0^\circ, 160.0^\circ)$. 

Therefore, for the 2:3:4 resonant chain, our numerically averaged Hamiltonian successfully predicts the librating centres of resonant angles, as well as the adiabatic evolution induced by the convergent migration. In fact, starting from circular orbits, asymmetric ACR solutions are also found in the 3:4:6 and the 12:15:20 resonant chains \citep[][]{siegel2021resonant}, and our calculations detected three and four stable ACR families starting from the circular configuration, respectively. For reference, we list in Table~\ref{tab:acrno} the basic properties of the ACR solutions in some low-order three-planet resonant chains.

\begin{table}[htbp]
	\centering
	\caption{The properties of ACR solutions in low-order three-planet resonant chain. The resonant chain is given in the first column. The degeneracies (see Appendix~\ref{appendix1}) of the resonant angles $(\theta_{12}, \Delta\varpi_{12}, \theta_{23}, \Delta\varpi_{23})$ defined as in Eq.\eqref{eq:critang} are in the second column. The third column gives the Laplace angle $\varphi_0=l_1\lambda_1+l_2\lambda_2+l_3\lambda_3$ by the coefficients $(l_1,l_2,l_3)$. The number of ACR families we found are given in the fourth column, with suffixes `S' and `A' indicating `symmetric' and `asymmetric', respectively. The libration centre(s) of $\varphi_0$ are in the fifth column.}
	\begin{tabular}{|c|c|c|c|c|}
		\hline
		Chain & Degeneracy &  $\varphi_0$ & ACRs & Centre \\
		\hline
		1:2:3 & (1,1,1,1) & $(1,-4,3)$ & 1S    & 180 \\
		%\hline 
		1:2:4 & (1,1,1,1) & $(1,-3,2)$ & 1S    & 180 \\
		%\hline
		2:3:4 & (2,2,2,1) & $(1,-3,2)$ & 2A &  $\pm82$\\
		%\hline 
		2:3:5 & (1,1,1,1) & $(4,-9,5)$ & 1S    & 180 \\
		%\hline
		2:3:6 & (2,2,2,1) & $(1,1,-2)$ & 2A    & $\pm 90$ \\
		%\hline
		3:4:5 & (1,1,1,1) & $(3,-8,5)$ & 1S    & 180 \\
		%\hline
		3:4:6 & (3,3,3,1) & $(1,-2,1)$ & 1S+2A & 180, $\pm51$ \\
		%\hline
		3:4:8 & (1,1,1,1) & $(3,-5,2)$ & 1S    & 180\\
		%\hline
		3:5:7 & (1,1,1,1) & $(3,-10,7)$ & 1S    & 0 \\
		%\hline
		3:6:8 & (1,1,1,1) & $(1,-5,4)$ & 1S    & 180 \\
		%\hline
		4:6:9 & (1,1,1,1) & $(2,-5,3)$ & 1S    & 180 \\
		%\hline
		12:15:20 & (4,4,4,1) & $(1,-2,1)$ & 4A & $\pm135, \pm45$ \\
		\hline
	\end{tabular}%
	\label{tab:acrno}
\end{table}%

From Table~\ref{tab:acrno}, we see the number of stable ACR solutions is equal to the highest degeneracy number of the corresponding resonant angles. This number, and the symmetry properties (symmetric or asymmetric) of these stable ACR families, might depend not only on the resonant orders, but also on the degeneracies of the involved two-body MMRs. A general model for this question is still needed. 

%\textcolor{blue}{We may conclude that the asymmetric ACRs starts from the circular configuration are mainly caused by the low-order MMR between the non-adjacent planes $(m_1,m_3)$ Strongly depend Generally, we can develop a simple criterion for the existence of asymmetric ACRs: if all the two-body MMRss share the same resonant order: i.e $q_{13}=q_{12}=q_{23}$ (see Eq.\eqref{eq:res_angles}), there exists the asymmmetric family which bifurcates from the circular family.  For the results reported (librating center of the three-body-Laplace angle $\varphi_0$) SOME COMMENTS ABOUT THE ACRS AND MOTION TYPES IN THIS RESONANT CHAIN ARE NEEDED HERE, TOO. }

\subsection{Remarks on final resonant configurations}

As we have shown above, starting from circular configuration ($e_1=e_2=e_3=0$), the stable families of periodic orbits (ACR solutions, either symmetric or asymmetric) always exist in the resonant chains. These solutions guide the formation and adiabatic evolution of the resonant chains during the planetary migration. Depending on the migration parameters ($\tau_{a,i}, \tau_{e,i}$) and initial conditions, a triple-planetary system finally attains different configurations of resonant chain, including quasi-periodic motion around the ACR solutions and the joint two-body MMRs. 

After the three planets have been trapped in a resonant chain, they may acquire high eccentricities as the system evolves along the path defined by the ACR solutions as the convergent migration continues. There are also some ACR solutions in the high eccentricity region that do not originate from the circular configuration \citep[][]{voyatzis2016}. Since our methods of finding the ACR solutions work for arbitrary eccentricities, these high-eccentricity configurations can be calculated, but they cannot be attained through the adiabatic process from circular orbits as described above. Considering the fact that the typical eccentricities in the resonant chain are small \citep[see e.g.][]{sun2017terrestrial}, we skip these configurations in this paper.

Another factor that can affect the final orbital configuration in a resonant chain is the planetary masses. For massive planets (e.g. Jupiter mass), the short-term perturbations will manifest their significances, and thus the formation and evolution of resonances in the system may not be smooth (adiabatic). One numerical example of the 1:2:4 resonant chain of massive planets can be found in \citet{voyatzis2016}. The system can be captured into resonances through chaotic (irregular) processes and finally attain configurations of high eccentricities. 
Last but not least, in a compact planetary system, the tidal effect may modify the planets' semimajor axes and eccentricities in an adiabatic way, thus it might play an important role in shaping the configuration of a resonant chain. As a matter of fact, the most well-known resonant chain, the literal `Laplace resonance' in the three Jovian moons is believed to have formed through the tidal evolution \citep[see e.g.][]{Malhotra1991}.

\section{Conclusion}

In this paper, we generalize the semi-analytical model of MMR between two planets to the resonant chain in the triple-planetary systems. By constructing and appropriately averaging the Hamiltonian that describes the dynamics of three planets in a resonant chain, we numerically compute the periodic orbits in the resonant chain. The periodic orbits in the resonance (a.k.a. apsidal corotational resonance, ACR solutions) are the stationary solutions to the averaged Hamiltonian equations and correspond to the extrema of the Hamiltonian function, which can be searched out numerically in the phase space. 

Adopt the 1:2:3 resonant chain in the Kepler-51 system as an example, we calculated these ACR solutions and then analysed their stabilities by checking the topologies of the Hamiltonian on different representative planes. 

The stable symmetric ACRs we found are in good agreement with the stable symmetric periodic orbits obtained previously through the continuation method \citep{antoniadou2022}. The symmetric ACR solutions lose their stabilities as the eccentricities increase, and they might bifurcate to asymmetric ACRs, which create stable configurations in the high-eccentricity region.

To investigate the orbital dynamics of the resonant chain near the stable ACRs, we construct dynamical maps on the representative $(e_1,e_2)$ plane and associate the orbital features with the Hamiltonian topology on the same plane. The dynamical maps show that the domain near the stable ACRs is always characterised by regular motion of quasi-periodic libration. The consistence between the topology and dynamical map simplifies the calculation of stability of planetary systems in the resonant chain.

The behaviour of resonant angles in the resonant chain might be so complicated that we cannot always determine the resonance state simply by checking whether the corresponding critical angle is librating or not. In some cases, the short-term perturbations muddle the libration of resonance angles even when the system is dominated by the resonance.  Thus, we calculate the deviation ($\chi^2$) of a resonant angle from a uniform distribution, and empirically define the $\chi^2$ criterion to distinguish the resonance.

In the vicinity of the ACRs,  we found two types of stable motion in the resonant chain. One is the periodic or quasi-periodic motion at/around the ACR solution, which is characterised by concurrent libration of the two-body resonant angles and the three-body Laplace angle. And the other is the quasi-periodic motion, in which only the two-body resonant angles but not the Laplace angle librate. We denote the latter motion as ``joint two-body MMRs''. Another difference between these two types of motion is that the centre (in terms of eccentricities and critical angles) of libration for the former is just the ACR solution while for the latter the ACR is not even ``reached'' by the libration. According to our experiences, in both cases, the two-body MMRs between planet pairs are the essential mechanism that stabilises the planetary resonant chain, and the pure three-body resonance contributes little to the stability. Last but not least, we note that the two-body MMRs in a resonant chain is not the same as the usual two-body resonance that is isolated from other planets in a planetary system. 

Just like the usual two-body MMR, the resonant chain could also be formed via convergent migration of planets. Setting carefully the initial conditions and migration velocities for each planet, we verify the formation and evolution of resonant chain. Starting from near circular orbits, the planets can be trapped to form the resonant chain. And all these systems are either in the quasi-periodic orbits around the ACR solution or in the joint two-body MMRs as predicted by our model. As the migration continues, the system will evolve to high-eccentricity region along the ACR solutions. 

In the high-eccentricity region, there exist some ACR solutions that are not arising from near circular orbits. These solutions can also be determined using the methods introduced in this paper, but these resonant configuration generally cannot be achieved through the convergent migration. In addition, such resonant chain of high eccentricity has relatively poor stability.

The configuration and stability of a resonant chain may be affected by the planetary mass. Not only the locations of the ACRs change, but also the stability region shrinks as the individual planet mass increases. In fact, the more massive the planets, the stronger the disturbances between planets. Thus the resonant chains, especially those compact ones, are more likely to occur in planetary systems of sub-Neptunes or even lighter planets.   

Our methods introduced in this paper can be applied to any resonant chains. We have shown briefly our calculations of the ACR solutions and formation for the 2:3:5, 3:5:7, 2:3:4, 3:4:6, and 12:15:20 resonant chains. Diversified orbital configurations are found in different systems. Particularly, both symmetric ACR solutions (e.g. in the 1:2:3, 2:3:5, 3:5:7 resonant chains) and asymmetric ones (e.g. in 2:3:4, 2:3:6, 3:4:6, and 12:15:20) emerge from circular orbits. Our methods can also be applied to longer resonant chains consisting of more than 3 planets. As an example we calculated in Appendix~\ref{appendix2} the periodic solutions of the four-planet 3:4:6:8 resonant chain in Kepler-223, and show the resonant configurations as the function of scaled AMD $\delta$.

\begin{acknowledgements}
This work has been supported by the National Key R\&D Program of China (2019YFA0706601) and National Natural Science Foundation of China (NSFC, Grants No.12373081, No.12150009 \& No.11933001). We also thank the supports from the science research grant from the China Manned Space Project with NO.CMS-CSST-2021-B08.  
\end{acknowledgements}

 \bibliographystyle{aa} % style aa.bst
 \bibliography{ResChain.bib} % your references Yourfile.bib

\begin{appendix}
\section{Degeneracy of the resonance}
\label{appendix1}
The literal expansion of perturbation part $\mathcal{H}_1$ of the Hamiltonian in Eq.\eqref{eq:hamil} takes the form 
\begin{equation}
\label{eq:a1}
\begin{aligned}
\mathcal{H}_1&=\sum_{1\le i<j\le n} \frac{p_ip_j}{m_0}- \mathcal{G}\frac{m_im_j}{r_{ij}}= \sum_{1\le i<j\le n}H_{1,ij}\left(\theta_{ij},\Delta\varpi_{ij},Q_{ij}\right) \\
&=\sum_{1\le i<j\le n}\sum_{l,m,k}H_{1,ij}^{lmk} \cos\left(l\theta_{ij}+m\Delta\varpi_{ij}+kQ_{ij}\right),
\end{aligned}
\end{equation}
where $l,m,k\in\mathcal{Z}$,  $\theta_{ij} = k_{ji} \lambda_j - k_{ij} \lambda_i - (k_{ji} - k_{ij}) \varpi_{i}, \Delta\varpi_{ij} = \varpi_i - \varpi_j, Q_{ij}= \lambda_i - \lambda_j$. After the numerically averaging process, terms with $k\neq 0$ in Eq.\eqref{eq:a1} will not appear in the averaged Hamiltonian $\bar{\mathcal{H}}$. Thus, the literal expansion of $\bar{\mathcal{H}_1}$ can be reduced to
\begin{equation}
\begin{aligned}
\bar{\mathcal{H}}_1&=\sum_{1\le i<j\le n}\frac{1}{T_{Q_{ij}}}
\int_0^{T_{Q_{ij}}}\left(\frac{p_ip_j}{m_0}-\mathcal{G}\frac{m_im_j}{r_{ij}}\right)\dif Q_{ij}\\
&=\sum_{1\le i<j\le n}H_{1,ij}\left(\theta_{ij},\Delta\varpi_{ij}\right)\\
&=\sum_{1\le i<j\le n}\sum_{l,m}H_{1,ij}^{lm}\cos\left(l\theta_{ij}+m\Delta\varpi_{ij}\right).
\end{aligned}
\end{equation}

There are $n(n-1)$ resonant angles $(\theta_{ij},\Delta\varpi_{ij}),1\le i<j\le n$ appearing in the final expression at first glance. %\textcolor{red}{Thus, two resonant configurations are identical requires that all the  two-body resonant angles are same (Mod[$360^\circ$])  }. 
But in fact, these angles are linearly dependent. If we choose the two-body resonant angles between the adjacent planet pair $\theta_{i,i+1}, \Delta\varpi_{i,i+1}$ to characterise the configuration, due to the resonance between the non-adjacent planets in the resonant chain, the coefficients of the angles $\theta_{i,i+1}, \Delta\varpi_{i,i+1}$ may be fractions rather than integers. As a result, the Hamiltonian may not be $2\pi$-periodic for all the resonant angles, e.g. the resonant configuration characterised by $(\theta_{12}, \Delta\varpi_{12}, \cdots, \theta_{n-1,n}, \Delta\varpi_{n-1,n})$ may not be identical to the one by $(\theta_{12}+2\pi, \Delta\varpi_{12}, \cdots, \theta_{n-1,n}, \Delta\varpi_{n-1,n})$.
For instance, the resonant angles for the 2:3:4 resonant chain are
\begin{equation}
	\left\{
	\begin{aligned}
		& \theta_{12}=3\lambda_2-2\lambda_1-\varpi_1, & \Delta\varpi_{12}=\varpi_1-\varpi_2, \\
		& \theta_{23}=4\lambda_3-3\lambda_2-\varpi_2, & \Delta\varpi_{23}=\varpi_2-\varpi_3.
	\end{aligned} \right.
\end{equation}
And the resonant angles between planets $m_1, m_3$ 
\begin{equation}
	\theta_{13}=2\lambda_3-\lambda_1-\varpi_1, ~~~\Delta\varpi_{13}=\varpi_1-\varpi_3
\end{equation}
can be linearly expressed by  $\theta_{12}, \Delta\varpi_{12}, \theta_{23}, \Delta\varpi_{23}$ as
\begin{equation}
	\theta_{13}=\frac{1}{2}\left(\theta_{12}+\theta_{23}-\Delta\varpi_{12}\right), ~~~\Delta\varpi_{13}=\Delta\varpi_{12}+\Delta\varpi_{23}.
\end{equation}
In this case, $\theta_{12}, \theta_{23}$, and $\Delta\varpi_{12}$ are not $2\pi$-periodic angles, but $4\pi$-periodic. 

\begin{figure}
   \centering
   \includegraphics[width=\hsize]{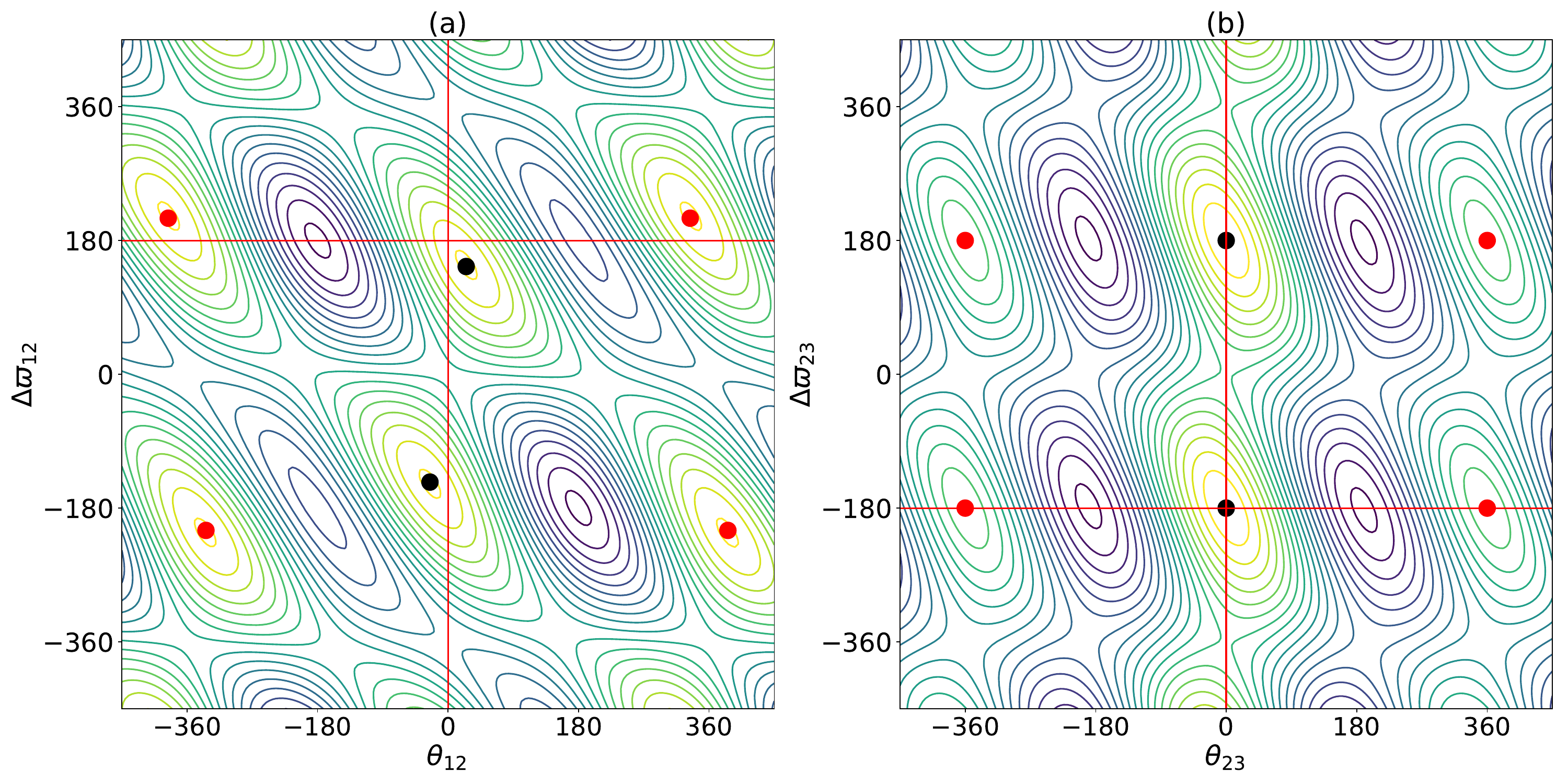}
      \caption{Topology of the averaged Hamiltonian for the 2:3:4 resonant chain on (a) $(\theta_{12},\Delta\varpi_{12})$ plane and (b) $(\theta_{23},\Delta\varpi_{23})$ plane. The black and red solid circles indicate the location of the stable and unstable ACRs, respectively.   }
         \label{fig:topo234}
   \end{figure}

Generally, we define the degeneracy of an angle $\phi$ ($\phi$ stands for any one of $\theta_{ij}, \Delta\varpi_{ij}$) as the smallest positive integer $q$ which satisfies the following condition:
\begin{equation}
\begin{aligned}
&  \bar{\mathcal{H}}_1\left(\theta_{12}, \cdots, \phi, \cdots, \theta_{n-1,n}, \Delta\varpi_{n-1,n}\right)  \\
 & ~~~ =\bar{\mathcal{H}}_1\left(\theta_{12},\cdots,\phi+2q\pi,\cdots,\theta_{n-1,n},\Delta\varpi_{n-1,n}\right).
\end{aligned}
\end{equation}
Therefore, for the 2:3:4 resonant chain, the degeneracy of the angles $(\theta_{12},\Delta\varpi_{12},\theta_{23},\Delta\varpi_{23})$ are $(2,2,2,1)$.
The topology of the averaged Hamiltonian with $e_1= e_2= e_3=10^{-4}$ on angular plane is shown in Fig.~\ref{fig:topo234}, in which the periodicity confirms the degeneracy of resonant angles. From the plots in Fig.~\ref{fig:topo234}, we also note that no stable symmetric configuration exists for near circular orbit.

%The symmetric condition for the three-planet system indicates that all the two-body MMRs must be in symmetric configuration in a symmetric three-body resonant chain. For 2:3:4 resonant chain, although the configuration $(\theta_{12},\Delta\varpi_{12},\theta_{23},\Delta\varpi_{23})=(0,\pi,0,\pi)$ is symmetric for two-body MMR between inner pair $m_1,m_2$ and outer pair $m_2,m_3$, the resonant angles of non-consective $m_1,m_3$ $\theta_{13}=-\pi/2$ which is different from $0,\pi$ indicates that such configuration is asymmetric. 
%We note that the degenercy $q$ also depends on the choice of the resonant angles.

According to the form of averaged Hamiltonian, the asymmetric ACRs always appear in pair. If an ACR locates at $(\theta_{12}, \Delta\varpi_{12}, \cdots,\theta_{n-1,n}, \Delta\varpi_{n-1,n})$, one must find an ACR at $(-\theta_{12},-\Delta\varpi_{12},\cdots,-\theta_{n-1,n},-\Delta\varpi_{n-1,n})$.

\section{Extension to four-planet resonant chain: Kepler-223 as an example}
\label{appendix2}
The methods introduced in this paper can be extended to resonant chains of more planets. As an example, we briefly present our analyses on the four-planet system Kepler-223, in which the orbital periods are close to $T_1:T_2:T_3:T_4 =3:4:6:8$.  
The following resonant angles are adopted to characterise the resonant configuration.
\begin{equation}
\left\{
\begin{aligned}
& \theta_{12}=4\lambda_2-3\lambda_1-\varpi_1, & \Delta\varpi_{12}=\varpi_1-\varpi_2, \\
& \theta_{23}=3\lambda_3-2\lambda_2-\varpi_2, & \Delta\varpi_{23}=\varpi_2-\varpi_3, \\
& \theta_{34}=4\lambda_4-3\lambda_3-\varpi_3, & \Delta\varpi_{34}=\varpi_3-\varpi_4. \\
\end{aligned} \right.
\end{equation}
%As shown in Appendix.\ref{appendix1},The final Hamiltonian are not $2\pi$ periodic all resonant angles. It's important to find out the degeneracy of the resonant angles first.
After some algebra, we obtain the degeneracy of these resonant angles as $(3,3,6,2,2,1)$, which means that the searching for resonant configuration will be done in $\mathcal{T}^6=[0,6\pi]\times[0,6\pi]\times[0,12\pi]\times[0,4\pi]\times[0,4\pi]\times[0,2\pi]$.
The numerical averaging over the synodic angle $Q$ reduces the final Hamiltonian to 6 DoF with two free parameters, the spacing parameter $L$ and the total angular momentum $G$,
\begin{equation}
\begin{aligned}
&L=\sum_{i=1}^4\frac{\Lambda_i}{k_i}, & G=\sum_{i=1}^4\Lambda_i-\Gamma_i,
\end{aligned}
\end{equation}
where $\Lambda_i$ and $\Gamma_i$ are the Poincar\'e variables as in Eq.\eqref{eq:poinvar}.

 \begin{figure*}
   \resizebox{\hsize}{!}
            {\includegraphics[width=12cm]{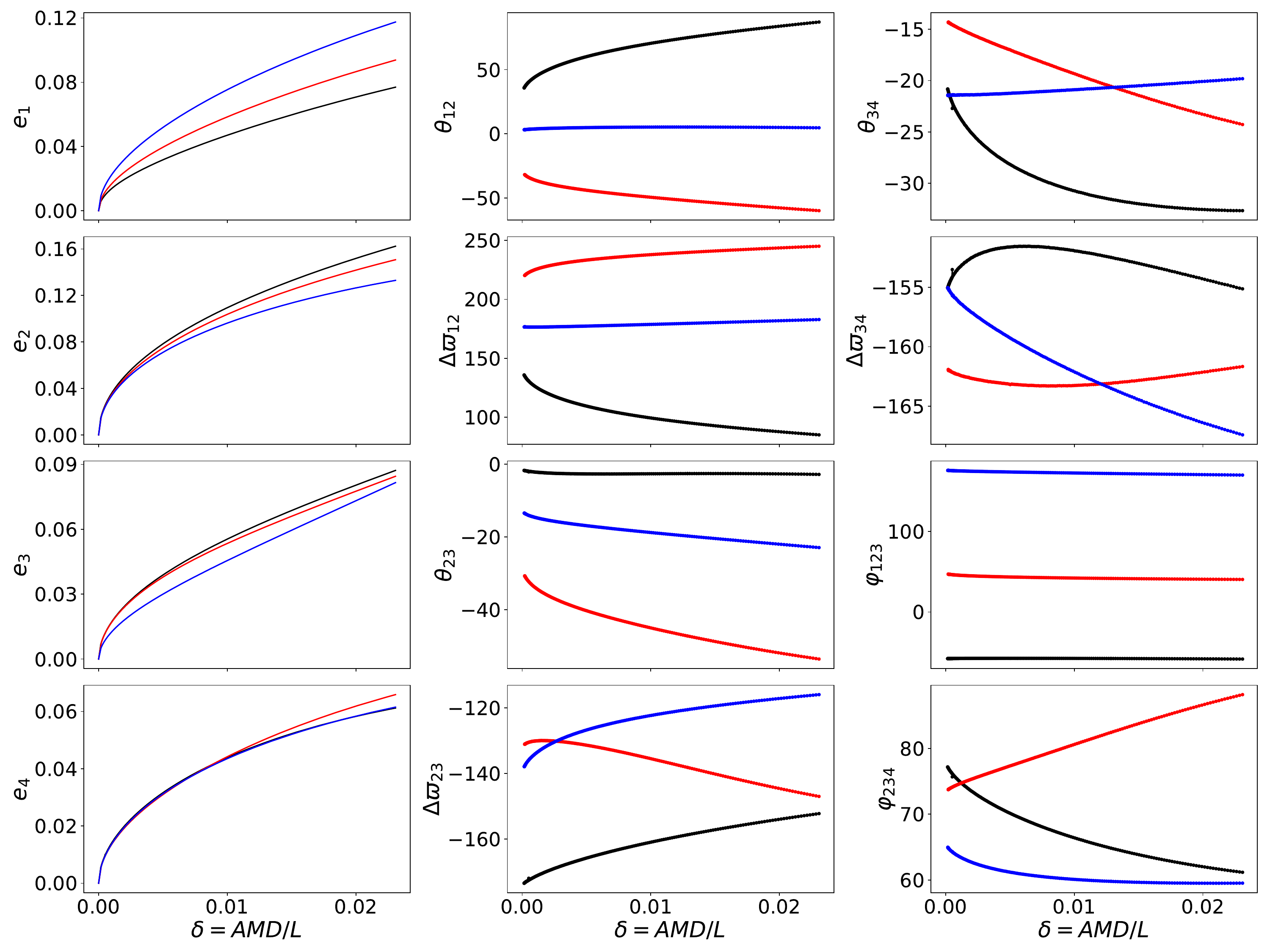}}
      \caption{Locations and resonant configurations of six families of stable ACR for the 4-planet resonant chain in Kepler-223. The variations of the eccentricities, the resonant angles, and the three-planet Laplace angles, are plotted against the scaled AMD ($\delta$). Each family together with its symmetric opposite (see text) is represented by a specific colour in all panels. }
         \label{fig:kepler223}
   \end{figure*}

We only consider here the stable ACRs that correspond to libration in the resonance. Taking the mass parameters of Kepler-223 from \citet{mills2016}, and following the geometric method described in the main text, we found 6 ($=2\times3$) families of stable ACR. In Fig.~\ref{fig:kepler223} we summarise the locations and configurations of these ACR families as the function of the scaled AMD, defined in the same way as Eq.\eqref{eq:delta}, $\delta=\sum_{i=1}^4\Gamma_i/L$. We also plot the Laplace angles $\varphi_{123}= \lambda_1+ \lambda_3- 2\lambda_2, \varphi_{234}= \lambda_2+ 2\lambda_4- 3\lambda_3$ for adjacent three planets. It's worth noting that the six asymmetric solutions we obtained actually take only three configurations, and each configuration contains two branches that are symmetric with respect to each other. That's why only three lines in each panel for eccentricities in Fig.~\ref{fig:kepler223} can be seen.

Our results obtained from the averaged Hamiltonian are consistent with the analytical results \citep{delisle2017}. Since the analytical model in \citet{delisle2017} is truncated at the first order of eccentricity, it gives constant critical angles at the ACR solutions. However, our method provides more precise results. Some resonant angles $\theta_{ij}, \Delta\varpi_{ij}$ are found to experience large variations along the ACR solutions, while the three-planet Laplace angle $\varphi_{123}$ remains almost constant.

\end{appendix}

\end{document}